\documentclass[twocolumn,letterpaper,aps,prc,longbibliography,superscriptaddress,showpacs,nofootinbib,floatfix]{revtex4-1} 

\usepackage{graphicx}	
\usepackage{xspace}     
\newcommand{\pt}{\mbox{$p_T$}\xspace}

\newcommand{\Npart}{\mbox{$N_{\rm part}$}\xspace}
\newcommand{\Ncoll}{\mbox{$N_{\rm coll}$}\xspace}
\newcommand{\Npartnospace}{\mbox{$N_{\rm part}$}}
\newcommand{\Ncollnospace}{\mbox{$N_{\rm coll}$}}

\newcommand{\sqsn}{\mbox{$\sqrt{s_{_{NN}}}$}\xspace}
\newcommand{\sqsntwo}{\mbox{$\sqrt{s_{_{NN}}}=200$~GeV}\xspace}
\newcommand{\pp}{\mbox{$p$$+$$p$}\xspace}
\newcommand{\dau}{\mbox{$d$$+$Au}\xspace}

\newcommand{\auau}{\mbox{Au$+$Au}\xspace}

\newcommand{\ee}{$e^+e^-$ }

\newcommand{\mee}{$m_{ee}$ }

\newcommand{\ccbar}{$c\overline{c} $ }

\newcommand{\sqnr}{$\sqrt{s_{_{NN}}}$~=~200~GeV }

\begin{document}

\title{Dielectron production in Au$+$Au collisions 
       at $\sqrt{s_{NN}}$=200~GeV}

\newcommand{\abilene}{Abilene Christian University, Abilene, Texas 79699, USA}
\newcommand{\augie}{Department of Physics, Augustana University, Sioux Falls, South Dakota 57197, USA}
\newcommand{\banaras}{Department of Physics, Banaras Hindu University, Varanasi 221005, India}
\newcommand{\barc}{Bhabha Atomic Research Centre, Bombay 400 085, India}
\newcommand{\baruch}{Baruch College, City University of New York, New York, New York, 10010 USA}
\newcommand{\bnlcoll}{Collider-Accelerator Department, Brookhaven National Laboratory, Upton, New York 11973-5000, USA}
\newcommand{\bnlphys}{Physics Department, Brookhaven National Laboratory, Upton, New York 11973-5000, USA}
\newcommand{\caucr}{University of California-Riverside, Riverside, California 92521, USA}
\newcommand{\charlesczech}{Charles University, Ovocn\'{y} trh 5, Praha 1, 116 36, Prague, Czech Republic}
\newcommand{\chonbuk}{Chonbuk National University, Jeonju, 561-756, Korea}
\newcommand{\ciae}{Science and Technology on Nuclear Data Laboratory, China Institute of Atomic Energy, Beijing 102413, P.~R.~China}
\newcommand{\cns}{Center for Nuclear Study, Graduate School of Science, University of Tokyo, 7-3-1 Hongo, Bunkyo, Tokyo 113-0033, Japan}
\newcommand{\colorado}{University of Colorado, Boulder, Colorado 80309, USA}
\newcommand{\columbia}{Columbia University, New York, New York 10027 and Nevis Laboratories, Irvington, New York 10533, USA}
\newcommand{\czechtech}{Czech Technical University, Zikova 4, 166 36 Prague 6, Czech Republic}
\newcommand{\dapnia}{Dapnia, CEA Saclay, F-91191, Gif-sur-Yvette, France}
\newcommand{\debrecen}{Debrecen University, H-4010 Debrecen, Egyetem t{\'e}r 1, Hungary}
\newcommand{\elte}{ELTE, E{\"o}tv{\"o}s Lor{\'a}nd University, H-1117 Budapest, P{\'a}zm{\'a}ny P.~s.~1/A, Hungary}
\newcommand{\ewha}{Ewha Womans University, Seoul 120-750, Korea}
\newcommand{\fsu}{Florida State University, Tallahassee, Florida 32306, USA}
\newcommand{\gsu}{Georgia State University, Atlanta, Georgia 30303, USA}
\newcommand{\hanyang}{Hanyang University, Seoul 133-792, Korea}
\newcommand{\hiroshima}{Hiroshima University, Kagamiyama, Higashi-Hiroshima 739-8526, Japan}
\newcommand{\howard}{Department of Physics and Astronomy, Howard University, Washington, DC 20059, USA}
\newcommand{\ihepprot}{IHEP Protvino, State Research Center of Russian Federation, Institute for High Energy Physics, Protvino, 142281, Russia}
\newcommand{\illuiuc}{University of Illinois at Urbana-Champaign, Urbana, Illinois 61801, USA}
\newcommand{\inrras}{Institute for Nuclear Research of the Russian Academy of Sciences, prospekt 60-letiya Oktyabrya 7a, Moscow 117312, Russia}
\newcommand{\instpasczech}{Institute of Physics, Academy of Sciences of the Czech Republic, Na Slovance 2, 182 21 Prague 8, Czech Republic}
\newcommand{\isu}{Iowa State University, Ames, Iowa 50011, USA}
\newcommand{\jaea}{Advanced Science Research Center, Japan Atomic Energy Agency, 2-4 Shirakata Shirane, Tokai-mura, Naka-gun, Ibaraki-ken 319-1195, Japan}
\newcommand{\jyvaskyla}{Helsinki Institute of Physics and University of Jyv{\"a}skyl{\"a}, P.O.Box 35, FI-40014 Jyv{\"a}skyl{\"a}, Finland}
\newcommand{\karoly}{K\'aroly R\'oberts University College, H-3200 Gy\"ngy\"os, M\'atrai\'ut 36, Hungary}
\newcommand{\kek}{KEK, High Energy Accelerator Research Organization, Tsukuba, Ibaraki 305-0801, Japan}
\newcommand{\korea}{Korea University, Seoul, 136-701, Korea}
\newcommand{\kurchatov}{National Research Center ``Kurchatov Institute", Moscow, 123098 Russia}
\newcommand{\kyoto}{Kyoto University, Kyoto 606-8502, Japan}
\newcommand{\labllr}{Laboratoire Leprince-Ringuet, Ecole Polytechnique, CNRS-IN2P3, Route de Saclay, F-91128, Palaiseau, France}
\newcommand{\lahorelums}{Physics Department, Lahore University of Management Sciences, Lahore 54792, Pakistan}
\newcommand{\lawllnl}{Lawrence Livermore National Laboratory, Livermore, California 94550, USA}
\newcommand{\losalamos}{Los Alamos National Laboratory, Los Alamos, New Mexico 87545, USA}
\newcommand{\lpc}{LPC, Universit{\'e} Blaise Pascal, CNRS-IN2P3, Clermont-Fd, 63177 Aubiere Cedex, France}
\newcommand{\lund}{Department of Physics, Lund University, Box 118, SE-221 00 Lund, Sweden}
\newcommand{\maryland}{University of Maryland, College Park, Maryland 20742, USA}
\newcommand{\mass}{Department of Physics, University of Massachusetts, Amherst, Massachusetts 01003-9337, USA}
\newcommand{\michigan}{Department of Physics, University of Michigan, Ann Arbor, Michigan 48109-1040, USA}
\newcommand{\muhlenberg}{Muhlenberg College, Allentown, Pennsylvania 18104-5586, USA}
\newcommand{\myongji}{Myongji University, Yongin, Kyonggido 449-728, Korea}
\newcommand{\nagasaki}{Nagasaki Institute of Applied Science, Nagasaki-shi, Nagasaki 851-0193, Japan}
\newcommand{\nara}{Nara Women's University, Kita-uoya Nishi-machi Nara 630-8506, Japan}
\newcommand{\natmephi}{National Research Nuclear University, MEPhI, Moscow Engineering Physics Institute, Moscow, 115409, Russia}
\newcommand{\newmex}{University of New Mexico, Albuquerque, New Mexico 87131, USA}
\newcommand{\nmsu}{New Mexico State University, Las Cruces, New Mexico 88003, USA}
\newcommand{\ohio}{Department of Physics and Astronomy, Ohio University, Athens, Ohio 45701, USA}
\newcommand{\ornl}{Oak Ridge National Laboratory, Oak Ridge, Tennessee 37831, USA}
\newcommand{\orsay}{IPN-Orsay, Univ. Paris-Sud, CNRS/IN2P3, Université Paris-Saclay, BP1, F-91406, Orsay, France}
\newcommand{\peking}{Peking University, Beijing 100871, P.~R.~China}
\newcommand{\pnpi}{PNPI, Petersburg Nuclear Physics Institute, Gatchina, Leningrad region, 188300, Russia}
\newcommand{\riken}{RIKEN Nishina Center for Accelerator-Based Science, Wako, Saitama 351-0198, Japan}
\newcommand{\rikjrbrc}{RIKEN BNL Research Center, Brookhaven National Laboratory, Upton, New York 11973-5000, USA}
\newcommand{\rikkyo}{Physics Department, Rikkyo University, 3-34-1 Nishi-Ikebukuro, Toshima, Tokyo 171-8501, Japan}
\newcommand{\saispbstu}{Saint Petersburg State Polytechnic University, St.~Petersburg, 195251 Russia}
\newcommand{\saopaulo}{Universidade de S{\~a}o Paulo, Instituto de F\'{\i}sica, Caixa Postal 66318, S{\~a}o Paulo CEP05315-970, Brazil}
\newcommand{\seoulnat}{Department of Physics and Astronomy, Seoul National University, Seoul 151-742, Korea}
\newcommand{\stonybrkc}{Chemistry Department, Stony Brook University, SUNY, Stony Brook, New York 11794-3400, USA}
\newcommand{\stonycrkp}{Department of Physics and Astronomy, Stony Brook University, SUNY, Stony Brook, New York 11794-3800, USA}
\newcommand{\tenn}{University of Tennessee, Knoxville, Tennessee 37996, USA}
\newcommand{\titech}{Department of Physics, Tokyo Institute of Technology, Oh-okayama, Meguro, Tokyo 152-8551, Japan}
\newcommand{\tsukuba}{Center for Integrated Research in Fundamental Science and Engineering, University of Tsukuba, Tsukuba, Ibaraki 305, Japan}
\newcommand{\vandy}{Vanderbilt University, Nashville, Tennessee 37235, USA}
\newcommand{\weizmann}{Weizmann Institute, Rehovot 76100, Israel}
\newcommand{\wigner}{Institute for Particle and Nuclear Physics, Wigner Research Centre for Physics, Hungarian Academy of Sciences (Wigner RCP, RMKI) H-1525 Budapest 114, POBox 49, Budapest, Hungary}
\newcommand{\yonsei}{Yonsei University, IPAP, Seoul 120-749, Korea}
\newcommand{\zagreb}{University of Zagreb, Faculty of Science, Department of Physics, Bijeni\v{c}ka 32, HR-10002 Zagreb, Croatia}
\affiliation{\abilene}
\affiliation{\augie}
\affiliation{\banaras}
\affiliation{\barc}
\affiliation{\baruch}
\affiliation{\bnlcoll}
\affiliation{\bnlphys}
\affiliation{\caucr}
\affiliation{\charlesczech}
\affiliation{\chonbuk}
\affiliation{\ciae}
\affiliation{\cns}
\affiliation{\colorado}
\affiliation{\columbia}
\affiliation{\czechtech}
\affiliation{\dapnia}
\affiliation{\debrecen}
\affiliation{\elte}
\affiliation{\ewha}
\affiliation{\fsu}
\affiliation{\gsu}
\affiliation{\hanyang}
\affiliation{\hiroshima}
\affiliation{\howard}
\affiliation{\ihepprot}
\affiliation{\illuiuc}
\affiliation{\inrras}
\affiliation{\instpasczech}
\affiliation{\isu}
\affiliation{\jaea}
\affiliation{\jyvaskyla}
\affiliation{\karoly}
\affiliation{\kek}
\affiliation{\korea}
\affiliation{\kurchatov}
\affiliation{\kyoto}
\affiliation{\labllr}
\affiliation{\lahorelums}
\affiliation{\lawllnl}
\affiliation{\losalamos}
\affiliation{\lpc}
\affiliation{\lund}
\affiliation{\maryland}
\affiliation{\mass}
\affiliation{\michigan}
\affiliation{\muhlenberg}
\affiliation{\myongji}
\affiliation{\nara}
\affiliation{\nagasaki}
\affiliation{\natmephi}
\affiliation{\newmex}
\affiliation{\nmsu}
\affiliation{\ohio}
\affiliation{\ornl}
\affiliation{\orsay}
\affiliation{\peking}
\affiliation{\pnpi}
\affiliation{\riken}
\affiliation{\rikjrbrc}
\affiliation{\rikkyo}
\affiliation{\saispbstu}
\affiliation{\saopaulo}
\affiliation{\seoulnat}
\affiliation{\stonybrkc}
\affiliation{\stonycrkp}
\affiliation{\tenn}
\affiliation{\titech}
\affiliation{\tsukuba}
\affiliation{\vandy}
\affiliation{\weizmann}
\affiliation{\wigner}
\affiliation{\yonsei}
\affiliation{\zagreb}
\author{A.~Adare} \affiliation{\colorado} 
\author{C.~Aidala} \affiliation{\losalamos} \affiliation{\michigan} 
\author{N.N.~Ajitanand} \affiliation{\stonybrkc} 
\author{Y.~Akiba} \affiliation{\riken} \affiliation{\rikjrbrc} 
\author{R.~Akimoto} \affiliation{\cns} 
\author{J.~Alexander} \affiliation{\stonybrkc} 
\author{M.~Alfred} \affiliation{\howard} 
\author{H.~Al-Ta'ani} \affiliation{\nmsu} 
\author{A.~Angerami} \affiliation{\columbia} 
\author{K.~Aoki} \affiliation{\kek} \affiliation{\riken} 
% ...
\collaboration{PHENIX Collaboration} \noaffiliation

\date{\today}

%------------------------------------------------------------------------------|

\begin{abstract}

%\linenumbers

We present measurements of $e^+e^-$ production at midrapidity in Au$+$Au 
collisions at $\sqrt{s_{_{NN}}}$ = 200~GeV. The invariant yield is studied 
within the PHENIX detector acceptance over a wide range of mass ($m_{ee} 
<$ 5~GeV/$c^2$) and pair transverse momentum ($p_T$ $<$ 5~GeV/$c$), for 
minimum bias and for five centrality classes. The \ee yield is compared to 
the expectations from known sources. In the low-mass region 
($m_{ee}=0.30$--0.76~GeV/$c^2$)  there is an enhancement that increases with 
centrality and is distributed over the entire pair \pt range measured. It 
is significantly smaller than previously reported by the PHENIX experiment 
and amounts to $2.3\pm0.4({\rm stat})\pm0.4({\rm syst})\pm0.2^{\rm model}$ 
or to $1.7\pm0.3({\rm stat})\pm0.3({\rm syst})\pm0.2^{\rm model}$ for 
minimum bias collisions when the open heavy flavor contribution is 
calculated with {\sc pythia} or {\sc mc@nlo}, respectively.  The inclusive 
mass and $p_T$ distributions as well as the centrality dependence are well 
reproduced by model calculations where the enhancement mainly originates 
from the melting of the $\rho$ meson resonance as the system approaches 
chiral symmetry restoration. In the intermediate-mass region ($m_{ee}$ = 
1.2--2.8~GeV/$c^2$), the data hint at a significant contribution in 
addition to the yield from the semileptonic decays of heavy flavor mesons.

\end{abstract}

\maketitle

%%%%%%%%%%%%%%%%%%%%%%%%%%%%%%%%%%%%%%%%%%%%%%%%%  INTRODUCTION
\section{INTRODUCTION} 
   \label{sec:introduction}
   
Dileptons are important diagnostic tools of the Quark Gluon Plasma (QGP) 
formed in ultra-relativistic heavy ion collisions~\cite{Shuryak:1978ij}. 
They are unique observables for their sensitivity to the chiral symmetry 
restoration phase transition expected to take place together with, or at 
similar conditions to, the deconfinement phase 
transition~\cite{Petreczky:2012rq,Dominguez:2012bs}. When chiral symmetry 
is restored, the chiral doublets, such as the $\rho$ and the $a_1$ mesons, 
become degenerate in mass. As the $a_1$ meson is very difficult to observe 
experimentally, the $\rho$ meson is the main observable in this context. 
Due to its very short lifetime ($\tau \sim$ 1.3 fm/$c$), the $\rho$ meson 
quickly decays after its formation and is therefore a sensitive probe of 
the medium where it is formed. The $\rho$ meson is mostly produced close 
to the phase boundary and possible modifications of its spectral function 
in the high temperature and density conditions prevailing there are thus 
imprinted in its decay products. The decay into dileptons, as opposed to 
hadrons, is of particular interest as they escape unaffected by the 
interaction region, thus carrying this information to the detectors.

Dileptons are sensitive to the thermal radiation emitted by the system, 
both the partonic thermal radiation (quark annihilation into virtual 
photons, ${\it q}{\it \overline{q}} \rightarrow \gamma^* \rightarrow 
{\it l}^+ {\it l}^-$) emitted in the early stage of the collisions as well 
as the thermal radiation emitted later in the collision by the hadronic 
system. The main channel of the latter is pion annihilation, mediated 
through vector meson dominance by the $\rho$ meson ($\pi^+\pi^- 
\rightarrow \rho \rightarrow \gamma^* \rightarrow l^+l^-$). Dileptons are 
produced by a variety of sources all along the entire history of the 
collision and it is necessary to know precisely all these sources in 
order 
to single out the interesting signals characteristic of the QGP related to 
chiral symmetry restoration or thermal radiation~\cite{reviews}.

The CERES experiment pioneered the study of dielectrons at the Super 
Proton Synchrotron (SPS). A strong enhancement of low-mass electron pairs 
(\mee $<$ 1~GeV/$c^2$) with respect to the cocktail of expected hadronic 
sources, was found in all nuclear systems studied, in S+Au collisions at 
200~AGeV~\cite{Agakishiev:1995xb}, in Pb+Au collisions at 
158~AGeV~\cite{Agakishiev:1997au,Adamova:2006nu} and in Pb+Au collisions 
at 40~AGeV~\cite{Adamova:2002kf}. The enhancement was confirmed and 
further studied by the high statistics NA60 experiment that measured 
dimuons in In+In collisions at 
160~AGeV~\cite{Arnaldi:2006jq,Arnaldi:2007ru,Arnaldi:2008er,Arnaldi:2008fw}. 
In both experiments, the low-mass dilepton enhancement is explained by 
in-medium modification of the $\rho$ meson spectral 
function~\cite{vanHees:2007th,vanHees:2006ng,Ruppert:2007cr,Dusling:2006yv,Bratkovskaya:2008bf,Linnyk:2011hz}. 
The data rule out the conjectured dropping mass of the $\rho$ meson as the 
system approaches chiral symmetry restoration 
\cite{Brown:1991kk,Brown:1995qt,Li:1995qm}.  Instead, the data are well 
reproduced by a scenario in which the $\rho$ meson copiously produced by 
$\pi^+ \pi^-$ annihilation is broadened by the scattering off baryons in 
the dense hadronic medium. The low-mass dilepton excess is thus identified 
as the thermal radiation signal from the hadron gas phase with a modified 
$\rho$ meson spectral function.  A recent paper shows that in-medium 
modifications of vector and axial vector spectral functions lead to 
degeneracy of the $\rho$ and $a_1$ meson masses providing a direct link 
between the broadening of the $\rho$ meson spectral function and the 
restoration of chiral symmetry~\cite{Hohler:2013eba}.

NA60 found also an excess at higher masses ($m_{l^+l^-}$=1--3~GeV/$c^2$). 
Using precise vertex information this excess was associated with a 
prompt source originating at the vertex, as opposed to semi-leptonic 
decays of D mesons that originate at displaced vertices. The excess can be 
explained as thermal radiation from the 
QGP~\cite{Arnaldi:2006jq,Arnaldi:2007ru,Arnaldi:2008er,Arnaldi:2008fw,Ruppert:2007cr} 
but other interpretations based on hadronic models, similar to those that 
explain the low mass excess~\cite{vanHees:2007th,vanHees:2006ng}, or on 
hadronic rates constrained by chiral symmetry considerations 
\cite{Dusling:2006yv} can also reproduce the data.
 
At the Relativistic Heavy Ion Collider (RHIC), the PHENIX experiment 
reported a strong enhancement of low mass pairs in \auau collisions at 
\sqsn = 200~GeV~\cite{Adare:2009qk}. In the 0\%--10\% most central 
collisions, where the excess is concentrated, the enhancement factor, 
defined as the ratio of the measured yield over the cocktail yield reaches 
an average value of $7.6\pm0.5({\rm stat})\pm1.3({\rm syst})\pm1.5$ (cocktail) in the 
mass range \mee = 0.15--0.75~GeV/$c^2$. All models that successfully 
reproduce the SPS results fail to explain the PHENIX 
data~\cite{Adare:2009qk, Linnyk:2011vx}.

The PHENIX result~\cite{Adare:2009qk} was characterized by a considerable 
hadron contamination of the electron sample and by a small signal to 
background ($S/B$) ratio. In an effort to improve upon this measurement, a 
hadron-blind detector (HBD) was developed and installed in the PHENIX 
experiment~\cite{Kozlov:2003zr, Fraenkel:2005wx,Anderson:2011jw}.  The 
HBD provides additional electron identification, additional hadron 
rejection and improves the signal sensitivity.

In this paper we present dielectron results obtained with the HBD in 2010 
for \auau collisions at \sqsn = 200~GeV. The paper is organized as 
follows. Section \ref{sec:phenix_detector} describes the PHENIX detector 
with special emphasis on the HBD. In Section \ref{sec:analysis} we give a 
detailed account of the various steps of the data analysis including 
electron identification, pair cuts and background subtraction that is the 
crucial step in this analysis. The raw mass spectra, efficiency 
corrections and systematic uncertainties of the data are also discussed in 
this section. Section \ref{sec:cocktail} describes the procedures used to 
calculate the expected dielectron yield from the known hadronic sources. 
The results, including invariant mass spectra, \pt distributions and 
centrality dependence, are presented in Section \ref{sec:results}. In the 
same section, the results are discussed with respect to previously 
published results and compared to available theoretical calculations. A 
summary is given in Section \ref{sec:summary}.

%%%%%%%%%%%%%%%%%%%%%%%%%%%%%%%%%%%%%%%%%%%%%%  PHENIX DETECTOR
\section{PHENIX DETECTOR}
     \label{sec:phenix_detector}
     
\begin{figure}[hbt!]
\includegraphics[width=1.0\linewidth]{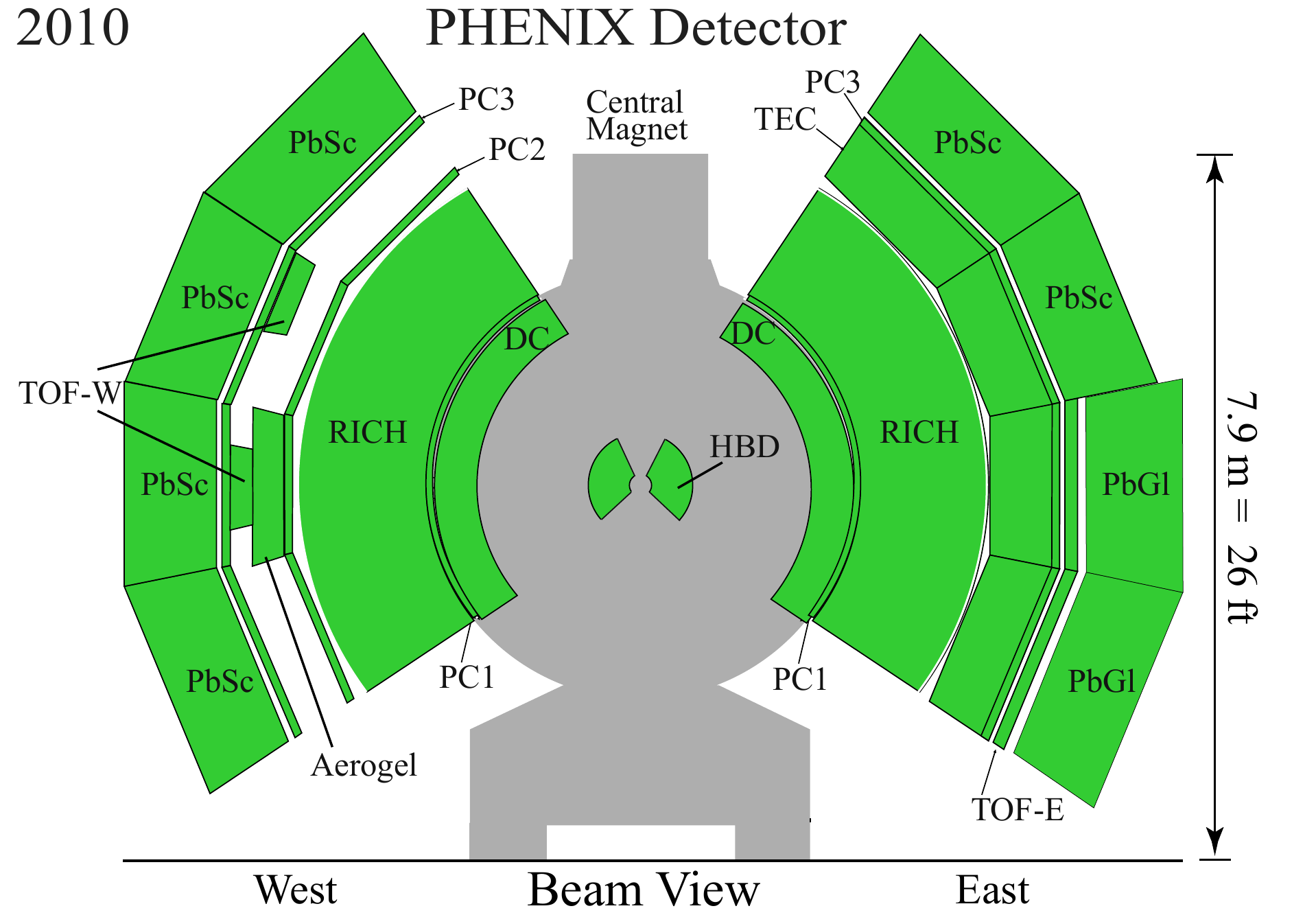}
\caption{(Color online) Beam view (at $z$ = 0) of the PHENIX central arm 
spectrometers during 2010 data taking.}
\label{fig:phenix_2010_beamview} 
\end{figure}

Figure~\ref{fig:phenix_2010_beamview} shows a schematic beam view of the 
PHENIX central arm detector, as used during 2010 data taking. A detailed 
description of the detector, except the HBD, can be found 
in~\cite{Adcox:2003zm}. In this section, we give only a brief description 
of the PHENIX sub-systems relevant for the present analysis: global 
detectors, central magnet, central arm detectors, including drift chambers 
(DC), pad chambers (PC), ring-imaging \v{C}erenkov (RICH) detectors, 
time-of-flight (TOF) detectors and electromagnetic calorimeters (EMCAL) 
and the HBD.

%%%%%%%%%%%%%%%%%%%%%%%%%%%%%%%%%%%%%%%%%%%%%%%%%%%%%%%%%%%% Fig_1

\subsection{Global detectors}
\label{sec:global_detectors}

The measurement of the collision-vertex position, time, and centrality, as 
well as the minimum-bias (MB) trigger, is provided by two beam-beam 
counters (BBC)~\cite{Allen:2003zt}. Each BBC comprises 64 quartz 
\v{C}erenkov counters, located at $\pm$144~cm along the beam axis from the 
center of PHENIX, with 2$\pi$ azimuthal coverage over the pseudorapidity 
interval $3.0<|\eta|<3.9$.  The collision-vertex position along the beam 
direction $z$ is determined from the difference of the average hit time of 
the photomultiplier tubes (PMTs) between the north and the south BBC. The 
$z$-vertex resolution ranges from $\sim$0.5~cm in central Au$+$Au 
collisions to $\sim$2~cm in \pp collisions. The MB trigger requires a 
coincidence between at least two hits in each of the BBC arrays thus 
capturing $92 \pm 3$\% of the total inelastic cross 
section~\cite{Adare:2013esx}.
 
\subsection{Central magnet}
\label{sec:central_magnet}

The PHENIX central magnet comprises two pairs of concentric coils, an 
inner coil pair and an outer coil pair, that can be operated independently 
and create an axial magnetic field parallel to the beam 
axis~\cite{Aronson:2003zn}. The coils are usually operated with current 
flowing in the same direction (the $++$ or $--$ configuration) so that 
their magnetic fields add together. For the dilepton measurement with the 
HBD in the 2010 run, the coils were operated with equal currents flowing 
in opposite directions. In this so called $+-$ configuration, the inner 
coil counteracts the action of the outer coil so that their magnetic 
fields cancel each other, creating an almost field free region in the 
inner space extending from the beam axis out to a radial distance of 
$\sim$60~cm where the inner coil is located (see Fig. 1 of 
Ref.~\cite{Anderson:2011jw}).  The field free region preserves the opening 
angle of \ee pairs and this is an essential pre-requisite for the 
operation of the HBD. The HBD exploits the fact that the opening angle of 
\ee pairs originating from $\gamma$ conversions or from $\pi^0$ Dalitz 
decays is very small. When only one of the two tracks is reconstructed in 
the central arms, the HBD can reject them by applying an opening angle cut 
or a double signal cut on the HBD hits (see Section \ref{sec:hbd}). In 
this configuration however, the total field integral is 
$\int B \cdot dl$ = 0.43 Tm, about 40\% of the value in the $++$ 
configuration.

\subsection{Central arm detectors}
\label{sec:ca_detectors}

PHENIX measurements at midrapidity are made with two central arm 
spectrometers, as shown in Fig.~\ref{fig:phenix_2010_beamview}.
Each central arm covers pseudorapidity $|\eta| <$ 0.35 and 
azimuthal angle $\Delta\phi =\pi/2$.

Charged-particle tracks are reconstructed using hit information from the 
DC, the first layer of PC (PC1) and the collision point along the 
z-direction~\cite{Adcox:2003zp}. The DCs are located outside the magnetic 
field in the radial distance 2.02--2.46~m from the beam axis. They provide 
an accurate measurement of the particle trajectory in the plane 
perpendicular to the beam axis. The PC1s are multiwire proportional 
chambers located just behind the DC at 2.47--2.52 m in radial distance 
from the beam axis~\cite{Adcox:2003en}. They provide a three dimensional 
space point that is used to determine the track origin along the beam 
axis. The transverse momentum ($p_T$) of each particle is determined from 
the bending of its trajectory in the azimuthal direction. The total 
momentum $p$ is determined by combining \pt with the polar angle 
information of PC1 and the vertex position $z$. The reconstructed tracks 
are projected onto the HBD (see next subsection) and onto the central-arm 
detectors that provide electron identification:  RICH, EMCal, and TOF.

The RICH is the primary central-arm detector used for electron 
identification in PHENIX~\cite{Akiba:1999rs}, and is located in the radial 
region of 2.5--4.1~m, just behind PC1.  The RICH uses CO$_2$ as the gas 
radiator at atmospheric pressure, and has a \v{C}erenkov threshold of 
$\gamma$ =~35. This corresponds to a momentum threshold of 18~MeV/$c$ for 
electrons and 4.7~GeV/$c$ for pions. Two spherical mirrors reflect the 
\v{C}erenkov light and focus it onto two arrays of 1280 PMTs each located 
outside the acceptance on each side of the RICH entrance window. The 
average number of hit PMTs per electron track is $\sim$5, and the average 
number of photo-electrons detected is $\sim$10. Below the pion threshold, 
the pion rejection is $\sim$$10^4$ in \pp or low multiplicity collisions. 
However, in high-multiplicity collisions, hadron tracks are misidentified 
as electrons when their trajectory is nearly parallel to that of a genuine 
electron.  This effect limits the $e/\pi$ separation to $\sim$$10^{-3}$ in 
central Au$+$Au collisions and requires special care as described below.
 
The EMCal measures the energy deposited by electrons and their shower 
shape~\cite{Aphecetche:2003zr}. It comprises eight sectors each covering 
$\Delta\phi\approx\pi/8$ in azimuth, where six sectors are made from 
lead-scintillator (PbSc) with an energy resolution 
$4.5\%\oplus8.3\%/\sqrt{E\ {\rm [GeV]}}$ and two are lead-glass (PbGl) 
with an energy resolution $4.3\%\oplus7.7\%/\sqrt{E\ {\rm [GeV]}}$. The 
radial distance from the beam axis is 5.10~m for PbSc and 5.50~m for PbGl 
(see Fig. \ref{fig:phenix_2010_beamview}). The matching of the measured 
energy to the track momentum is used to identify electrons. The latter are 
all relativistic in the accepted momentum range (\pt $>$ 0.2~GeV/$c$), 
hence the energy-to-momentum ratio is close to unity.

To further separate electrons and hadrons we use the time-of-flight 
information from the PbSc part of the EMCal which covers 75\% of the 
acceptance but has a valid time response for 64\% of the acceptance. In 
addition, we use the time-of-flight information from the TOF-east 
detector (TOF-E)~\cite{Aizawa:2003zq} covering an additional 16\% of the 
acceptance. The former has a time resolution of $\sim$450 ps, while the 
latter has a resolution of $\sim$150 ps. The rest of the acceptance, 9\%, 
does not have a usable TOF coverage, because the time resolution of 
$\sim$700 ps provided by PbGl detectors is not sufficient for an effective 
separation of electrons and hadrons.

            \subsection {The Hadron Blind Detector}
            \label{sec:hbd}

The HBD was installed in PHENIX prior to 2010.  A detailed description of 
the concept, construction and performance of the HBD is given in 
Ref.~\cite{Anderson:2011jw}. Only a brief account is given here with 
emphasis on the specific aspects relevant to the present analysis.

\begin{figure}[hbt!]
\includegraphics[width=1.0\linewidth]{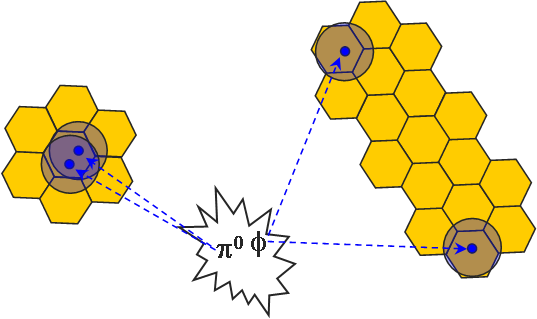}
\caption{(Color online) Sketch illustrating the HBD response to an \ee 
pair from $\pi^0$ Dalitz decay and from a $\phi$ meson decay. The circles 
represent the \v{C}erenkov blobs whereas the hexagons are the hexagonal 
pads of the HBD readout plane.}
\label{fig:hbd_idea}
\end{figure}

\begin{figure}[hbt!]
\includegraphics[width=1.0\linewidth]{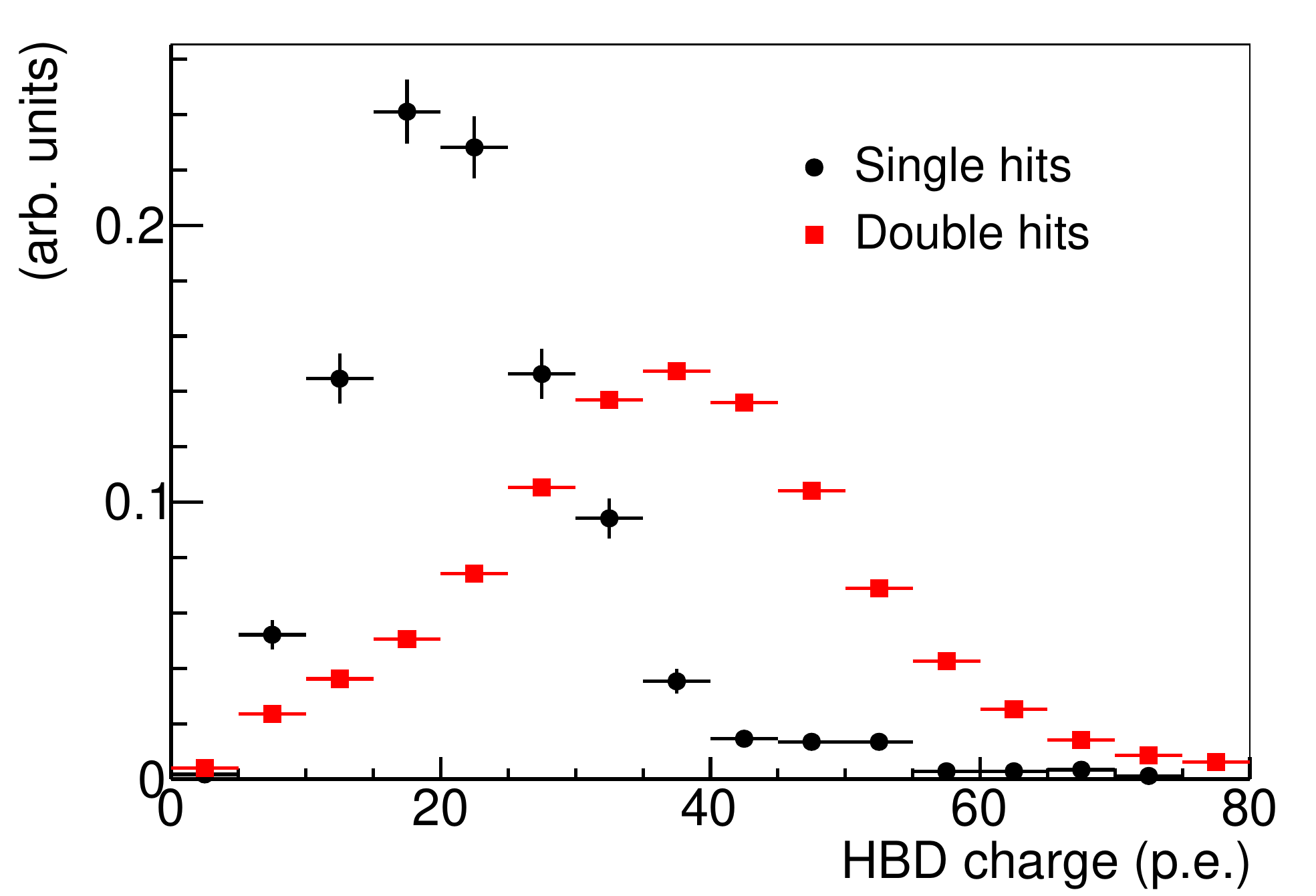}
\caption{(Color online) HBD response to single electron hits and double 
electron hits in the 60\%--92\% centrality bin. The two distributions are normalized to give an integral yield of one.} 
\label{fig:hbd_single_double}
\end{figure}

The HBD provides additional electron identification and additional hadron 
rejection to the central arm detectors. Its main task is to recognize and 
reject $\gamma$ conversions and $\pi^0$ Dalitz decays which are the 
dominant sources of the combinatorial background. Very often, only one of 
the two tracks of an \ee pair from these sources is detected in the 
central arm, whereas the second one is lost because it falls out of the 
acceptance, is curled by the magnetic field or is not detected due to the 
inability to reconstruct low momentum tracks with \pt $<$ 200~MeV/$c$. The 
HBD exploits the fact that most of these pairs have a very small opening 
angle and thus produce two overlapping hits in the HBD, resulting in a 
charge response with an amplitude double the one corresponding to a single 
hit.  Being sensitive to electrons down to very low momentum (see below), 
the HBD can detect both tracks and can effectively reject them by applying 
a double hit cut on the HBD signal.  On the other hand, decays with a large 
opening angle between the electron and positron produce two well separated 
single hits on the HBD pad plane as illustrated in Fig. 
\ref{fig:hbd_idea}. The ability to distinguish single from double hits is 
one of the main performance parameters of the HBD. This is illustrated in 
Fig. \ref{fig:hbd_single_double}, which shows the HBD response to single 
and double electron hits in real data. Single and double hits are selected 
from reconstructed low-mass pairs with large ($>$ 100 mrad) and small ($<$ 
50 mrad) opening angles, respectively.

%%%%%%%%%%%%%%%%%%%%%%%%%%%%%%%%%%%%%%%%%%%%%%%%%%%%%%%%%%%% Fig_2

%%%%%%%%%%%%%%%%%%%%%%%%%%%%%%%%%%%%%%%%%%%%%%%%%%%%%%%%%%%% Fig_3

\begin{figure}[hbt!]
\includegraphics[width=0.8\linewidth]{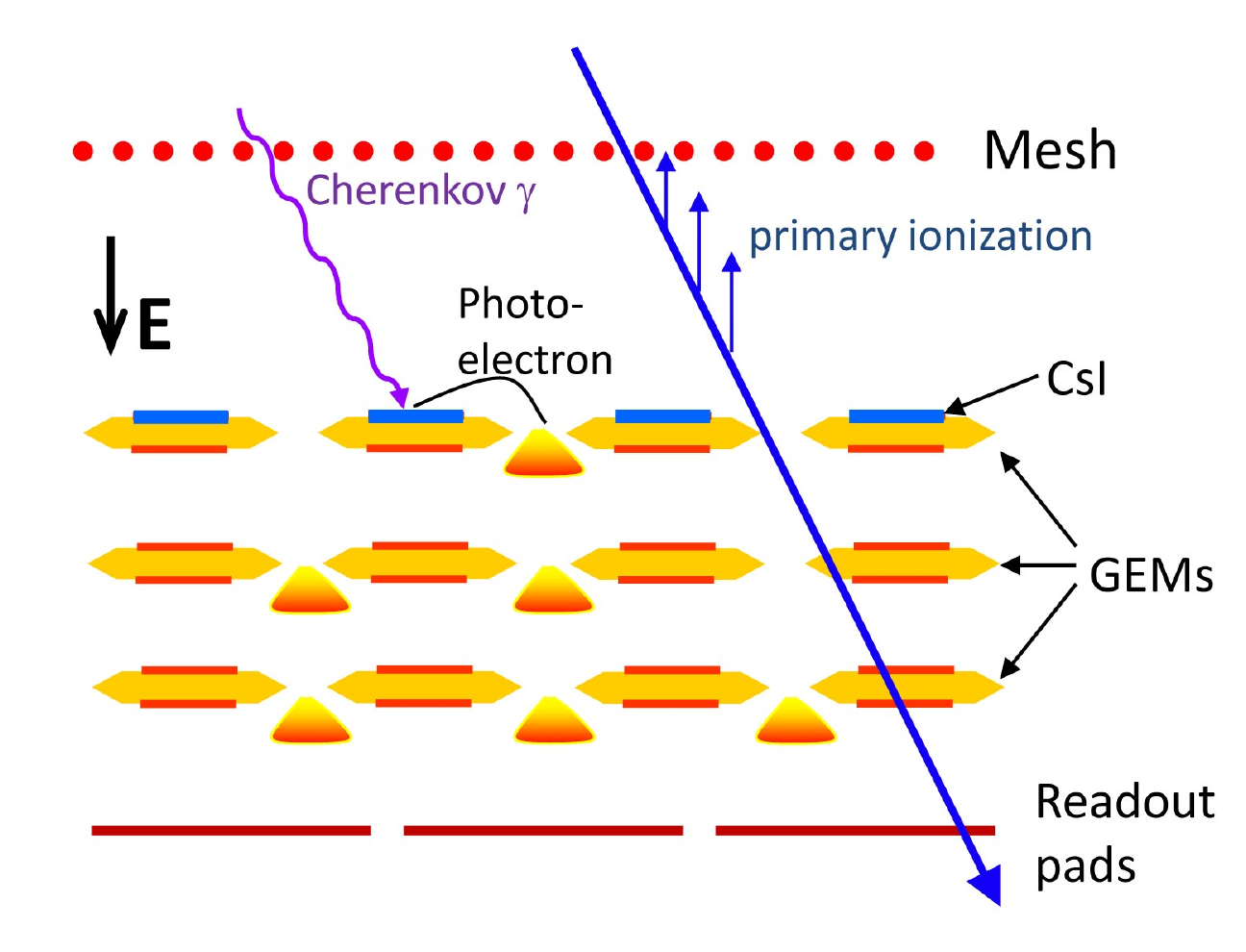}
\caption{(Color online) Triple GEM stack operated in reverse bias mode 
where ionization electrons produced by a charged particle are repelled 
toward the mesh.}
\label{fig:gem_stack} 
\end{figure}

The HBD is a \v{C}erenkov detector. It has a 50~cm long radiator directly 
coupled, in a windowless configuration, to a triple 
gas-electron-multiplier (GEM) detector~\cite{Sauli:1997qp} which has a CsI 
photocathode evaporated on the top face of the upper-most GEM foil and pad 
readout at the bottom of the GEM stack (see Fig.~\ref{fig:gem_stack}). The 
HBD uses pure CF$_4$ at atmospheric pressure that has an average 
\v{C}erenkov threshold of $\gamma$ = 28.8 over the detector bandwidth, 
corresponding to a momentum threshold of $\sim$ 15~MeV/$c$ for electrons 
and $\sim$ 4.0~GeV/$c$ for pions. In this scheme, \v{C}erenkov radiation 
from particles passing through the radiator is directly collected on the 
photocathode forming a circular blob image rather than a ring as in a RICH 
detector. The pad readout plane comprises hexagonal cells with a hexagon 
side of 1.55 cm. One cell subtends an opening angle of approximately 50 
mrad and has an area of 6.2 cm$^2$, comparable to the blob size which has 
a maximum area of 10 cm$^2$. The electron response of the HBD is thus 
typically distributed over a maximum of 3 readout cells and subtends a 
maximum opening angle of 75 mrad.

%%%%%%%%%%%%%%%%%%%%%%%%%%%%%%%%%%%%%%%%%%%%%%%%%%%%%%%%%%%% Fig_4

The hadron blindness property of the HBD is achieved by operating the 
detector in reverse bias mode where the mesh defining the detection volume 
is set at a lower voltage with respect to the CsI 
photocathode~\cite{Kozlov:2003zr, Fraenkel:2005wx} (see Fig. 
\ref{fig:gem_stack}). Consequently, the ionization electrons produced by 
charged particles in the drift region defined by the entrance mesh and the 
photocathode are mostly repelled towards the mesh. Only the ionization 
electrons created in a thin layer of $\sim$$100~\mu$m above the 
photocathode are collected and amplified by the GEM stack leading to a 
very small signal, equivalent to a few p.e., localized in one single cell 
of the pad plane.

The choice of CF$_4$ in a windowless configuration as the common gas for 
the radiator and the detector amplification medium, results in a large 
bandwidth of UV photon sensitivity from 6.2 eV (the threshold of the CsI 
photocathode) up to 11.1 eV (the CF$_4$ cut-off). This translates into an 
average yield of 20 photo-electrons (p.e.) per electron, as shown in Fig. 
\ref{fig:hbd_single_double}, corresponding to a measured figure of merit 
$N_0$ of 330~cm$^{-1}$, very high for a gas \v{C}erenkov 
detector~\cite{Anderson:2011jw}.

The HBD is located close to the interaction vertex, in the field-free 
region, starting immediately after the beam pipe at r = 5~cm and extending 
up to r = 60~cm. The detector comprises two identical arms, each covering 
112.5$^\circ$ in azimuth and $\pm$0.45 units of pseudorapidity. The active 
area of each arm is subdivided into 10 detector modules, 5 along the 
azimuthal axis and 2 along the $z$ axis. With this segmentation, each 
detector module is $\sim 23 \times 27$ cm$^2$ in size. The material budget 
(See Table \ref{tab:hbd_material}) in front of the GEM detectors is 0.62\% 
of a radiation length dominated by the CF$_4$ contribution of 0.56\%. To 
this, one has to add the contribution of the GEM stack, the vessel 
back plane and the front-end electronics attached to the vessel to give a 
total of 2.4\% of a radiation length for the entire detector.

%===================================================== Table_I
\begin{table}[hbt!]
\caption{Material budget of the HBD within the central arm 
acceptance~\protect\cite{Anderson:2011jw}. }
\begin{ruledtabular} \begin{tabular}{clcc}
& Component & Radiation length & \\
&  & (\%) & \\
\hline
& Window (aclar/kapton) & 0.04 & \\
& Gas (CF$_4$) & 0.56 & \\
& GEM stack & 0.42 & \\
& Vessel back plane + front-end electronics & 1.4 & \\
\\
& Total & 2.4 &\\
\end{tabular} \end{ruledtabular}
\label{tab:hbd_material}
\end{table}
 
Good gain calibration is crucial to achieve the best possible separation 
between single and double hits in the HBD. Gain variations occur as a 
function of time due to two main factors: (i) variations of temperature 
and pressure and (ii) charging effects of the GEM foils produce an initial 
rise of the gain after switching on the HV, that can last for several 
hours before stabilizing~\cite{gem_charge_up}.  These gain variations are 
taken into account by performing a gain calibration of each module every 
three minutes during data collection. This is done by exploiting the 
scintillation light produced by charged particles traversing the CF$_4$ 
radiator.  The scintillation signal is easily identified by the 
characteristic exponential shape of single electrons in the HBD pulse 
height distribution of low-multiplicity \auau 
collisions~\cite{Anderson:2011jw}. Furthermore, the average cell charge 
per event was found to slowly decrease by 10\%--15\% over the 10 week 
duration of the run for some of the modules. This is attributed to a slow 
deterioration of the quantum efficiency of the photocathodes. This effect 
was noticed in $\sim$40\% of the modules, the others did not show any sign 
of aging although all photocathodes were produced under identical 
procedures. An additional time dependent correction factor is applied to 
account for this effect.

In high multiplicity \auau collisions, a large amount of scintillation 
light is produced by charged particles traversing the CF$_4$ gas, 
resulting in a large detector occupancy. The number of photoelectrons per 
cell can be as high as $\sim$10 in the most central collisions. This 
underlying event background is subtracted on an event-by-event basis. For 
each event and for each module the average charge per unit area $\langle 
Q\rangle$ is calculated as:
\begin{equation}
\langle Q\rangle = \sum Q_{cell}/ \sum a_{cell},
\label{eq:unit_area}
\end{equation}
where $Q_{cell}$ and $a_{cell}$ are the cell charge and area, 
respectively. The summation is carried out over all the cells of a given 
module, excluding the cells that are matched to an electron track and 
their first neighbors.  The cell charge used for further analysis 
$Q_{cell}^{*}$, is then given by:

\begin{equation}
Q_{cell}^{*} = Q_{cell} - \langle Q\rangle \times a_{cell}
\end{equation}

\begin{figure}[hbt!]
\includegraphics[width=0.49\linewidth]{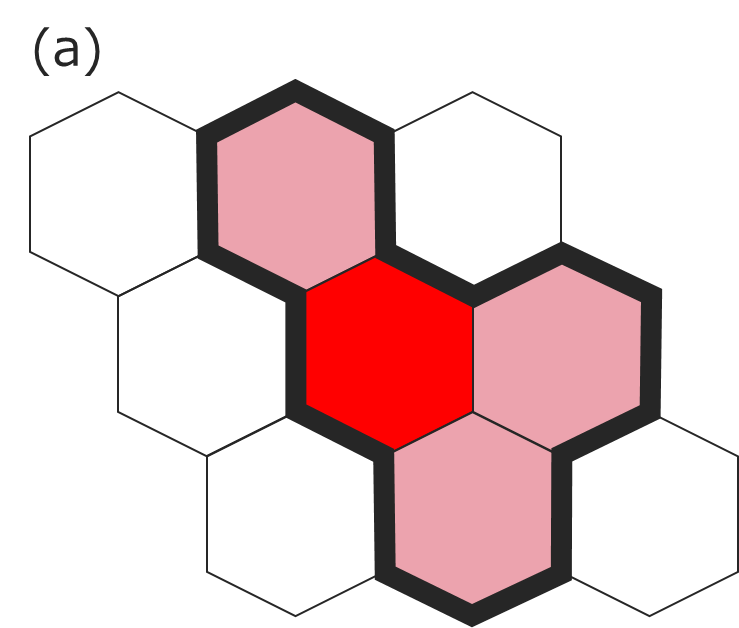}
\includegraphics[width=0.49\linewidth]{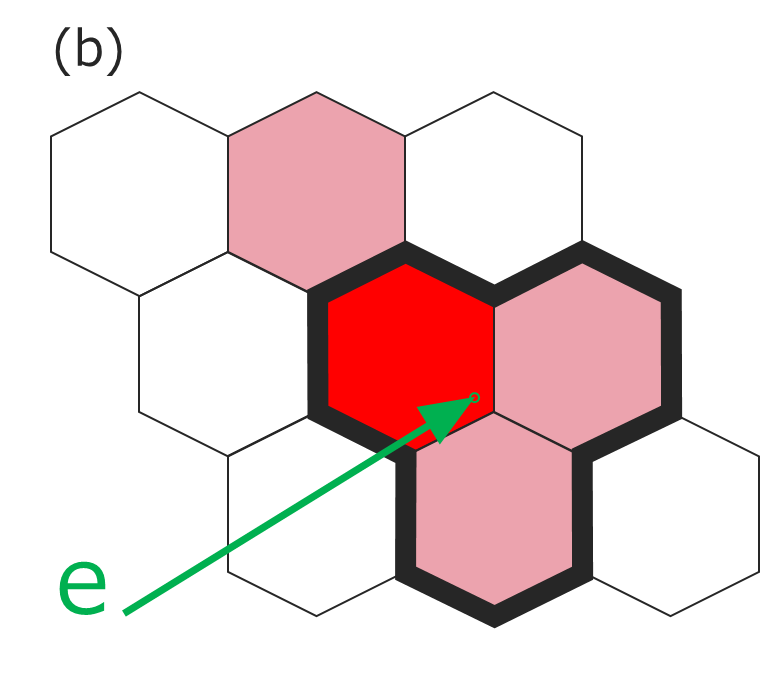}
\caption{(Color online) (a) Stand-alone cluster formed by a seed cell (red) 
and three of its first neighbors resulting in a four cell cluster. Fired 
cells are colored. (b) The same pattern results in a three cell cluster 
with the projection-based algorithm that uses the projection point of an 
electron track onto the pad plane.} 
\label{fig:clusterizer}
\end{figure}

After subtraction of the underlying event charge, two independent 
algorithms are used for the HBD hit recognition. The first is a 
stand-alone algorithm in which a cluster is formed by a seed cell with 
$Q_{cell}^{*}>$ 3 p.e. together with the fired cells (defined as 
$Q_{cell}^{*} >$ 1 p.e.) among its first six neighbors. Such clusters can 
have up to seven cells.  A central arm electron track projected onto the 
HBD readout plane is then matched to the closest cluster. This algorithm 
works very well in \pp or peripheral \auau collisions producing a typical 
single electron response with an average of 20 p.e.. In higher 
multiplicity events, this algorithm yields a higher charge per electron 
and a higher fraction of fake hits as it picks up more charge from the 
fluctuations of the underlying event background.  
Figure~\ref{fig:clusterizer}(a) shows an example of a seed cell and three 
of its first neighbors forming a four cell cluster.

The second algorithm uses the track projection point onto the HBD to form 
a cluster around it. The pointing resolution of a track to HBD is $\sim$3 
mm at $p_{T} \sim 0.5$~GeV/$c$ which is much smaller than the size of a 
pad. The algorithm allows only up to three cells in a cluster, depending 
on the track projection position within the cell. If the track projection 
points to the middle part of the cell, only that cell is used, but if it 
points to the edge of a cell one or two additional neighboring cells are 
summed up in the cluster~\cite{Makek:2011zz}. The same pattern of fired 
cells shown in Fig.~\ref{fig:clusterizer}(a) would result in a three cell 
cluster in the projection-based algorithm as illustrated in 
Fig.~\ref{fig:clusterizer}(b).  The projection-based algorithm results in 
a more precise selection of the true hit, less fake hits and less pick up 
of charge from underlying event fluctuations.

This is especially important in the most central collisions. On the other 
hand, the limited cluster size truncates the charge information, resulting 
in a somewhat reduced efficiency and less power to discriminate between 
single and double hits. Therefore both algorithms are utilized in a 
complementary way, the stand alone providing a higher efficiency and better 
single to double hit separation and the projection-based providing a 
better rejection of fake hits.

%%%%%%%%%%%%%%%%%%%%%%%%%%%%%%%%%%%%%%%%%%%%%%%%%%%%%%%%%%%% Fig_5

           \subsection{Acceptance}
           \label{sec:acceptance}
           
\subsubsection{Acceptance during 2010 run}
\label{sec:acceptance_run10}

As mentioned in Section \ref{sec:central_magnet}, the PHENIX central arm 
magnets were operated in the $+-$ configuration during the 2010 run.  
Compared to the standard $++$ magnetic field configuration of PHENIX, the 
$+-$ configuration has an increased acceptance for low \pt tracks of about 
20\%.

Charged particles are bent in the azimuthal direction, $\phi$, by the 
magnetic field. Because the DC and RICH are needed to reconstruct the tracks 
and select the electron candidates, the azimuthal electron acceptance 
depends on their charge and \pt and on the radial location of each 
detector subsystem. We define the ideal track acceptance of the PHENIX 
detector in the $+-$ field configuration by the following set of 
conditions:

\begin{eqnarray}
\phi_{\rm min} \leq \phi_0 + q \frac{k_{\rm DC}}{p_T} \leq \phi_{\rm max} \\
\phi_{\rm min} \leq \phi_0 + q \frac{k_{\rm RICH}}{p_T} \leq \phi_{\rm max} \\
\theta_{\rm min} \leq \theta_0 \leq \theta_{\rm max}
\end{eqnarray}
for tracks originating at $z$=0 with charge $q$, transverse momentum \pt 
and emission angles $\phi_0$ and $\theta_0$.  $k_{\rm DC}=0.060$ 
rad$\times$GeV/$c$ and $k_{\rm RICH}=0.118$ rad$\times$GeV/$c$ are the 
effective azimuthal bends to the DC and the RICH, respectively.  The  
polar angle boundaries of $\theta_{\rm min}$=1.23 rad and 
$\theta_{\rm max}$=1.92 rad are defined by the PHENIX central-arms 
pseudorapidity acceptance $|\eta| <$ 0.35.  One of the arms covers the 
azimuthal range from $\phi_{\rm min}=~-\frac{3}{16}\pi$ to $\phi_{\rm 
max}=~\frac{5}{16}\pi$ and the other from $\phi_{\rm 
min}=~\frac{11}{16}\pi$ to $\phi_{\rm max}=~\frac{19}{16}\pi$. The results 
shown in Section \ref{sec:results}, indicated as ``in the PHENIX 
acceptance'', refer to the results filtered according to this 
parametrization of the ideal acceptance.

\subsubsection{Fiducial cuts
\label{fiducial_cuts}}

Several fiducial cuts are applied to remove inactive areas of subsystems 
or areas with intermittent response, in order to homogenize the detector 
response over sizable fractions of the run time. Regarding the operation 
of the drift chamber, the entire 200~GeV \auau data set is divided into 
five groups, with fiducial cuts applied to each group separately such that 
inside each group the drift chamber has a stable active area. The 
nonactive DC areas correspond to 19\%--31\% of the total DC acceptance, 
depending on the run group.

Fiducial cuts are also applied to the HBD to exclude tracks pointing to 
one inactive module out of the 20 modules of the HBD. Another fiducial cut 
removes conversion electrons originating from the HBD support structure, 
which are strongly localized in $\phi$ near the edges of the acceptance. 
Other fiducial cuts are applied to remove inactive or low efficiency areas 
in PC1 and EMCal.

In summary, the ideal PHENIX acceptance is reduced by the fiducial cuts by 
an amount that varies between 32\% and 42\%, depending on the run group, 
with an average of 36\% for all selected runs.

\section{ANALYSIS}
     \label{sec:analysis}
     
This section describes the basic steps of the \auau data analysis. It is 
organized as follows. The data set and event selection cuts are presented 
in subsection \ref{sec:data_set_and_event_selection}. Subsection 
\ref{sec:track_reconstruction} describes the track reconstruction. The 
methods applied to identify electrons are presented in detail in 
subsection \ref{sec:electron_identification} and the cuts applied to 
electron pairs are explained in subsection \ref{sec:pair_cuts}. A detailed 
account of the various background sources and their subtraction is 
provided in subsection \ref{sec:background}. Next we present the raw 
spectra and corrections (subsection \ref{sec:raw_spectra}) and discuss the 
systematic uncertainties (subsection \ref{sec:systematic_uncertainties}).  
In the final subsection \ref{sec:cross_checks} we discuss a second 
independent analyses used as a cross check of the main analysis.

\subsection{Data set and event selection
\label{sec:data_set_and_event_selection}}

The \auau collision data at \sqsntwo were collected during 2010. 
Collisions were triggered using the beam-beam counters, with the MB 
trigger condition (see subsection \ref{sec:global_detectors}).

The centrality is determined for each \auau collision from the sum of the 
measured charge in both BBCs combined with a Glauber model of the 
collision~\cite{Miller:2007ri} as described in Ref.~\cite{Adler:2013aqf}. 
In this analysis, the data sample is divided into five centrality classes: 
0\%--10\%, 10\%--20\%, 20\%--40\%, 40\%--60\% and 60\%--92\%. The average 
number of participants $\langle$\Npartnospace$\rangle$ and collisions 
$\langle$\Ncollnospace$\rangle$ together with their systematic 
uncertainties associated with each centrality bin are summarized in 
Table~\ref{tab:glauber}.

%===================================================== Table_II
\begin{table}[hbt!]
\caption{Average values of the number of participants 
$\langle$\Npartnospace$\rangle$ and number of collisions 
$\langle$\Ncollnospace$\rangle$ for \auau collisions at \sqsntwo with the 
corresponding uncertainties. The values are derived from a Glauber 
calculation \protect\cite{Miller:2007ri,Adler:2013aqf}.}
\begin{ruledtabular} \begin{tabular}{clccc}
&Centrality & $\langle$\Npart$\rangle$({\rm syst}) 
& $\langle$\Ncollnospace$\rangle$({\rm syst}) & \\
\hline
& 0\%--10\%   & 324.0 (5.7) & 951.1 (98.6) & \\
& 10\%--20\% & 231.0 (7.3) & 590.1 (61.1)  & \\
& 20\%--40\% & 135.6 (7.0) & 282.4 (28.4)  & \\
& 40\%--60\% & 56.0 (5.3) & 82.6 (9.3)     & \\
& 60\%--92\% & 12.5 (2.6) & 12.1 (3.1)     & \\
& 0\%--92\%   & 106.3 (5.0) & 251.1 (26.7) & \\
\end{tabular} \end{ruledtabular}
\label{tab:glauber}
\end{table}

The data were recorded with an online vertex selection of either 
$\pm$20~cm (narrow vertex) or $\pm$30~cm (wide vertex). The former 
selection was applied to the data recorded at the beginning of each store, 
when the luminosity was relatively high.  For the latter selection, an 
additional-offline vertex cut of $30<z<25$~cm was applied. This 
asymmetric cut is needed to avoid the increased yield of conversion 
electrons originating from the side panels of the HBD. These cuts resulted 
in $1.8 \times 10^9$ events with the narrow-vertex selection, $3.8 \times 
10^9$ events with the wide-vertex selection, and a total of $5.6 \times 
10^9$ MB events.

\subsection{Track reconstruction
\label{sec:track_reconstruction}}

Charged particle tracks are reconstructed in the central arms using the DC 
and PC1~\cite{Adcox:2003zp}.  The procedure assumes that all tracks 
originate from the collision vertex. Each reconstructed track is then 
projected onto the other detectors, RICH, EMCal, TOF and HBD, and the 
projection points are associated to reconstructed hits in these detectors.

After a track is reconstructed, the initial momentum vector of the track 
at the $z$ vertex is calculated. The transverse momentum \pt is determined 
by measuring the angle $\alpha$ between the reconstructed particle 
trajectory and a line that connects the $z$-vertex point to the particle 
trajectory at a reference radius R = 220~cm. The angle $\alpha$ is 
approximately proportional to charge/\pt. In the reverse field 
configuration used in the 2010 run, the momentum resolution is found to be 
1.6\% at \pt = 0.5~GeV/$c$.

\subsection{Electron identification
\label{sec:electron_identification}}

\subsubsection{Detectors and variables used for electron identification
\label{sec:eid_variables}}  

For electron identification, the present analysis uses the HBD along with 
the central arm detectors RICH and EMCal and the time-of-flight 
information from the TOF-E detector and the EMCal. The relevant 
variables for electron identification from these detectors are:

\begin{itemize}
\setlength\itemsep{0mm}

\item[] \textbf{n0:} number of hit PMTs in the RICH in the expected range 
of a \v{C}erenkov ring.

\item[] \textbf{disp:} distance between a track projection and its 
associated ring center in the RICH.

\item[] \textbf{chi2/npe0:} a $\chi^{2}$-like shape variable of the RICH 
ring associated with the track per $npe0$, the number of photoelectrons 
measured in the ring.

\item[] \textbf{emcsdr:} distance between the track projection point onto 
the EMCal and the associated EMCal cluster, measured in units of standard 
deviation of the momentum dependent matching distribution.

\item[] \textbf{prob:} probability that the EMCal cluster is of 
electromagnetic origin, based on the shower shape.

\item[] \textbf{dep:} variable quantifying the energy-momentum matching 
for electrons. It is defined as $dep=\frac{E/p-1}{\sigma_{E/p}}$, where 
$E$ is the energy measured by the EMCal, $p$ is the track momentum and 
$\sigma_{E/p}$ is the momentum-dependent standard deviation of the 
Gaussian-like $E/p$ distribution.

\item[] \textbf{stof(PbSc) and stof(TOF-E):} time-of-flight deviation 
from the one expected for electrons measured by either the EMCal-PbSc or the 
TOF-E detector, converted in units of standard deviation of the 
Gaussian-like time-of-flight distribution.

\item[] \textbf{hbdcharge(P), hbdsize(P):} cluster charge and size from 
the HBD projection-based algorithm.

\item[] \textbf{hbdid:} reduced cluster charge threshold from the 
projection-based algorithm. This is the threshold of the hbdcharge(P) 
variable, that has been tuned to reduce the number of the nongenuine HBD 
hits by a fixed factor. E.g. by requiring hbdid$\geq$10, the number of the 
nongenuine HBD hits is reduced to 1/10 of the initial number. These 
thresholds are tuned depending on event multiplicity and HBD cluster size.

\item[] \textbf{maxpadcharge(S):} charge of the single pad with largest 
charge in the cluster of the stand-alone algorithm.

\item[] \textbf{hbdcharge(S), hbdsize(S):} cluster charge and size from 
the stand-alone algorithm.

\end{itemize}

First, electron candidates are selected from the total sample of tracks 
that contains mostly hadrons. This is accomplished by applying very loose 
cuts such as n0 $>$ 0, which requires at least one fired PMT around the 
track projection in the RICH and $E/p>$ 0.4 which rejects the tracks that 
strongly deviate from the expected $E/p$ of $\sim$~1. The sample of 
electron candidates selected in such a way comprises the signal electrons, 
background electrons (mostly conversions from the HBD back plane), and a 
relatively large number of misidentified hadrons.

\subsubsection{Exclusion of RICH photo-multipliers}
\label{sec:tagging_rich}

\begin{figure}[hbt!]
\includegraphics[width=1.0\linewidth]{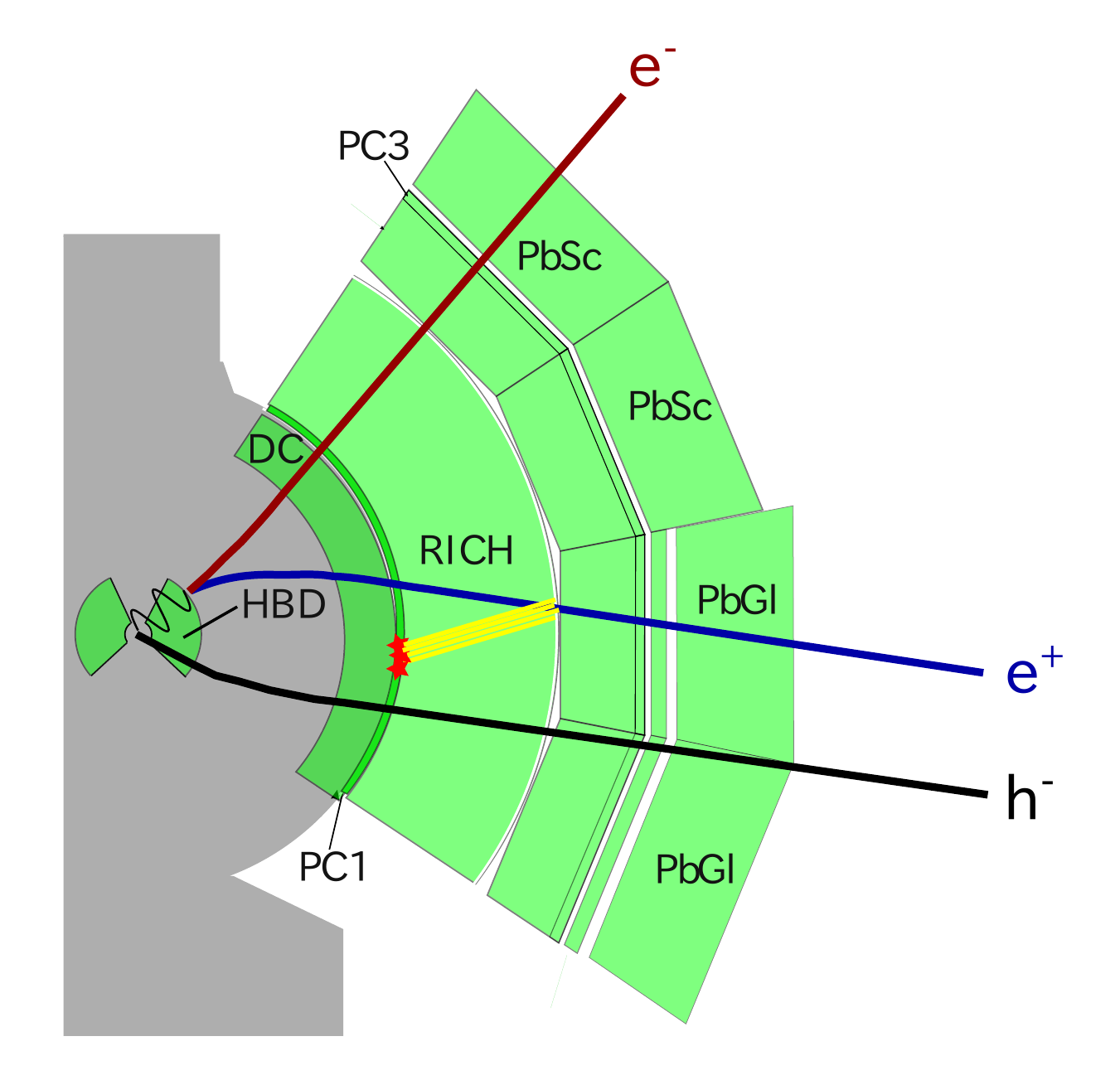}  
\caption{(Color online) Illustration of a case leading to ring sharing in 
the RICH detector. The hadron track parallel to the positron track after 
the magnetic field will be misidentified as an electron.}
\label{fig:eh_view}
\end{figure}

The RICH detector in PHENIX uses spherical mirrors to project the 
\v{C}erenkov light created by electrons in the radiator gas onto the PMT 
plane. As a consequence of this mirror geometry, parallel tracks after the 
field are projected to the same point in the PMT plane. In other words, if 
a hadron track is parallel to an electron track that produces a genuine 
response in the RICH, the hadron will appear to have the same response as 
the electron and thus it will be misidentified as an electron. Figure 
\ref{fig:eh_view} shows a typical example of this ring sharing effect. In 
this example, an electron-positron pair is generated by a photon 
conversion in the HBD backplane. After the magnetic field, a hadron track 
is parallel to the positron track. Consequently, the hadron and the 
positron share the same photomultipliers in the RICH detector and the 
hadron is misidentified as an electron.

%%%%%%%%%%%%%%%%%%%%%%%%%%%%%%%%%%%%%%%%%%%%%%%%%%%%%%%%%%%% Fig_6

This ring sharing effect occurs because the RICH reconstruction algorithm 
allows multiple use of fired PMTs by different tracks. The ring sharing is 
a significant effect.  In the 2010 run, the majority of electrons are 
generated by $\gamma$ conversion in the HBD backplane. Although these 
conversions can successfully be rejected by the HBD, their response in the 
RICH remains and there is some probability that the misidentified hadron 
will also remain in the pool of electron candidates.
 
To reduce PMT sharing by different tracks in the RICH, the 
original RICH algorithm is modified. The PMTs fired by electrons that are 
clearly identified as background electrons, are removed, the ring 
reconstruction algorithm is re-applied and new n0, npe0, disp, $\chi^2$ 
variables are derived. These background electrons are mainly conversion 
electrons from the HBD backplane, electron tracks pointing outside the HBD 
acceptance, electrons produced by conversion on the HBD support structure 
or low \pt electrons with \pt $<$ 200~MeV/$c$.

\subsubsection{The neural networks}
\label{sec:neural_networks}

After the initial rejection of nonsignal electrons and the reduction of 
the ring sharing effect, the sample of electron candidates is still highly 
contaminated by background electrons and misidentified hadrons.  A 
standard procedure to increase the purity of the electron sample would be 
to apply a sequence of one-dimensional cuts on all or some of the fourteen 
variables listed above. However, such a procedure results in a large 
efficiency loss that becomes significant in the \ee pair analysis where 
the pair efficiency is approximately equal to the single track efficiency 
squared. In this analysis we implement instead a multivariate approach 
that is based on the neural network package TMultilayerPerceptron from 
{\sc root}~\cite{neural_network}.

The neural network comprises three layers: the input layer, the hidden 
layer and the output layer. The input layer is composed of all the input 
variables normalized to have their values between 0 and 1. The hidden 
layer comprises a selected number of neurons and the output layer 
comprises a single output variable. The number of neurons in the hidden 
layer determines the ability of the neural network to distinguish between 
the signal and the background, but this ability saturates with increasing 
number of neurons. For each neural network, we make sure that the number 
of neurons is sufficiently large to provide the best possible performance, 
typically 10--15 neurons.  In addition, we make sure that a sufficient 
number of tracks is selected for the training sample, such that the 
performance of the neural network does not depend on the training 
statistics.The neural network output is a single probability-like 
variable, in which values closer to 1 mostly correspond to signal, while 
values closer to 0 mostly correspond to background (examples of the neural 
network output distributions will be shown below). By selecting the tracks 
above a certain threshold, we can reject most of the background while 
keeping a large fraction of the signal.

We use three different neural networks specially trained on subsets of the 
large list of eID variables to reject (i) hadrons misidentified as 
electrons in the central arms (${\rm NN_h}$), (ii) background electrons 
which are mostly HBD backplane conversions (${\rm NN_e}$) and (iii) double 
hits in the HBD (${\rm NN_d}$).  In this way we basically have three 
handles to separately treat each type of background. The neural networks 
learn to distinguish the signal and the background on well defined 
samples. The first two neural networks, ${\rm NN_h}$ and ${\rm NN_e}$, are 
trained on {\sc hijing} events. The third neural network ${\rm NN_d}$ is 
trained on a sample of single particle event simulations, 
$\phi\rightarrow$\ee decays for single response and 
$\pi^0\rightarrow\gamma$\ee Dalitz decays for double response. The 
training is done separately for each centrality bin in order to properly 
treat the multiplicity effects. For centralities $>$ 40\%, we use the 
neural network trained for the 20\%--40\% centrality bin, where the 
statistics of the training sample is higher. This is justified because 
already in the 20\%--40\% centrality bin, multiplicity effects are 
unimportant and the separation between signal and background is good. The 
training is also done separately for the three cases of time-of-flight 
information (TOF-E, PbSc-TOF, no time-of-flight information).

The simulated events are passed through a {\sc geant} simulation of the 
PHENIX detector and through the same reconstruction code that is used for 
the data analysis. They are divided into two samples. One is used for 
training purposes and the other one to monitor the neural network output. 
The simulated events are not used to determine absolute efficiencies 
(those are determined from simulation as discussed later in Section 
\ref{sec:raw_spectra}. They are used only for training and monitoring 
purposes and the {\sc hijing} events are particularly valuable in this 
respect. They allow us to assess the origin and relative magnitude of the 
various background sources at each step of the electron identification 
chain, as well as the neural network performance in its ability to reject 
the background while preserving the signal. Details of the three neural 
networks are given below.

\subsubsection{Hadron rejection}
\label{sec:nn_h}

The first neural network, ${\rm NN_h}$, aims at reducing the hadron 
contamination. It exploits the information from all the relevant 
detectors, HBD, RICH, EMCal and TOF-E. The signal (S) for the training of 
${\rm NN_h}$ comprises electron tracks originating at the collision 
vertex, whereas the background (B) comprises all the remaining 
misidentified hadron tracks in the sample.
 
%%%%%%%%%%%%%%%%%%%%%%%%%%%%%%%%%%%%%%%%%%%%%%%%%%%%%%%%%%%% Fig_7
\begin{figure}[hbt!]
\includegraphics[width=1.0\linewidth]{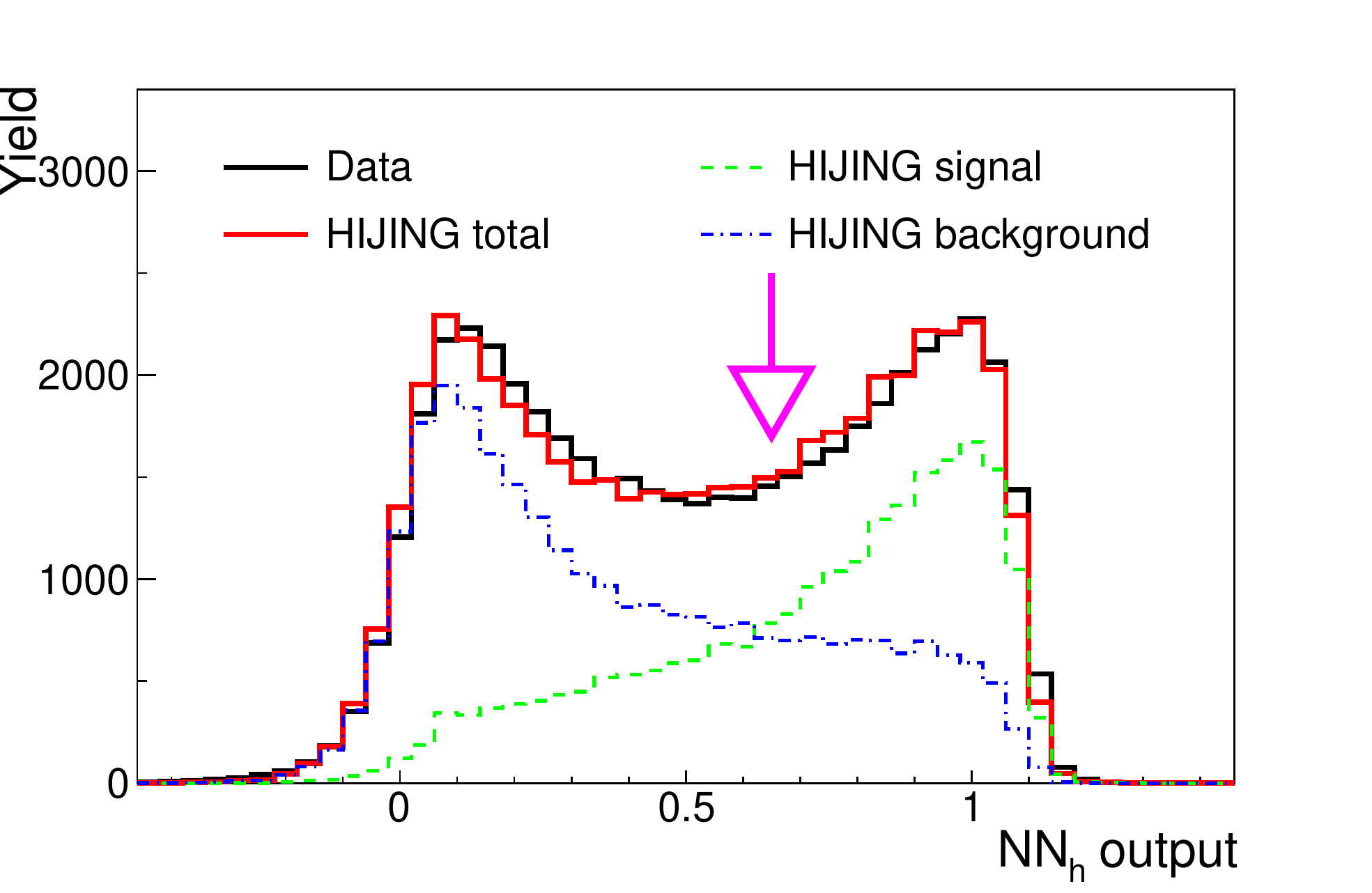}
\caption {(Color online) Comparison of the output values of the neural 
network ${\rm NN_h}$ for the 0\%--10\% centrality bin applied to the 
{\sc hijing} monitoring sample (red line) and to real data (black line). 
The figure also shows the signal (green) and the background (blue) 
components of the {\sc hijing} simulation. The arrow represents the 
average final cut selected by the cut optimization procedure. See text in 
Section \ref{sec:nn_cuts}.}
\label{fig:NN_h} 
\end{figure}

Figure \ref{fig:NN_h} shows the output values of ${\rm NN_h}$ for the 
{\sc hijing} monitoring sample (red line) and also shows the 
output of ${\rm NN_h}$ applied on real data (black line). The truth 
information from the {\sc hijing} events in terms of signal and background 
is shown separately. It should be noted that in the {\sc hijing} 
monitoring sample, all electron tracks are considered. The signal 
comprises the genuine electrons excluding the HBD backplane conversions 
and the background is all remaining tracks.
 
\subsubsection{Background electron rejection}
\label{sec:nn_e}

After rejecting hadrons in the previous step, the dominant background in 
the electron sample comes from the conversions in the HBD backplane that 
were not rejected by the conservative process described 
in~\ref{sec:tagging_rich}. Because these conversions do not leave a signal 
in the HBD they can be recognized and rejected if the tracks do not have a 
matching HBD response. The rejection capability is however limited by 
fluctuations remaining after the underlying event subtraction in the HBD. 
To provide the optimal rejection of the remaining backplane 
conversions we use a neural network, ${\rm NN_e}$, which is based on the 
HBD information reconstructed by both the stand-alone and the 
projection-based algorithms. The signal tracks for the training of 
${\rm NN_e}$ comprise all signal electrons remaining after the previous 
step, while the background sample includes only the electrons originating 
from the HBD backplane.

\begin{figure}[hbt!]
\includegraphics[width=1.0\linewidth]{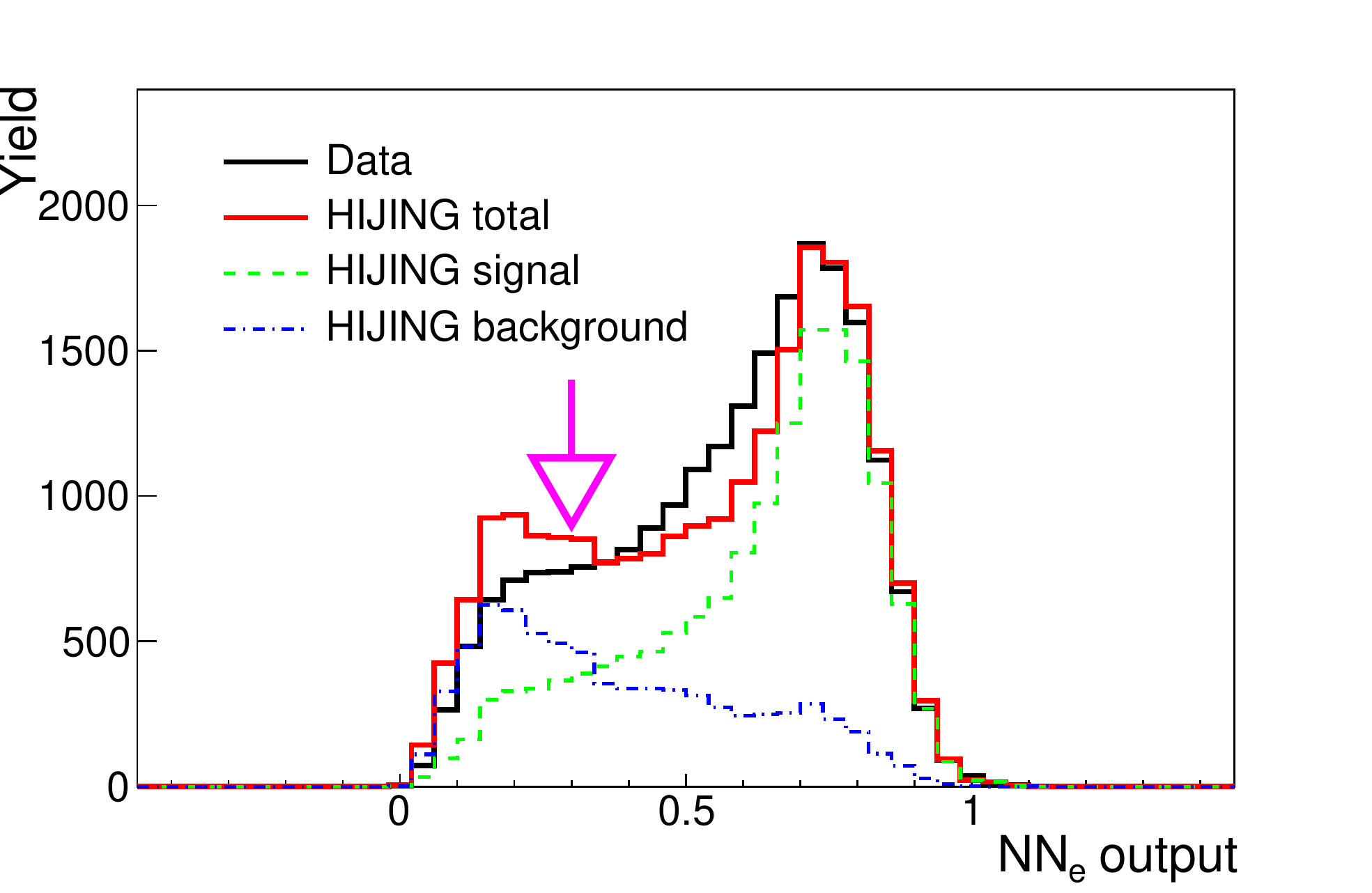}
\caption {(Color online) Comparison of the output values of the neural 
network ${\rm NN_e}$ for the 0\%--10\% centrality bin applied to the 
{\sc hijing} monitoring sample (red line) and to real data (black line). 
The figure also shows the signal (green) and the background (blue) 
components of the {\sc hijing} simulation. The arrow represents the 
average final cut selected by the cut optimization procedure. See text in 
Section \ref{sec:nn_cuts}.}
\label{fig:NN_e} 
\end{figure}

Figure \ref{fig:NN_e} shows the distribution of output values of 
${\rm NN_e}$ applied to the {\sc hijing} monitoring sample (red line) 
and to data (black line).  The signal and background components of the 
{\sc hijing} simulation are shown separately.

%%%%%%%%%%%%%%%%%%%%%%%%%%%%%%%%%%%%%%%%%%%%%%%%%%%%%%%%%%%% Fig_8

\subsubsection{Double-hit rejection in the HBD
\label{sec:nn_d}}

After removing hadrons and backplane conversions as much as possible, the 
major sources of background are the beam-pipe and radiator conversions and 
electrons from $\pi^0$ Dalitz decays where only one track is reconstructed 
in the central arms. These electrons have a zero or very small opening 
angle and most of them lead to a double hit in the HBD. Double hits can be 
recognized using the HBD response reconstructed in parallel by both the 
stand-alone and the projection-based algorithms. The response is coupled in 
a neural network, ${\rm NN_{d}}$ separately optimized for different HBD 
cluster sizes as well as centrality classes. The ${\rm NN_d}$ cut is an 
implicit small opening angle cut given by the maximum cluster size which 
is of the order of 75 mrad.

%%%%%%%%%%%%%%%%%%%%%%%%%%%%%%%%%%%%%%%%%%%%%%%%%%%%%%%%%%%% Fig_9
\begin{figure}[hbt!]
\includegraphics[width=1.0\linewidth]{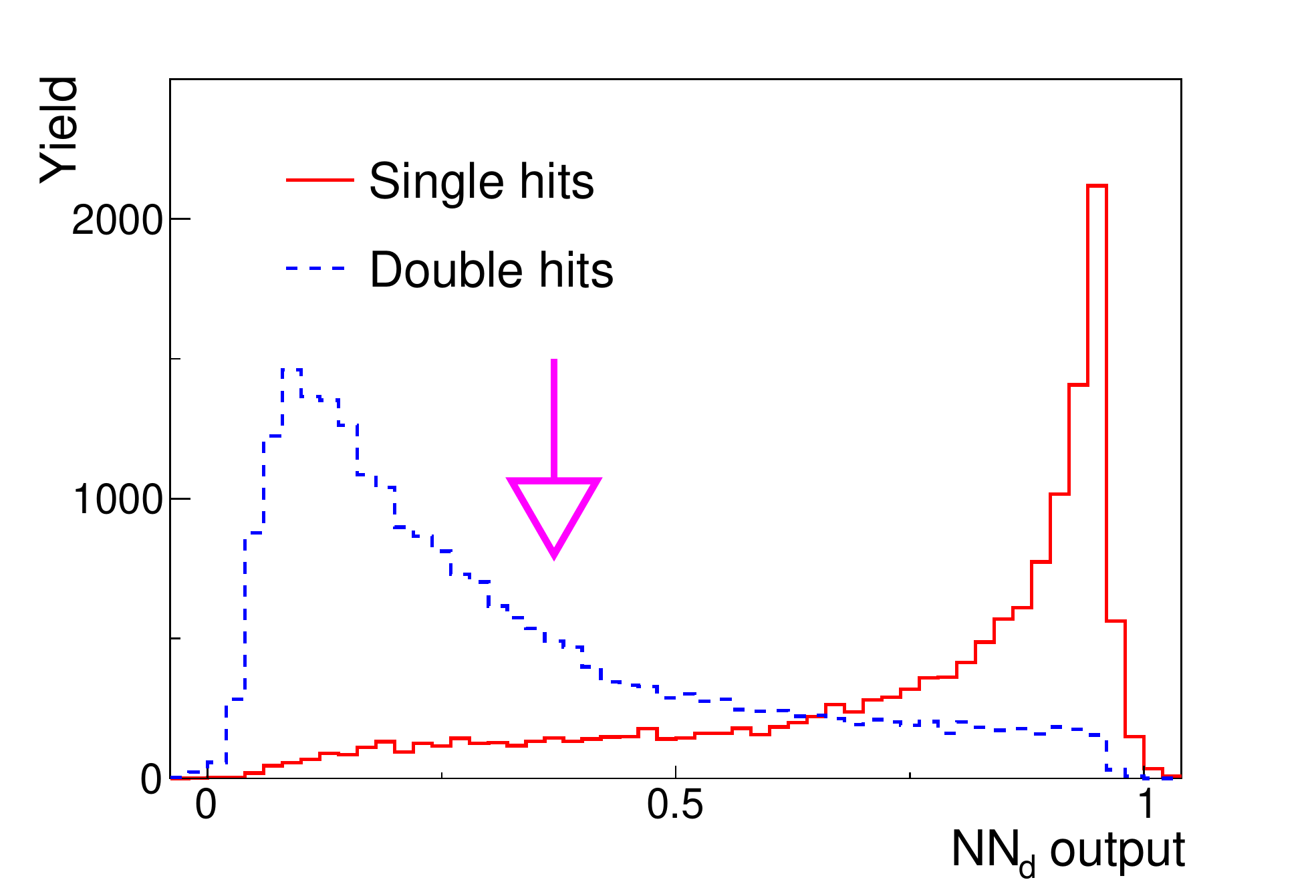}
\caption {(Color online) The output of the neural network ${\rm NN_d}$ 
for the recognition of single and double hits in the HBD. Single response 
(solid line) is provided by electrons from simulated $\phi\rightarrow$\ee 
decays and double response (dashed line) by electrons from 
$\pi^0\rightarrow\gamma$\ee Dalitz decays.  This example is for 30\%--40\% 
centrality and for a three cell cluster size. The arrow represents the 
average final cut selected by the cut optimization procedure. See text in 
Section \ref{sec:nn_cuts}.}
\label{fig:NN_d} 
\end{figure}

Figure \ref{fig:NN_d} shows the distribution of the output variable of the 
neural network ${\rm NN_{d}}$ for the separation of single and double 
hits in the HBD.  The single response is provided by electrons from 
simulated $\phi\rightarrow$\ee decays and the double response by electrons 
from $\pi^0\rightarrow\gamma$\ee Dalitz decays. The simulations are 
embedded into real HBD background events in order to take into account 
centrality dependent occupancy effects.

\subsubsection{Cut optimization}
\label{sec:nn_cuts}

The final selection of cuts on each neural network output variable is 
optimized using {\sc hijing} events. The thresholds are varied separately 
to maximize the effective signal, $S/\sqrt{B}$. Because the statistics of 
the {\sc hijing} samples are by far insufficient for a pair analysis, for 
the signal $S$ we use the number of single electrons from charm decay per 
event, which is an easily identified signal in {\sc hijing}, and for the 
background $B$ we use the total number of electrons per event. The cut 
optimization is done separately for each centrality class, for two $p_T$ 
ranges ($p_T$ $<$ 300~MeV/$c$ and $p_T$ $>$ 300~MeV/$c$), for each cluster 
size, and for each TOF configuration. The effective signal for each setup 
is maximized subject to the following conditions:

\begin{itemize}

\item The three types of TOF configuration (with PbSc timing information, 
with TOF-east timing information and without any timing information), have 
similar efficiencies with differences of less than 15\%.

\item Hadron contamination less than 5\% for TOF-E and PbSc-TOF and 
less than 10\% for the no-TOF case.

\end{itemize}

The arrows in Figs. \ref{fig:NN_h}-\ref{fig:NN_d} represent the average 
final cuts selected by the cut optimization procedure for these particular 
cases. The final cuts produce an electron sample with small hadron 
contamination, of less than 5\%, for all centralities. Strong cuts on the 
HBD are needed to achieve this small hadron contamination, resulting in a 
single electron efficiency of 25\%--40\% depending on centrality, at \pt 
$>$ 0.5~GeV/$c$ (See Section \ref{sec:raw_spectra}).

\subsection{Pair cuts
\label{sec:pair_cuts}}

The track selection criteria described above provide an electron sample 
with high purity. However, besides these criteria which are applied on a 
track-by-track basis, this analysis implements a series of dielectron 
cuts, based on the pair properties. These cuts are needed in order to 
remove ghost pairs i.e. pairs correlated by the close proximity of tracks 
in one of the detectors. Such correlations cannot be described by the 
mixed background, by definition, therefore this part of the phase-space 
must be removed from both the foreground and the mixed background. In the 
present analysis we remove the whole event, if such a pair is found, as 
was done in Ref.~\cite{Adare:2009qk}. This procedure removes only 
$\sim$2\% more of the total pair yield than discarding the pairs, because 
the average pair multiplicity is relatively low.

The most prominent detector correlation comes from the ring sharing effect 
in the RICH detector, discussed in Section \ref{sec:tagging_rich}, which  
arises when two tracks are parallel after the magnetic field, with at 
least one of them being an electron.

As mentioned above, the detector-correlated pairs are identified by 
applying a cut on the physical proximity of the tracks forming a pair in 
every detector and the cut value is determined by the corresponding double 
hit resolution. In the RICH detector, the cut selects pairs whose rings 
are closer than 36 cm, which is twice the diameter of the RICH ring 
($\sim$16.8 cm).  In the EMCal, the cut removes a region of $2.5\times 
2.5$ towers around the hit. In PC1 the pairs are selected for removal if 
their tracks are within 5 cm in $z$ or 0.02 rad in $\phi$.

\begin{figure}[hbt!]
\includegraphics[width=1.0\linewidth]{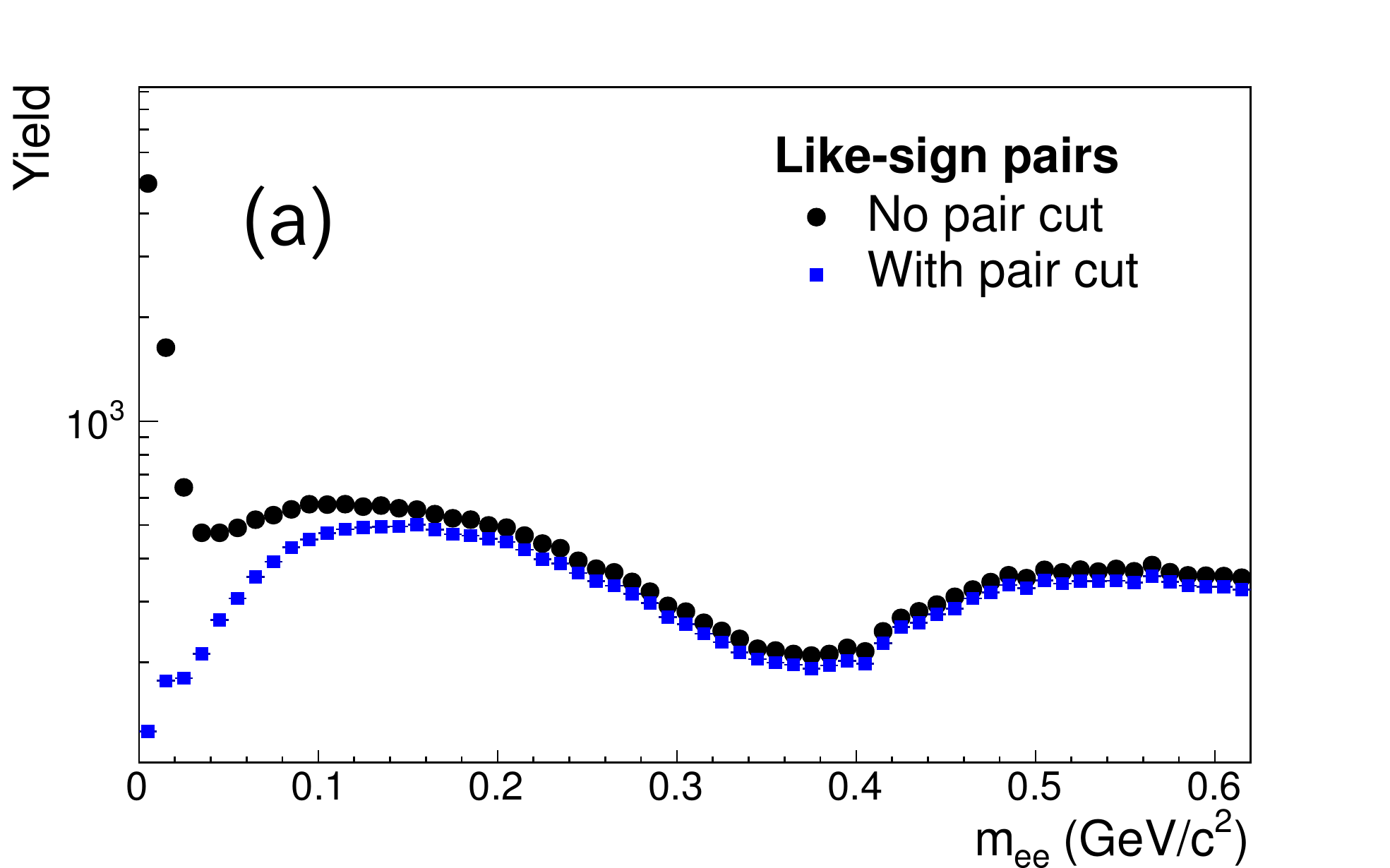}
\includegraphics[width=1.0\linewidth]{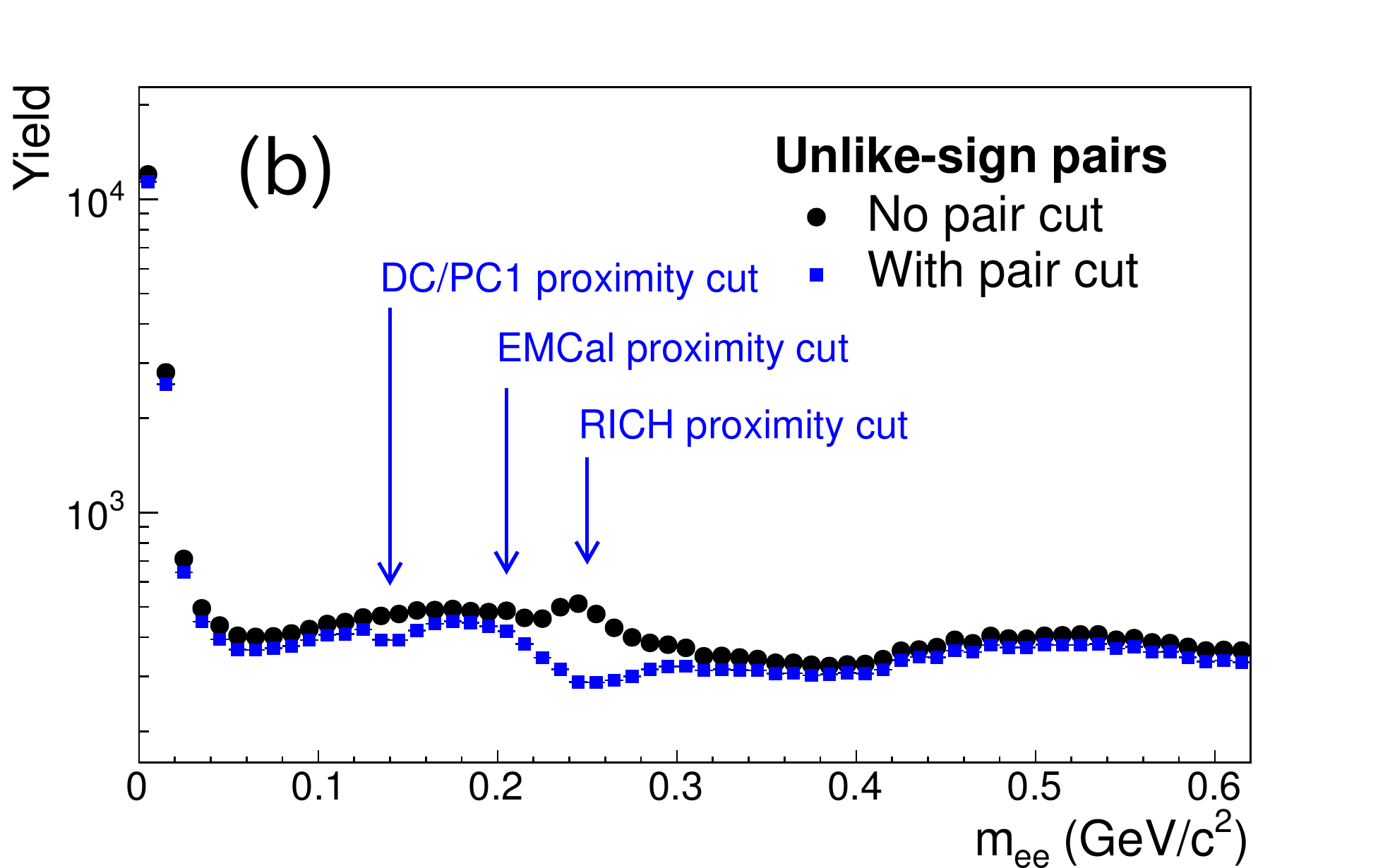}
\caption{\label{fig:cut_mass} (Color online) (a) Like-sign and (b) 
unlike-sign foreground spectra without any pair cuts (Black) and with 
RICH, EMCal and PC1 pair proximity cuts (Blue) for MB events.}
\end{figure}

The effect of these three pair cuts on the like-sign and unlike-sign mass 
spectra is shown in Fig. \ref{fig:cut_mass}. The like-sign yield close to 
$m_{ee}\sim0$~GeV/$c^2$ is affected by all cuts. On the other 
hand, in the unlike-sign foreground spectrum, the cuts affect well 
localized regions producing two clearly visible dips.  The dip at 
$m_{ee}\sim0.25$~GeV/$c^2$ is created by the RICH pair cut and the dip at 
$m_{ee}\sim0.15$~GeV/$c^2$ is created by the PC1 pair 
cut. The EMCal pair cut removes yield around 
0.20~GeV/$c^2$, but the effect is small compared to the other two cuts.

In addition to the RICH, EMCal and DC/PC1 ghost cuts, a 100 mrad opening 
angle cut is applied to remove ghost pairs in the HBD. This is a proximity 
cut that translates to a distance of two cells in the pad readout and 
roughly corresponds to the double hit separation of the HBD. This cut 
affects the yield at 
\mee $\sim0$~GeV/$c^2$ in both the like-sign and 
unlike-sign mass spectra.

%%%%%%%%%%%%%%%%%%%%%%%%%%%%%%%%%%%%%%%%%%%%%%%%%%%%%%%%%%%% Fig_10

          \subsection{Background Pair Subtraction}
          \label{sec:background}

Because the origin of the electron track candidates is not known, all 
electrons and positrons in the same event are paired to form the 
unlike-sign ($FG_{+-}$) and like-sign ($FG_{++}$ and $FG_{--}$) foreground 
mass spectra. This gives rise to a large combinatorial background that 
increases quadratically with the event multiplicity. In addition to that, 
there are several background sources of correlated pairs. The evaluation 
and subtraction of the background is the crucial step in the analysis of 
dileptons in particular in situations, like the present one, where the 
$S/B$ is at the sub-percent level. In this section, we describe in detail 
the various sources contributing to the background and the methodology 
used to evaluate each of them.

     \subsubsection{Background sources}

The unlike-sign foreground spectrum $FG_{+-}$ contains, in addition to the 
physical signal ($S$), a large background comprising the following 
sources:

\begin{itemize}
\item 

Uncorrelated combinatorial background ($CB$): It arises from the random 
combinations of electrons and positrons originating from different parent 
particles and is an inherent consequence of pairing all electrons with all 
positrons in the same event. The combinatorial background accounts for 
most of the total background, more than 99\% in the most central 
collisions and more than 90\% in peripheral collisions. The two electron 
tracks of combinatorial pairs are uncorrelated. However, they carry a 
global modulation induced by the collective flow of each individual 
collision. The evaluation of the combinatorial background together with 
the flow modulation is described in detail in the following subsection. 
(See Section \ref{sec:combinatorial_background}.)

\item 

Correlated background pairs. There are three different sources of 
correlated background pairs:

      \begin{itemize}

\item Cross pairs ($CP$): A cross pair can be produced when there are two 
$e^+e^-$ pairs in the final state of a single meson decay. One such case 
is $\pi^0 \rightarrow e^+e^- \gamma \rightarrow e^+e^-e^+e^-$.  The pair 
formed by an electron directly from $\pi^0$ and a positron from $\gamma$ 
conversion does not come from the same parent particle but it is a 
correlated pair through the same primary particle. (See Section 
\ref{sec:cross_pairs}.)

\item Jet pairs ($JP$): The jet pairs are produced by two electrons 
generated in the same jet or in back-to-back jets.  (See Section 
\ref{sec:jet_pairs}.)

\item Electron-hadron pairs ($EH$): Whereas the previous two sources of 
correlated pairs are of physics origin, the electron-hadron pairs are an 
artifact that results from residual detector correlations that cannot be 
handled by the pair cuts.  (See Section \ref{sec:electron_hadron_pairs}.)

	\end{itemize} 

\end{itemize}

One can then write:
\begin{equation}
\label{eq:fg12}
FG_{+-}  =  S + CB_{+-} + CP_{+-} + JP_{+-} +EH_{+-}
\end{equation}

All the background sources listed above form the yield of the like-sign 
foreground mass spectra $FG_{++}$ and $FG_{--}$. There is no signal in 
these spectra with the exception of a very small contribution of $e^+e^+$ 
and $e^-e^-$ pairs from $b\bar{b}$ decays ($BB$). So one can write:

\begin{equation}
\label{eq:fg11}
FG_{++}  =  CB_{++} + CP_{++} + JP_{++} +EH_{++} + BB_{++} \\
\end{equation}
\begin{equation}
\label{eq:fg22}
FG_{--}   =  CB_{--} + CP_{--} + JP_{--} +EH_{--} + BB_{--} \\ 
\end{equation}

Usually the like-sign pairs are subtracted from the unlike-sign pairs to 
obtain the signal. This is a convenient approach in a detector with 2$\pi$ 
azimuthal coverage, which ensures that the uncorrelated background is 
charge symmetric, under the assumption that the correlated background is 
also charge symmetric, i.e. it produces the same yield and mass 
distribution of like and unlike pairs. These conditions are not met in the 
present situation. The two central arm configuration of the PHENIX 
detector results in a substantial acceptance difference between like and 
unlike-sign pairs. Furthermore, the like-sign pairs contain a small signal 
component from $b\bar{b}$ decays that needs to be calculated separately. 
Finally, as shown below, the electron-hadron pairs are not charge 
symmetric. For these reasons, in this analysis we adopt a different 
approach in which each source is evaluated separately for a quantitative 
understanding of the like-sign yield. Once this is demonstrated, the 
background sources, $CB, CP, JP$ and $EH$ are subtracted from the 
inclusive foreground unlike-sign spectrum in order to obtain the mass 
spectrum of the signal pairs. The following subsections outline the 
evaluation of the various background sources.

The $BB$ contribution which is part of the signal is needed only for the 
quantitative evaluation of the like-sign spectra. The contribution is 
calculated using {\sc mc@nlo} (See Section \ref{sec:cocktail} for 
details), which generates both like-sign and unlike-sign contributions 
from $B\bar{B}$. The small like-sign contribution from $D\bar{D}$ is 
neglected.

\subsubsection{Combinatorial background (CB)}
       \label{sec:combinatorial_background}

The combinatorial background is determined using the event mixing 
technique, in which tracks from different events but with similar 
characteristics are combined into pairs. In this analysis, all events are 
classified into 11 bins in $z$ vertex between $-$30 cm and +25 cm, and 10 
bins in centrality between 0\% and 92\%.

In principle, the event mixing technique is expected to reproduce the 
shape of the combinatorial background with great statistical accuracy, 
because one can mix as many events as needed to reduce the statistical 
uncertainty to a negligible level. In fact it does not reproduce the 
shape. There is a small difference between the foreground combinatorial 
background and the mixed event background. The former is affected by the 
elliptic flow which is intrinsic to heavy ion collisions, whereas the 
latter is obtained by randomly picking up two tracks from different events 
and thus on the average does not have any flow effect.

To take into account the effect of flow in the mixed-events, one 
could make reaction plane bins, in addition to the vertex and centrality 
bins, so that only events with similar reaction plane are mixed. However, 
the method is limited by the reaction plane resolution and in PHENIX, the 
latter is not sufficient to reproduce the shape of the foreground 
combinatorial background. Instead, in the present analysis, a weighting 
method, based on an analytical calculation of the flow modulation, is used 
to account for the flow effects in the mixed events.

If particles are generated according to the following distribution 
function:

\begin{equation}
\label{eq:v2}
1 + 2 v_{2}\cos 2(\phi-\psi),
\end{equation}
where $\phi$ is the particle emission angle in azimuth, $\psi$ is the 
reaction plane angle and $v_2$ is the elliptic flow coefficient, then 
random pairs formed from these particles are distributed as (See 
Appendix~\ref{app:flow} for the derivation):
\begin{equation}
\label{eq:v2_weight}
P(\phi_a-\phi_b) = 1 + 2 v_{2,a}v_{2,b}\cos 2(\phi_a-\phi_b),
\end{equation}
where $\phi_{a(b)}$ is the azimuthal emission angle and $v_{2,a(b)}$ the 
elliptic flow of the two particles forming the pair.

In the weighting method, each mixed background pair is weighted by Eq. 
(\ref{eq:v2_weight}). The $v_2$ values of inclusive electrons are 
determined from the present data prior to the pair analysis as a function 
of centrality and electron $p_{T}$ using the reaction plane 
method~\cite{Adler:2005ab}. Exactly the same cuts as in the data analysis 
are used in the $v_2$ calculation. The obtained $v_2$ values are in very 
good agreement with the inclusive electron $v_2$ values reported in 
Ref.~\cite{Adare:2010de}.

We use a Monte-Carlo (MC) simulation to evaluate the method. The 
simulation generates electrons and positrons following a Poisson 
distribution with a mean value of three \footnote{There is not much 
meaning to the mean value of 3 of the Poisson distribution. It is a 
convenient choice to have one pair per event with a high probability.}. 
The particles are uniformly distributed in pseudorapidity between 
$\pm$0.35 and their momentum distribution is taken from data. The 
azimuthal emission angle $\phi$ is determined according to the 
distribution $1 + 2 v_{2}\cos 2(\phi-\psi)$, where $\psi$ is the reaction 
plane angle, which is uniformly distributed between $\pm \frac{\pi}{2}$. 
The $v_2$ values are taken from the 20\%--40\% centrality bin. The tracks 
that pass the PHENIX acceptance filter are used in the pair analysis.

%%%%%%%%%%%%%%%%%%%%%%%%%%%%%%%%%%%%%%%%%%%%%%%%%%%%%%%%%%%% Fig_11
\begin{figure}[hbt!]
\includegraphics[width=1.0\linewidth]{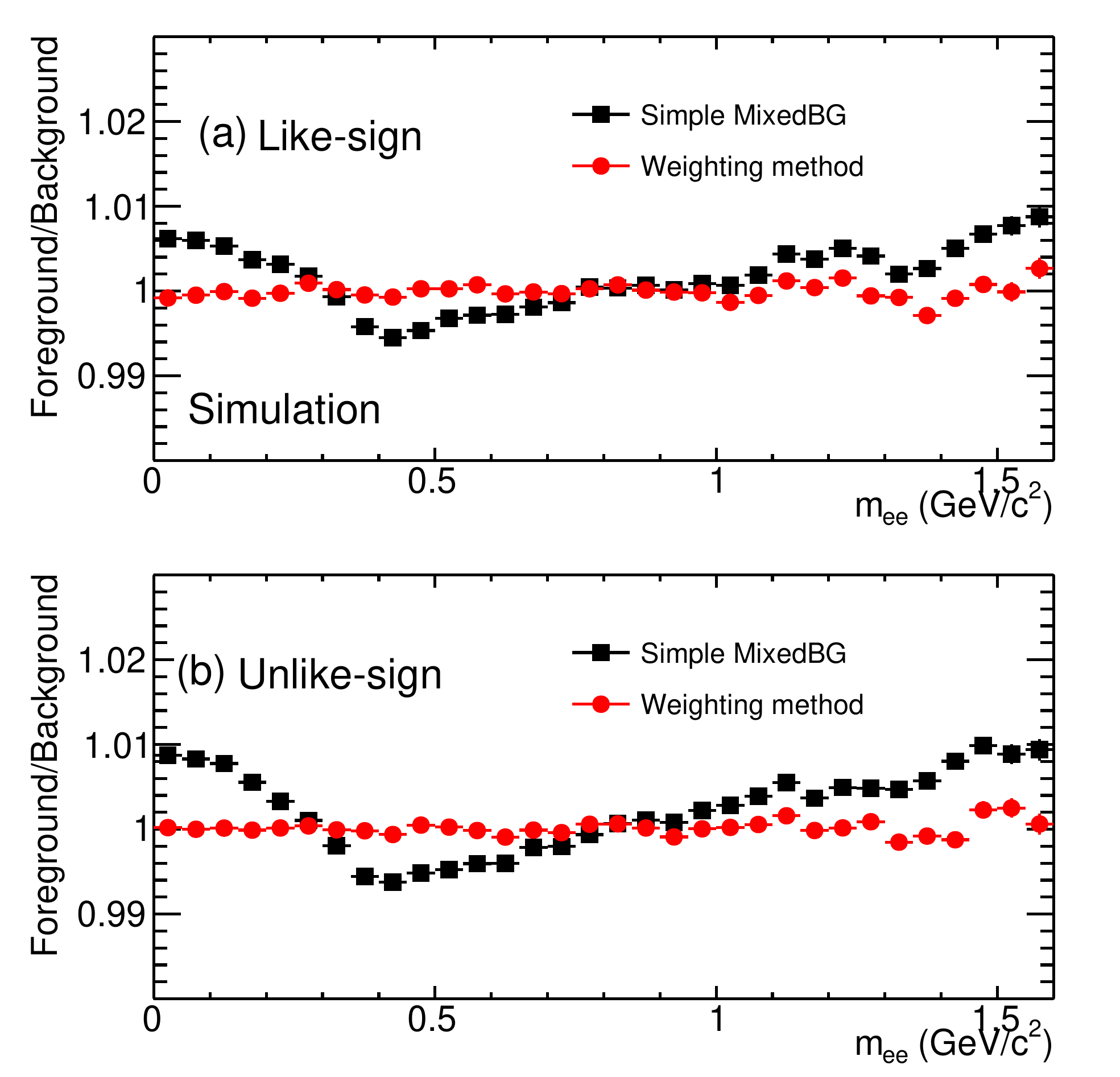}
\caption{(Color online) 
Foreground to mixed background ratio of (a) like-sign and (b) unlike-sign 
mass spectra ratio in a MC simulation. The foreground is generated with 
flow, whereas the mixed events are produced without flow i.e. using a 
simple mixed-event technique (squares) and with flow modulation using the 
weighting method (circles).
}
\label{fig:toy_v2}
\end{figure}

Figure \ref{fig:toy_v2} shows the ratio of the foreground to mixed 
background mass spectra. The squares correspond to the simple mixed-event 
technique without correcting for flow. We can see that in this approach 
the ratio is not flat, i.e. the foreground shape is not reproduced by the 
mixed background shape. The circles correspond to the weighting method. 
The ratio is completely flat over the entire mass range demonstrating that 
the weighting method properly accounts for the flow modulation.

A similar MC study was performed to evaluate whether triangular 
flow $v_3$ also induces shape distortion of the mass spectrum.  For the 
most central collisions, where $v_3$ is comparable to $v_2$ at high 
$p_T$~\cite{Adare:2011tg}, the simulations show that the $v_3$ effect is 
at least one order of magnitude smaller than for $v_2$ and we thus ignore 
triangular flow in the determination of the combinatorial background 
shape.

                   \subsubsection{Cross pairs (CP)}
                   \label{sec:cross_pairs}

Cross pairs can be produced when a hadron decay produces two $e^+e^-$ 
pairs in the final state. The following hadron decays and subsequent 
photon conversions lead to cross pairs:
\begin{equation}
\label{eqn:pi0eeg} \vspace{-0.4cm}
\pi^0 \rightarrow e^+_1e^-_1\gamma\rightarrow e^+_1e^-_1 e^+_2e^-_2  
\end{equation} 
\begin{equation}
\label{eqn:pi0gg} \vspace{-0.4cm}
\pi^0 \rightarrow \gamma_1\gamma_2 \rightarrow e^+_1e^-_1 e^+_2e^-_2
\end{equation} 
\begin{equation}
\label{eqn:etaeeg} \vspace{-0.4cm}
\eta \rightarrow e^+_1e^-_1\gamma \rightarrow e^+_1e^-_1 e^+_2e^-_2
\end{equation} 
\begin{equation}
\label{eqn:etagg}
\eta \rightarrow \gamma_1\gamma_2 \rightarrow e^+_1e^-_1 e^+_2e^-_2
\end{equation} 

The cross combinations give rise to two unlike-sign pairs ($e^+_1e^-_2$ 
and $e^+_2e^-_1$) as well as two like-sign pairs ($e^+_1e^+_2$ and 
$e^-_1e^-_2$) that are not purely combinatorial, but correlated via the 
$\pi^0$ or $\eta$ mass and momentum. Therefore, this contribution is not 
reproduced by the event-mixing technique.

To calculate the cross pairs, we use EXODUS (see Section 
\ref{sec:cocktail}) to generate $\pi^0$ and $\eta$ with the following 
input parameters:

\begin{itemize}

\item Flat-vertex distribution within $|z| < 30$ cm. The final results are 
weighted to restore the measured vertex distribution.

\item Flat pseudorapidity distribution within $|\eta| < 0.6$ and uniform 
in $\phi$ within $0<\phi<2\pi$.

\item Momentum distributions based on PHENIX measurements (see Section 
\ref{sec:cocktail}).

\end{itemize}

%%%%%%%%%%%%%%%%%%%%%%%%%%%%%%%%%%%%%%%%%%%%%%%%%%%%%%%%%%%% Fig_12
\begin{figure*}[hbt!]
  \includegraphics[width=0.49\linewidth]{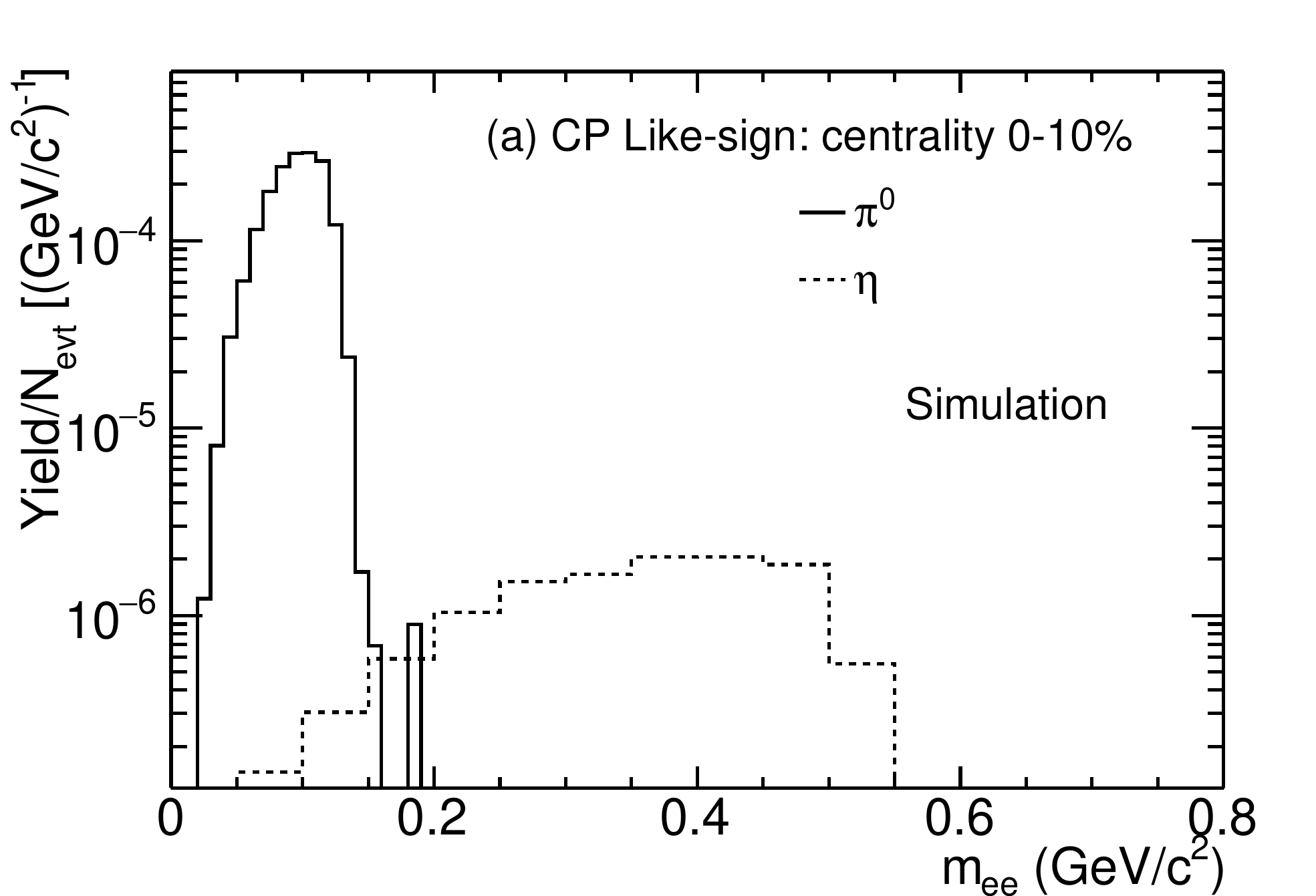}
  \includegraphics[width=0.49\linewidth]{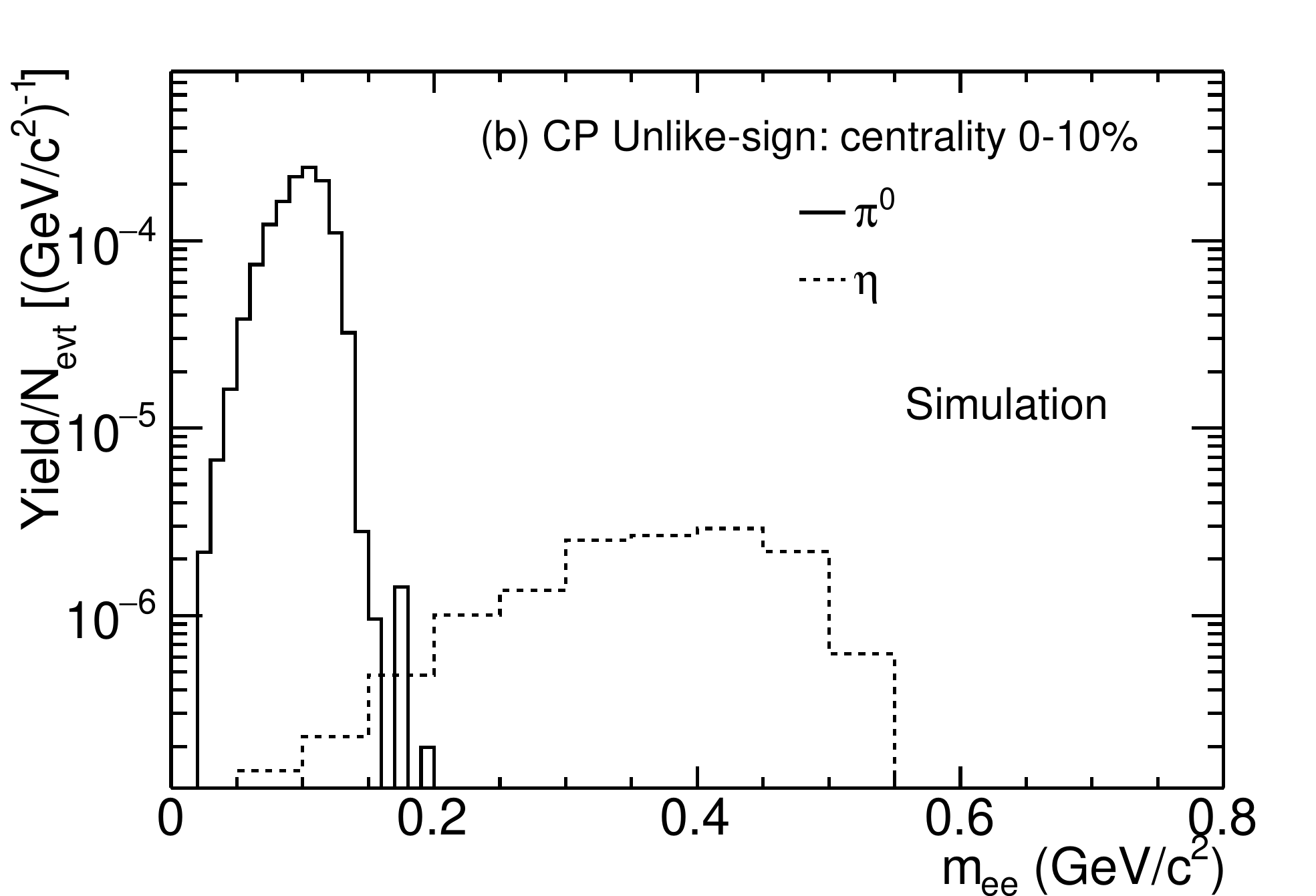}
\caption{Absolutely normalized (a) like-sign and (b) unlike-sign spectra 
of cross pairs ($CP$) from {\sc exodus} and {\sc geant} simulations for 
the 0\%--10\% centrality bin.  The $\pi^0$ and $\eta$ contributions are 
shown separately.}
    \label{fig:cross_pairs}
\end{figure*}

The generated $\pi^0$ and $\eta$ are passed through a {\sc geant} 
simulation of the PHENIX detector.  By selecting reconstructed cross 
pairs, one can determine the shape of the cross-pair invariant mass 
spectrum. The spectra are then absolutely normalized using the rapidity 
density values $dN_{\pi^0}$/$dy$ and $dN_{\eta}$/$dy$ as a function of 
centrality, summarized in Section \ref{sec:cocktail}. The absolutely 
normalized mass spectra of cross pairs for the 0\%--10\% centrality bin 
are shown in Fig. \ref{fig:cross_pairs}.
                
                   \subsubsection{Jet pairs (JP)}
                   \label{sec:jet_pairs}

The jet pairs are produced using the {\sc pythia} 6.319 code 
with {\sc cteq{\footnotesize 5}l} parton distribution 
functions~\cite{Sjostrand:2000wi}. The following hard 
quantum-chromodynamics (QCD) processes are 
activated~\cite{Adare:2009qk}:

\begin{itemize}
\item MSUB 11: $f_i f_j \rightarrow f_i f_j$
\item MSUB 12: $f_i \overline{f}_i \rightarrow f_k \overline{f}_k$
\item MSUB 13: $f_i \overline{f}_i \rightarrow gg$
\item MSUB 28: $f_i g \rightarrow f_i g$
\item MSUB 53: $gg \rightarrow f_k \overline{f}_k$
\item MSUB 68: $gg \rightarrow gg$
\end{itemize}

where $g$ denotes a gluon, $f_{i, j, k}$ are fermions with flavor $i$, 
$j$, $k$ and $\overline{f}_{i, j, k}$ are the corresponding antiparticles. 
A Gaussian width of 1.5~GeV/$c$ for the primordial $k_T$ distribution 
(MSTP(91)=1, PARP(91)=1.5) and 1.0 for the K-factor (MSTP(33)=1, 
PARP(31)=1.0) are used. The minimum parton $p_T$ is set to 2~GeV/$c$ 
(CKIN(3)=2.0). The $z$ coordinate of the vertex position is produced 
uniformly between $\pm$30~cm and then weighted to reproduce the measured 
distribution. From the {\sc pythia} output, $\pi^0$ and $\eta$ are 
extracted and passed through the {\sc geant} simulator of PHENIX in order 
to generate the inclusive \ee pairs.

\begin{figure}[hbt!]
\includegraphics[width=1.0\linewidth]{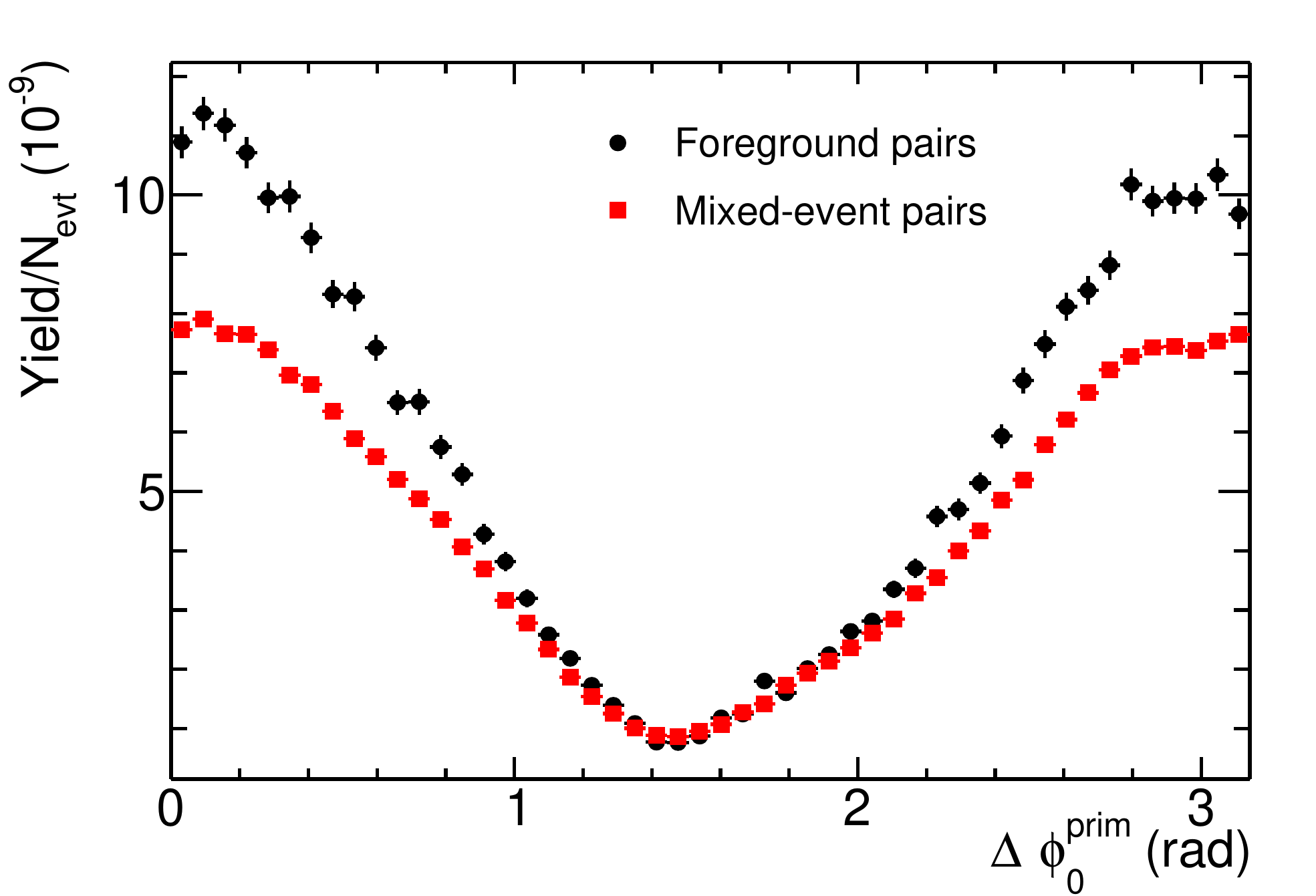}  
\caption{\label{fig:jet_sub_pp} (Color online)  $\Delta \phi_0^{prim}$ 
(difference in the azimuthal angle of the primary particles, $\pi^0$ or 
$\eta$) distributions of foreground and normalized mixed-event background 
like-sign pairs as obtained from the {\sc pythia} simulations.
}
\end{figure}

In addition to the jet pairs we are interested in, the foreground pairs 
from {\sc pythia} events contain also ``physical" pairs, cross pairs and 
combinatorial pairs. The ``physical'' pairs and cross pairs are excluded 
from the foreground pairs by requiring that the two electrons or positrons 
of the pair do not share the same particle in their history. The 
combinatorial background is statistically subtracted using the 
event-mixing technique. The mixed event like-sign pairs are normalized to 
the foreground like-sign pairs in the range $\Delta \phi^{prim}_0 \sim 
\pi/2$, where $\Delta \phi_0^{prim}$ is the difference in the azimuthal 
angle of the primary particles, $\pi^0$ or $\eta$. Figure 
\ref{fig:jet_sub_pp} shows the $\Delta \phi_0^{prim}$ distributions of the 
foreground pairs and the normalized mixed-event pairs. The excess yield 
around $\Delta \phi^{prim}_0 \sim 0 $ represents the dileptons from the 
same jet whereas the excess yield at $\Delta \phi^{prim}_0 \sim \pi$ 
corresponds to the dileptons from opposite or back-to-back jets.

%%%%%%%%%%%%%%%%%%%%%%%%%%%%%%%%%%%%%%%%%%%%%%%%%%%%%%%%%%%% Fig_13

%%%%%%%%%%%%%%%%%%%%%%%%%%%%%%%%%%%%%%%%%%%%%%%%%%%%%%%%%%%% Fig_14
\begin{figure*}[t]
 \includegraphics[width=0.49\linewidth]{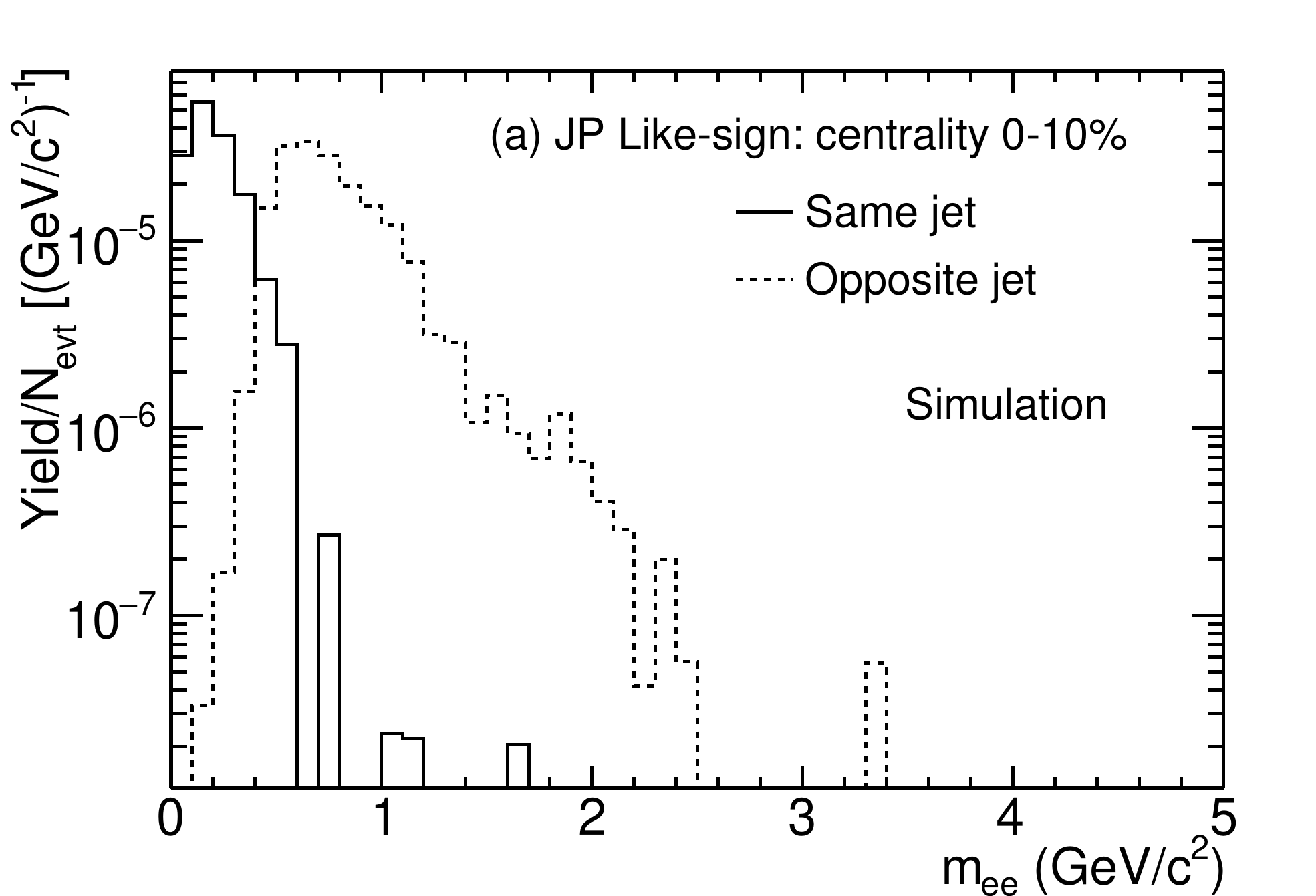}
 \includegraphics[width=0.49\linewidth]{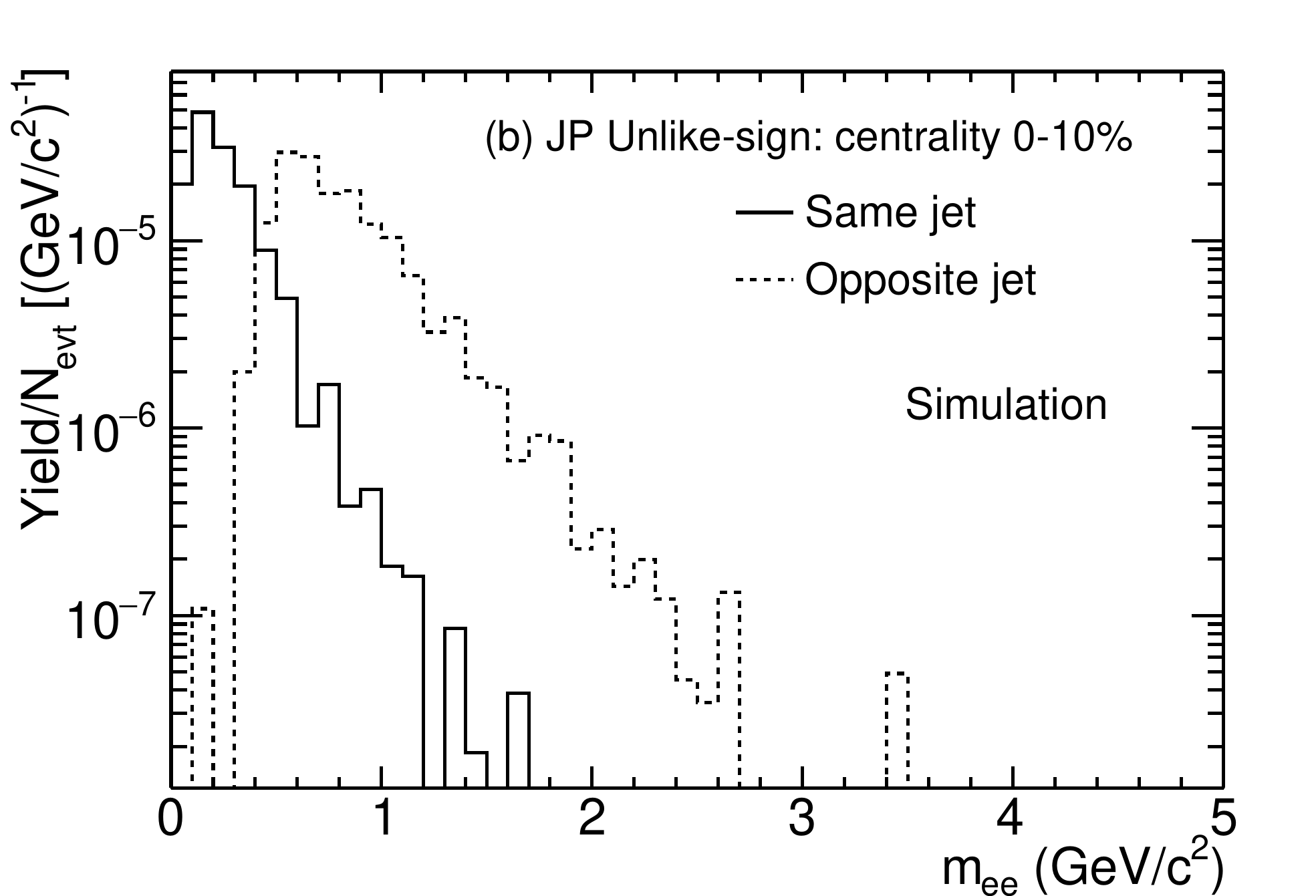}
\caption{Absolutely normalized (a) like-sign and (b) unlike-sign spectra 
of jet pairs ($JP$) simulated by {\sc pythia} and {\sc geant} for the 
0\%--10\% centrality bin. The near-side and away-side contributions are 
shown separately.
}
\label{fig:jets_0_10}
\end{figure*}

After subtracting the combinatorial background, the {\sc pythia} spectra 
are scaled to give the pion yield per \pp MB event . The scaling 
factor is determined such that the $\pi^0$ yield in the {\sc pythia} 
simulation matches the measured $\pi^{0}$ yield in \pp 
collisions~\cite{Adare:2007dg} and found to be 1/3.9.

The spectra need to be further scaled to obtain the jet contribution 
in \auau collisions for each centrality bin. This scaling is done 
following Ref.~\cite{Adare:2008ae}: an $ee$ jet pair originating from 
primary particles with momenta $p_{T,1}$ and $p_{T,2}$ is scaled by the 
average number of binary collisions $\langle$\Ncollnospace$\rangle$ for 
each centrality bin, times $R_{AA}(p_{T,1})$, times $I_{AA}(p_{T,2})$. The 
same jet or opposite jet $I_{AA}(p_{T,2})$ values are applied depending on 
the pair opening angle. The absolutely normalized jet pair spectra for the 
0\%--10\% centrality bin are shown in Fig. \ref{fig:jets_0_10}.

                 \subsubsection{Electron-hadron pairs (EH)}
                  \label{sec:electron_hadron_pairs}

Even after applying the pair cuts described in Section 
\ref{sec:pair_cuts}, electron-hadron pairs correlated through detector 
effects remain in the foreground pairs. An example of such an 
electron-hadron pair can be illustrated with the sketch of Figure 
\ref{fig:eh_view} discussed in Section \ref{sec:tagging_rich}. In this 
example, if both the positron and the mis-identified hadron are detected, 
the pair is identified as a RICH ghost pair and the entire event is 
rejected by the RICH ghost pair cut as described in Section 
\ref{sec:pair_cuts}. However, if the positron is not detected due to 
detector dead areas or reconstruction inefficiency, the pair formed by the 
electron and the mis-identified hadron is not rejected and remains in the 
sample. This pair is not a combinatorial pair but correlated through the 
positron. Although the mis-identification of hadrons via hit sharing 
occurs in all detectors, the RICH detector is the dominant contributor to 
these electron-hadron pairs.  Therefore, only the RICH detector is 
considered as the source of such correlated pairs.

We simulate electron-hadron pairs using electrons from $\pi^0$ and $\eta$ 
simulations and hadrons from real events. The $\pi^0$ and $\eta$ 
simulations are the same ones that are used for the cross pair simulation. 
The hadrons from real events are all the reconstructed tracks that fail 
the eID cuts.
 
%%%%%%%%%%%%%%%%%%%%%%%%%%%%%%%%%%%%%%%%%%%%%%%%%%%%%%%%%%%% Fig_15
\begin{figure*}[t]
 \includegraphics[width=0.49\linewidth]{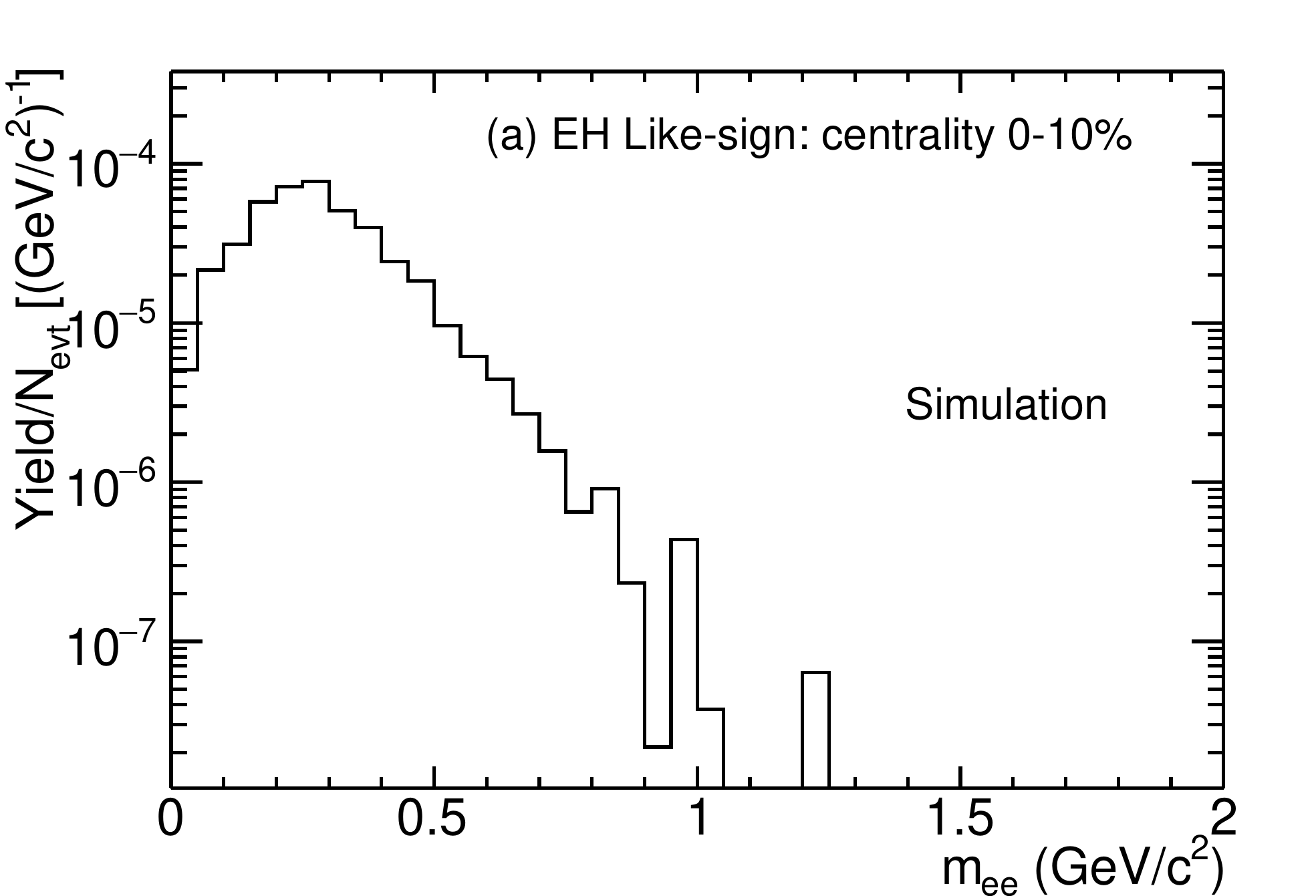}
 \includegraphics[width=0.49\linewidth]{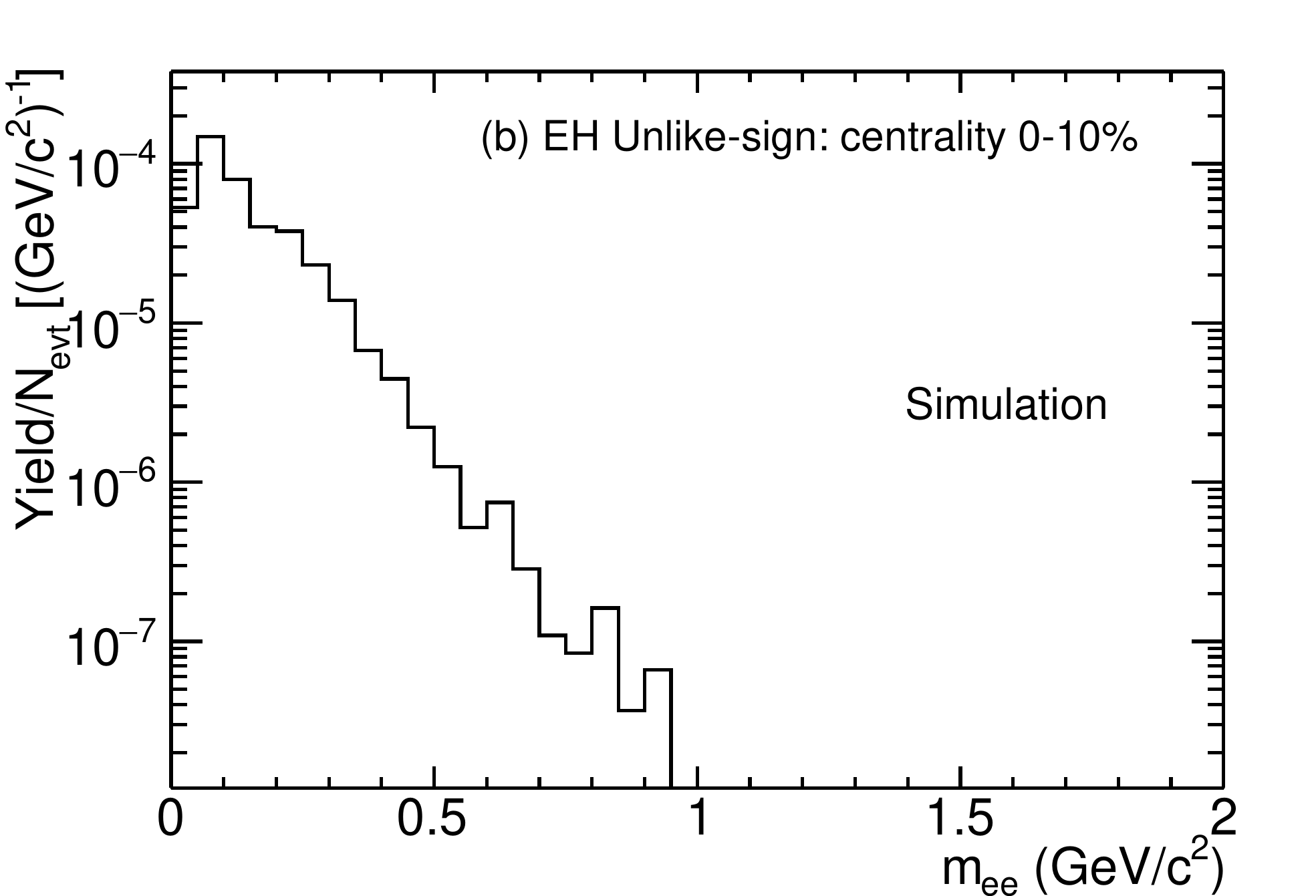}
 \caption{
Absolutely normalized (a) like-sign and (b) unlike-sign spectra of 
simulated electron-hadron pairs ($EH$) for the 0\%--10\% centrality bin. 
See text for details.
 }
   \label{fig:eh_0_10}
\end{figure*}

The simulation is performed in the following way:  First, a combined event 
is formed using electrons from one Dalitz decay of $\pi^0$ or $\eta$ 
generated with {\sc exodus} and hadrons from a real event.  Second, the 
information from their associated fired PMTs is merged and new rings are 
reconstructed.  Using the new RICH ring variables, the regular analysis 
procedure, including eID cuts and pair cuts, is performed on the combined 
event.  Finally, the pairs formed by the combination of an electron track 
from simulation and a hadron track from data are extracted.  The spectra 
are absolutely normalized using the $\pi^0$ $dN$/$dy$ values shown in 
Section \ref{sec:cocktail}. The absolutely normalized electron-hadron pair 
spectra for the 0\%--10\% centrality bin are shown in 
Fig.~\ref{fig:eh_0_10}. Contrary to the cross pairs and the jet pairs 
where the like- and unlike-sign spectra have a very similar shape, the 
electron-hadron pairs exhibit a sizable difference between the like- and 
unlike-sign spectra. The yield of electron-hadron pairs has a strong 
centrality dependence. It increases by a factor of $\sim$50 from 
peripheral to central collisions with respect to the $\pi^0$ rapidity 
density. This increase is mainly due to the expected scaling of the 
electron-hadron pairs with the square of the event multiplicity.

      \subsubsection{Background normalization}
           \label{sec:background_normalization}

The cross pairs, jet pairs, electron-hadron pairs and $b\bar{b}$ decay 
pairs are absolutely normalized. The mixed event technique provides only 
the shape of the combinatorial background. It needs to be normalized in 
order to be able to subtract the background and extract the signal. The 
only free parameters of the entire procedure are thus the normalization 
factors of the mixed event background like-sign spectra $nf_{++}$ and 
$nf_{--}$. They are determined by normalizing the mixed event background 
yield ($N_{MIX_{++(--)}}$) to the foreground yield ($N_{FG_{++(--)}}$), 
integrated over a selected region of phase space, after subtracting the 
correlated pairs integrated over the same region:
\begin{eqnarray}
nf_{++}=\frac{N_{FG_{++}} - N_{CP_{++}}- N_{JP_{++}} - N_{EH_{++}} - N_{BB_{++}}} {N_{MIX_{++}}} \nonumber \\ 
nf_{--}=\frac{N_{FG_{--}} - N_{CP_{--}}- N_{JP_{--}} - N_{EH_{--}} - N_{BB_{--}}} {N_{MIX_{--}}} \nonumber 
\end{eqnarray}
where $N_{CP_{++(--)}}$, $N_{JP_{++(--)}}$, $N_{EH_{++(--)}}$ and 
$N_{BB_{++(--)}}$ are the integral yields of each source in the 
normalization region. The normalization region is a window in the 
azimuthal angular distance of the two tracks $\Delta\phi_0$. It needs to 
satisfy two competing conditions.  On the one hand, a small normalization 
window containing only combinatorial pairs is preferred to avoid being 
affected by any residual yield (and systematic uncertainties) from the 
correlated background sources. On the other hand, a wide normalization 
window is required to reduce statistical uncertainty.  The normalization 
windows used in this analysis for each centrality bin are shown in 
Table~\ref{tab:norm_window} together with the corresponding number of 
like-sign pairs ($N_{LS} = N_{FG++} + N_{FG--}$). The region of small 
opening angles that correspond to small masses where the correlated pairs 
$CP$, $JP$ and $EH$ mostly contribute, is excluded in all centrality bins.

%===================================================== Table_III
\begin{table}[hbt!]
\caption{Normalization window for each centrality bin. The number of 
like-sign pairs $N_{LS}$ in the window is also shown.
}
\begin{ruledtabular} \begin{tabular}{ccccc}
& Centrality & Normalization window   & $N_{LS}$ & \\
&            &        $\Delta\phi_0 $ &          & \\
\hline
& 0\%--10\%   & 0.7 - 3.14 & 5.1M & \\
& 10\%--20\% & 0.7 - 2.1   & 1.1M & \\
& 20\%--40\% & 0.7 - 2.1  & 660K  & \\
& 40\%--60\% & 0.9 - 2.1  & 48K   & \\
& 60\%--92\% & 0.9 - 2.1  & 3K    & \\
\end{tabular} \end{ruledtabular}
\label{tab:norm_window}
\end{table}

The combinatorial background in Eqs. (\ref{eq:fg11}) and (\ref{eq:fg22}) 
is thus given by the normalized mixed-event background:

\begin{eqnarray}
CB_{++} (m_{ee}) = nf_{++} \cdot MIX_{++} (m_{ee}) \\
CB_{--} (m_{ee}) = nf_{--} \cdot MIX_{--} (m_{ee})
\end{eqnarray}

As long as electrons and positrons are produced in pairs and these pairs 
are uncorrelated, the total unlike-sign combinatorial background yield is 
the geometric mean of the total like-sign combinatorial yield, independent 
of single electron efficiency and acceptance~\cite{Adare:2009qk}:
\begin{equation}
\label{eq:2sqrt}
CB_{+-} = 2 \sqrt{CB_{++} \cdot CB_{--}}
\end{equation}
A similar relation holds true for the integral yields of the mixed-event 
background:
\begin{equation}
\label{eq:2sqrt_mix}
MIX_{+-} = 2 \sqrt{MIX_{++} \cdot MIX_{--}}
\end{equation}
The normalization factor $nf_{+-}$ of the unlike-sign mixed event 
background is thus deduced from the normalization factors of the like-sign 
mixed background, $nf_{++}$ and $nf_{--}$ as:
\begin{equation}
\label{eq:nf12}
nf_{+-}=\sqrt{nf_{++}\cdot nf_{--}}
\end{equation}

In the present analysis, the square root relation, Eq. (\ref{eq:2sqrt}), 
is violated by two independent factors. First, the relation does not hold 
true when pair cuts are applied to the spectra because pair cuts affect 
differently the unlike-sign and like-sign spectra. Second, elliptic flow 
induces an inherent distortion of the square root relation. Flow does not 
create or destroy particles. It only affects their azimuthal distribution 
and therefore in a perfect 2$\pi$ detector there is no effect and Eq. 
(\ref{eq:2sqrt}) is obeyed. However, in the case of the PHENIX detector, 
which is not a 2$\pi$ detector, the relation is violated as demonstrated 
in Appendix \ref{app:sqrt_relation}. Relation (\ref{eq:nf12}) can still be 
used provided that the violation is the same in the data and the mixed 
events. In the present analysis, we make sure that this is the case.  We 
start from a situation in which the mixed events satisfy Eq. 
(\ref{eq:2sqrt_mix}). We then apply to the mixed events the pair cuts, 
exactly as to the foreground events, and the flow modulation using a 
weighting factor procedure that is based on an exact analytical 
calculation. Thus we make sure that Eq. (\ref{eq:nf12}) is still valid.

       \subsubsection{Quantitative understanding of the background}
         \label{sec:quantitative_background}
         
\begin{figure}[hbt!]
\includegraphics[width=1.0\linewidth]{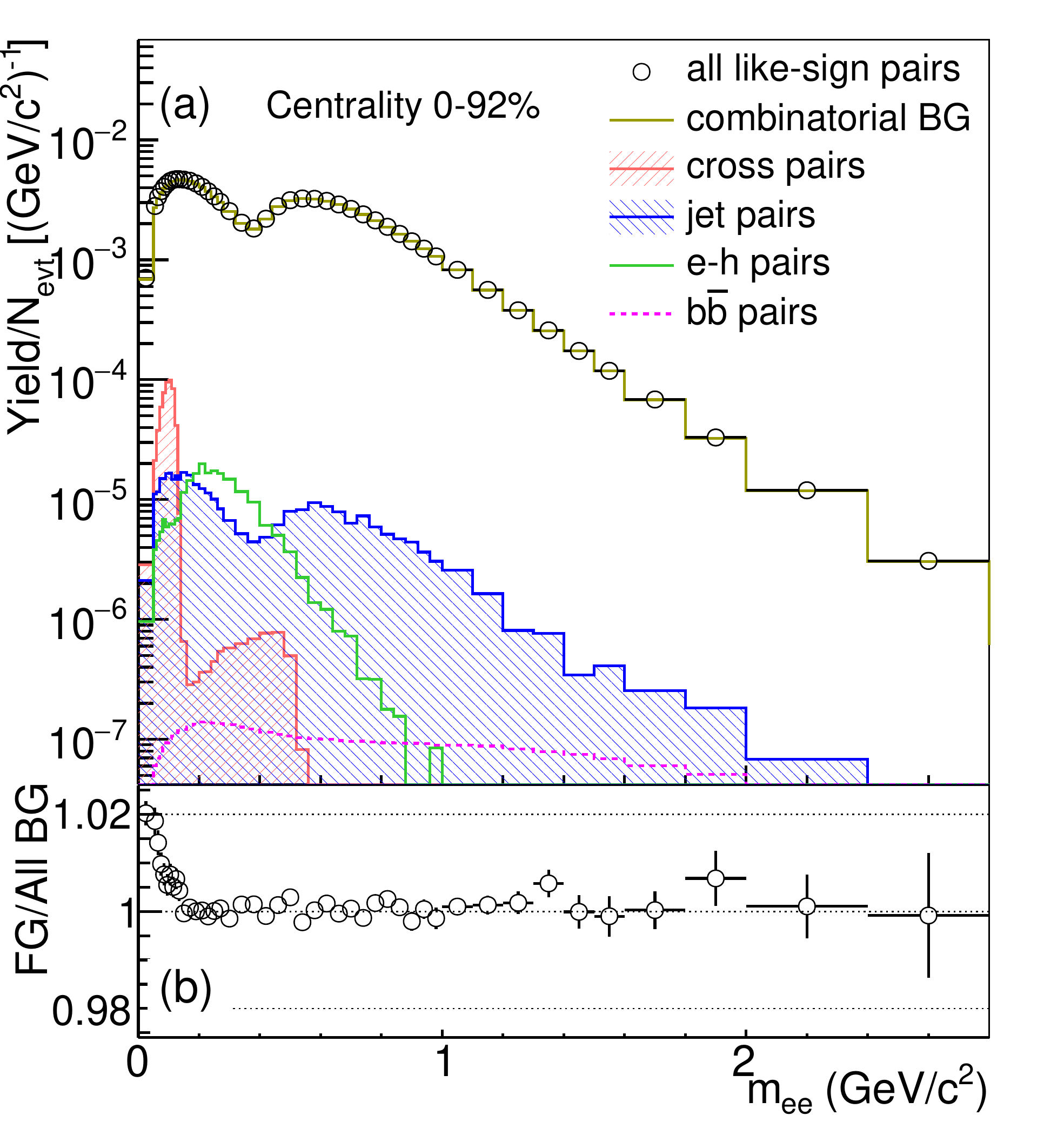}  
\caption{ (Color online) (a) Measured like-sign spectrum (open circles) 
together with the calculated background components (histograms) for 
MB events. (b) Ratio of the like-sign spectrum to the sum of all 
the background components.
}
\label{fig:like-sign_0-92}
\end{figure}

To illustrate our understanding of the background in quantitative terms, 
Fig. \ref{fig:like-sign_0-92} shows a comparison of the MB mass 
spectra for the foreground and the calculated background like-sign pairs.

%%%%%%%%%%%%%%%%%%%%%%%%%%%%%%%%%%%%%%%%%%%%%%%%%%%%%%%%%%%% Fig_16

\begin{figure}[hbt!]
\includegraphics[width=1.0\linewidth]{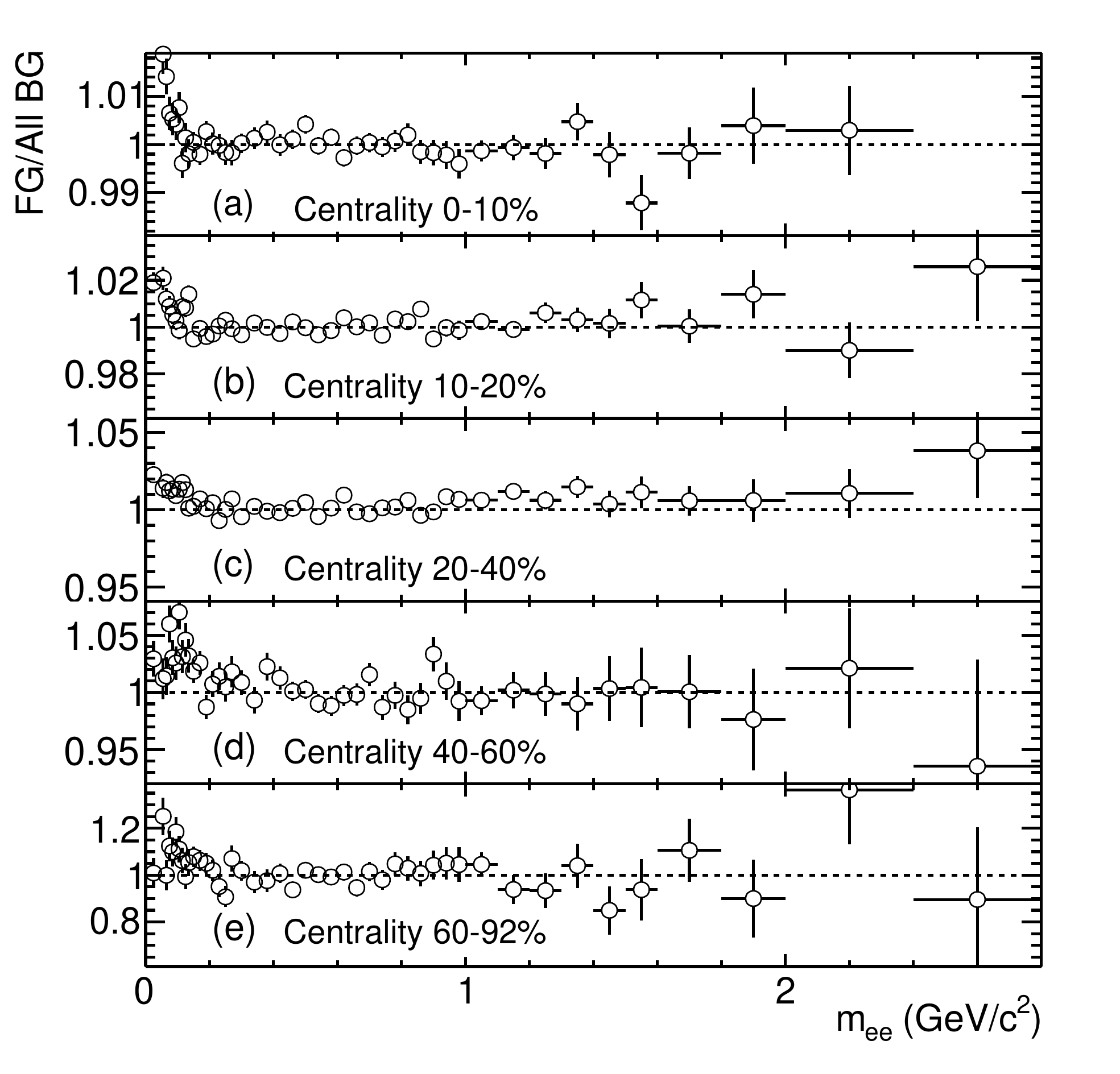}  
\caption{
Ratios of the like-sign foreground spectrum to the sum of all the 
background components for the five centrality bins used in this 
analysis.
}
\label{fig:like-sign}
\end{figure}

The top panel shows the foreground like-sign mass spectrum (open circles) 
together with the various background components discussed above (the 
normalized combinatorial background, and the absolutely calculated cross 
pairs, jet pairs and $e$-$h$ pairs) and the $b\bar{b}$ pairs calculated as 
described in Section \ref{sec:cocktail}. The bottom panel shows the ratio 
of the foreground like-sign spectrum to the sum of all the background 
components. Similar comparisons for the five centrality bins used in this 
analysis are shown in Fig. \ref{fig:like-sign}.

\begin{figure}[hbt!]
\includegraphics[width=0.9\linewidth]{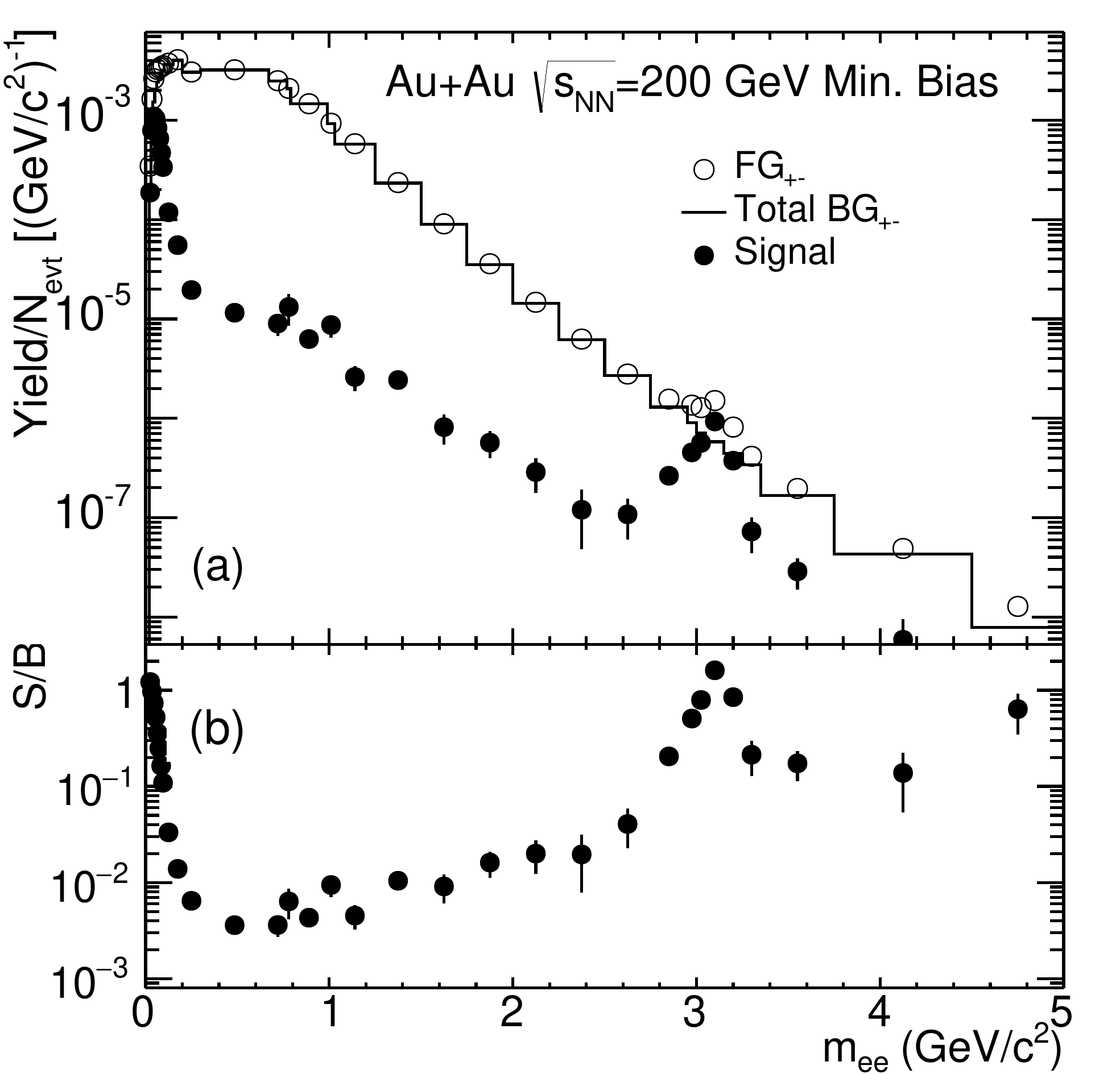}
\caption {
(a) MB mass spectra of the unlike sign foreground events (FG$_{+-}$), 
the calculated total background (BG$_{+-}$) and the raw signal S. 
(b) The signal to background ratio.
}
\label{fig:signal_to_background} 
\end{figure}

In general the background is well reproduced both in shape and magnitude. 
In particular, for the most central bins, the background is reproduced 
with sub-percent accuracy. There are, however, a couple of regions where 
the ratio foreground/background is different from one. There is a 
deviation of the order of a few percent at masses $m_{ee} < $~100 
MeV/$c^2$.  This is clearly visible in the three most central bins. A 
number of factors could be responsible for this deviation, such as scale 
errors in the cross pairs or the jet pairs. However, in this mass region 
the signal to background ratio is relatively good as shown in Fig. 
\ref{fig:signal_to_background} and a deviation of the order of a few 
percent in the background is negligible. There also seems to be a 
deviation at $m_{ee} > $ 1~GeV/$c^2$ for the 10\%--20\% and 20\%--40\% 
centrality bins. This deviation could indicate underestimations of the 
flow or the back-to-back jet contributions, due to the precision in these 
measurements, or the existence of an additional correlation that is not 
taken into account in any of the calculated background components. To be 
conservative, this deviation is considered as evidence of unsubtracted 
background and its magnitude is assigned as a mass dependent systematic 
uncertainty of the signal.

%%%%%%%%%%%%%%%%%%%%%%%%%%%%%%%%%%%%%%%%%%%%%%%%%%%%%%%%%%%% Fig_17

Figure \ref{fig:signal_to_background} shows the MB mass spectra 
of the foreground unlike sign events (FG$_{+-}$), the calculated total 
background (BG$_{+-}$) and the raw signal obtained by their subtraction. 
The signal to background ratio is shown in the bottom panel. This result 
will be discussed in reference to previously published PHENIX results in 
Section \ref{sec:comparison_run4}.

%%%%%%%%%%%%%%%%%%%%%%%%%%%%%%%%%%%%%%%%%%%%%%%%%%%%%%%%%%%% Fig_18
 
          \subsection{Raw Spectra and Efficiency Corrections}
          \label{sec:raw_spectra}
          
\begin{figure}[hbt!]
\includegraphics[width=1.0\linewidth]{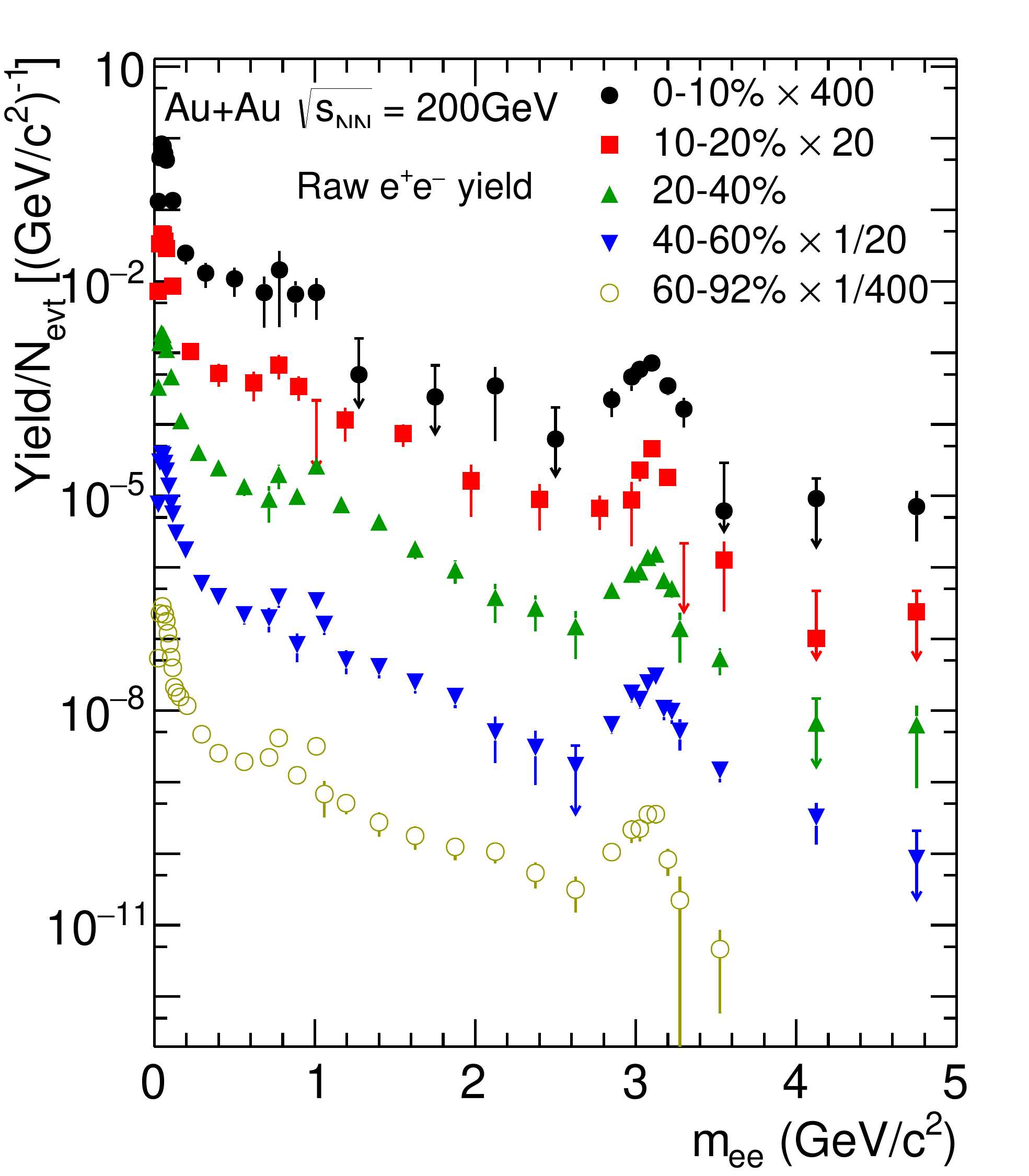}
\caption{(Color online) Raw mass spectra for the five centrality bins.}
\label{fig:raw_mass_spectra} 
\end{figure}

Figure \ref{fig:raw_mass_spectra} shows the raw mass spectra, obtained 
after subtracting the pair background, for the five centrality bins of 
this analysis.

%%%%%%%%%%%%%%%%%%%%%%%%%%%%%%%%%%%%%%%%%%%%%%%%%%%%%%%%%%%% Fig_19

To obtain the invariant mass spectrum inside the ideal PHENIX 
acceptance, the \ee raw mass yield is corrected for reconstruction 
efficiency effects according to:
\begin{eqnarray}
\frac{{\rm d}N}{{\rm d}m_{ee}} = \frac{1}{N_{\rm evt}}\frac{N(m_{ee})}{\Delta m_{ee}} \frac{1}{\epsilon^{total}_{\rm pair}}  
\end{eqnarray}
where $N_{\rm evt}$ is the number of events, $N(m_{ee})$ is the number of \ee 
pairs with invariant mass $m_{ee}$ and $\Delta m_{ee}$ is the mass bin 
width. $\epsilon^{total}_{\rm pair}$ is the total pair reconstruction 
efficiency that includes the eID efficiency of the neural networks, losses 
incurred by dead or inactive areas in the detector, pair cut losses and 
detector occupancy effects. The total pair reconstruction efficiency 
$\epsilon^{total}_{\rm pair}$ can thus be written as:
\begin{equation}
\label{eqn:effcor}
\epsilon^{total}_{\rm pair} = \epsilon^{eID}_{\rm pair}\cdot\epsilon^{live}_{\rm pair}\cdot\epsilon^{ghost}_{\rm pair}\cdot\epsilon^{mult}_{\rm pair}
\end{equation}
where $\epsilon^{eID}_{\rm pair}$ is the $e^+e^-$ pair reconstruction 
efficiency including the efficiency of all the electron identification 
cuts and the HBD double-hit rejection cut, $\epsilon^{live}_{\rm pair}$ is the 
pair efficiency from the detector active area with respect to the ideal 
PHENIX detector acceptance, $\epsilon^{ghost}_{\rm pair}$ reflects the 
efficiency loss due to the pair cuts that remove ghost pairs in the 
various detectors (see Section \ref{sec:pair_cuts}) and 
$\epsilon^{mult}_{\rm pair}$ is the multiplicity dependent efficiency loss 
discussed below in this subsection.

\begin{figure}[hbt!]
\includegraphics[width=1.0\linewidth]{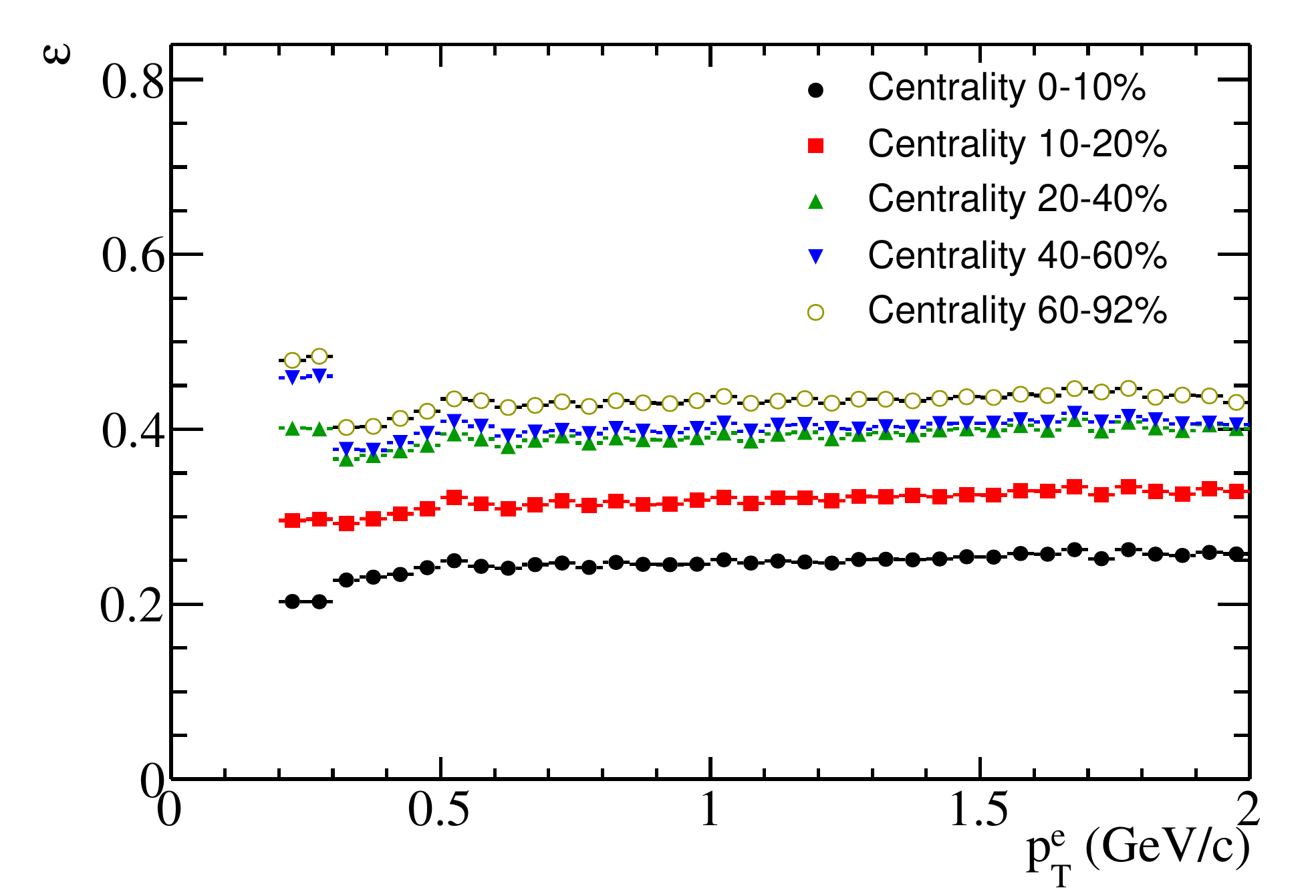}
\caption{(Color online) Single electron reconstruction efficiency vs. \pt 
for the five centrality bins.}
\label{fig:single_electron_eff} 
\end{figure}

The single electron reconstruction efficiency, defined as $\epsilon$ = 
$\sqrt{\epsilon^{eID}_{\rm pair}\cdot\epsilon^{mult}_{\rm pair}}$ is shown in Fig. 
\ref{fig:single_electron_eff} vs \pt for the five centrality bins. This 
efficiency is not actually used in the analysis.  It is shown here for 
illustration purposes. The change of efficiency below 0.3~GeV/$c$ arises 
from the cut optimization in two $p_T$ ranges (see Section 
\ref{sec:nn_cuts}).

%%%%%%%%%%%%%%%%%%%%%%%%%%%%%%%%%%%%%%%%%%%%%%%%%%%%%%%%%%%% Fig_20

The product 
$\epsilon^{eID}_{\rm pair}\cdot\epsilon^{live}_{\rm pair}\cdot\epsilon^{ghost}_{\rm pair}$ 
is determined as follows. A cocktail of all the known hadronic sources 
contributing to the \ee pair spectrum is generated within $|\eta| <$ 0.6 
and 2$\pi$ in azimuthal angle. Details about the various sources of the 
cocktail are given in Section \ref{sec:cocktail}.  The cocktail is passed 
through a full {\sc geant} simulation of the PHENIX detector~\cite{geant} 
and analyzed in the same way as the data, including eID cuts, fiducial 
cuts and pair cuts. The resulting output is referred to as the 
reconstructed cocktail. The ratio of this reconstructed cocktail to the 
generated cocktail filtered through the ideal PHENIX acceptance (but 
without momentum smearing), gives the product 
$\epsilon^{eID}_{\rm pair}\cdot\epsilon^{live}_{\rm pair}\cdot\epsilon^{ghost}_{\rm pair}$. 
This correction is derived in the two dimensional space of mass-pair \pt.

\begin{figure}[hbt!]
\includegraphics[width=0.49\linewidth]{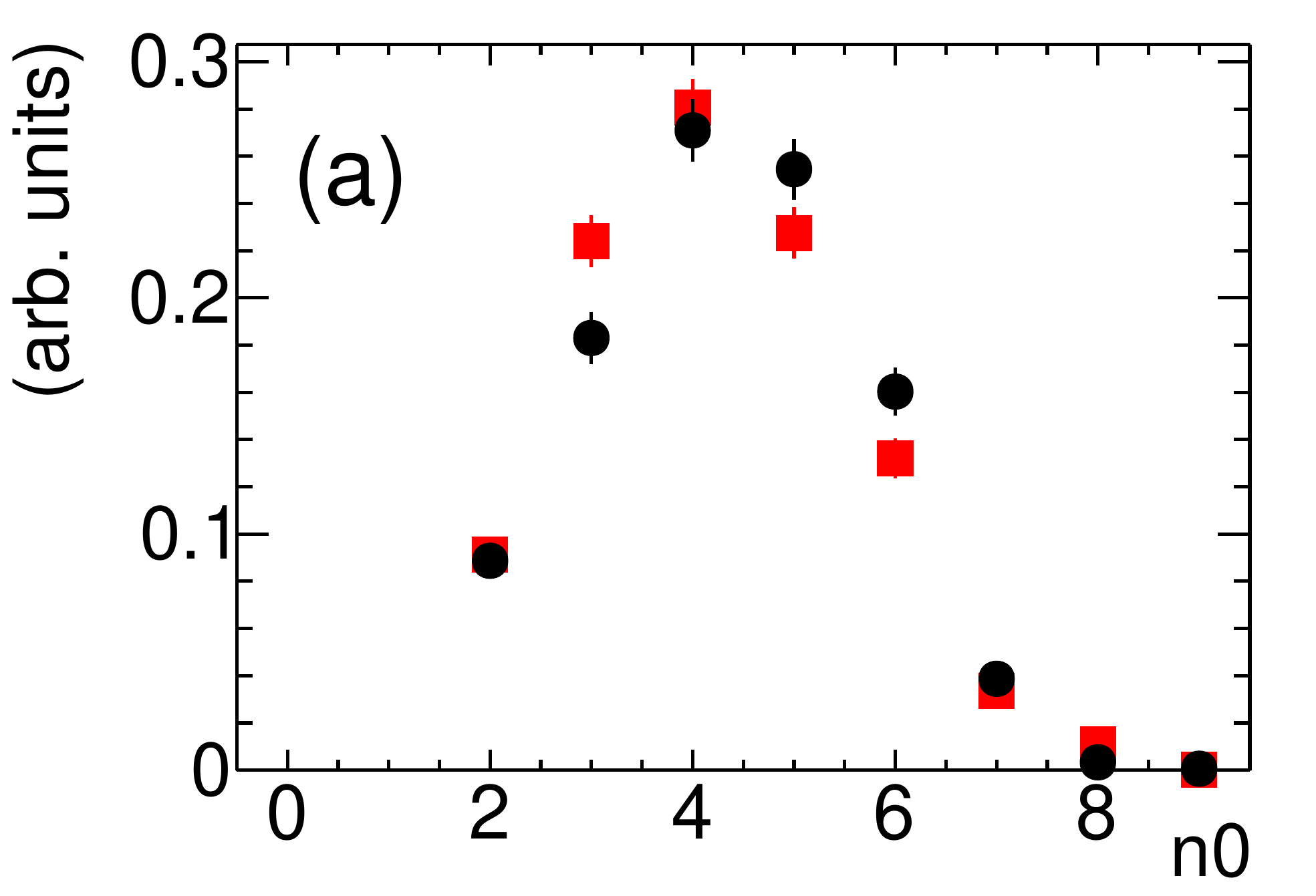}
\includegraphics[width=0.49\linewidth]{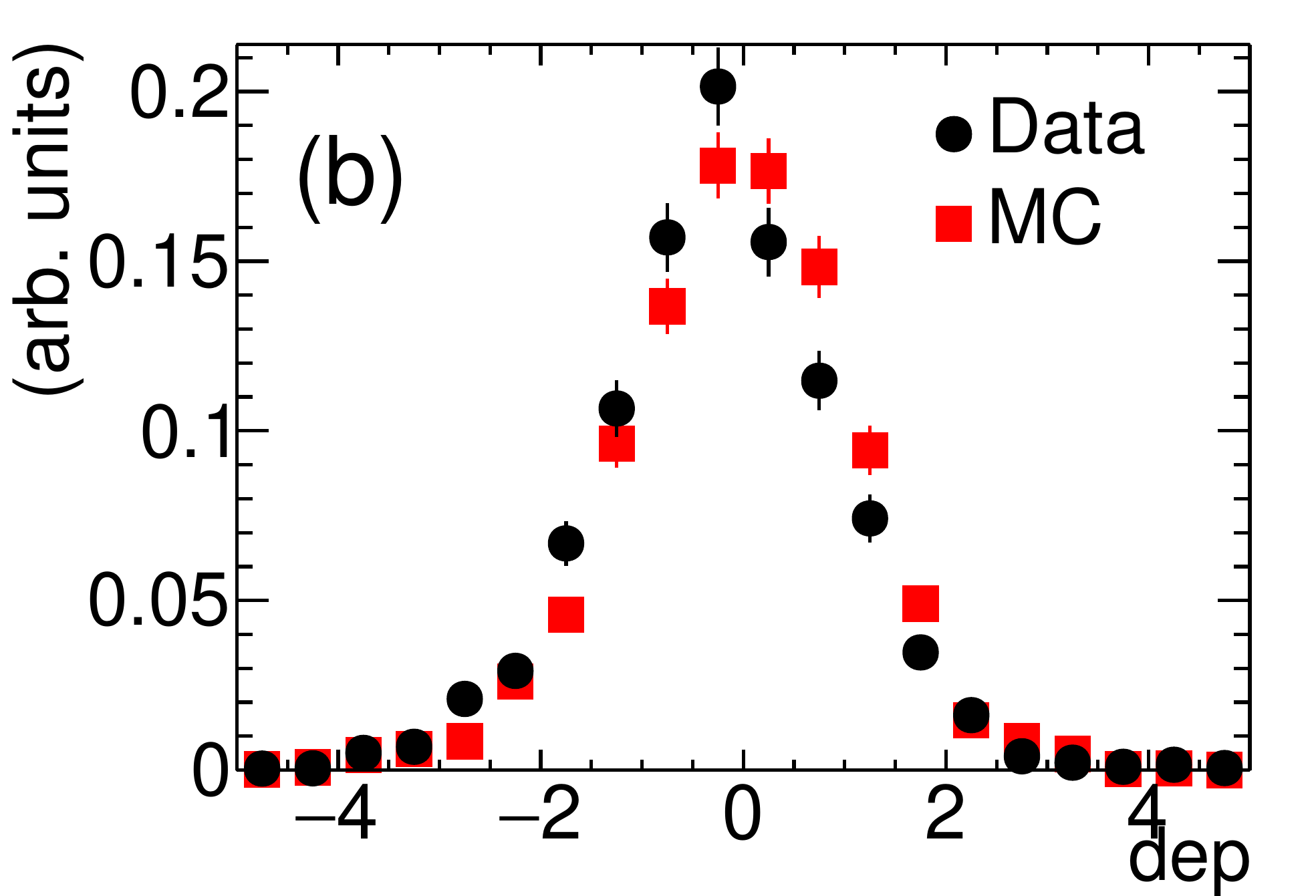}
\includegraphics[width=0.49\linewidth]{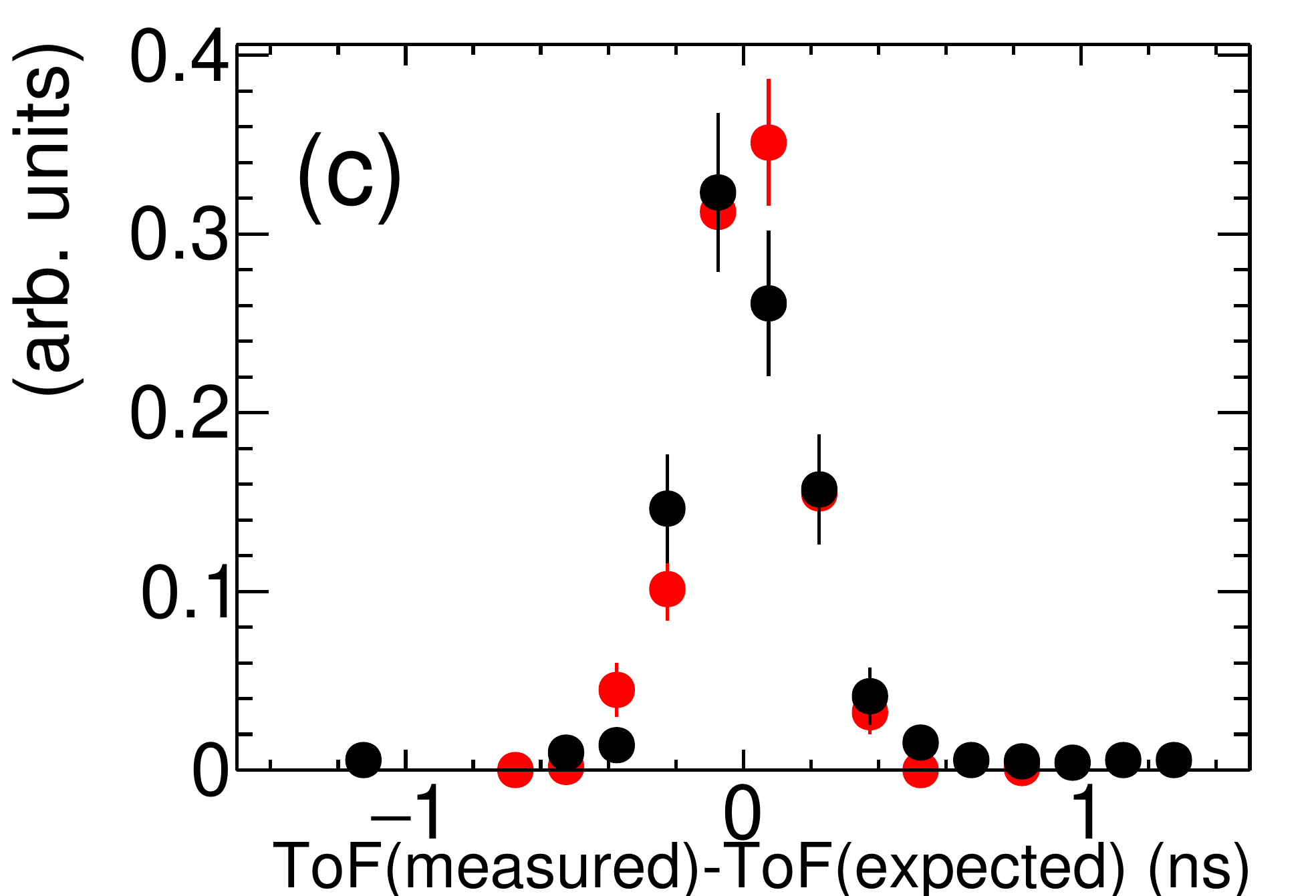}
\includegraphics[width=0.49\linewidth]{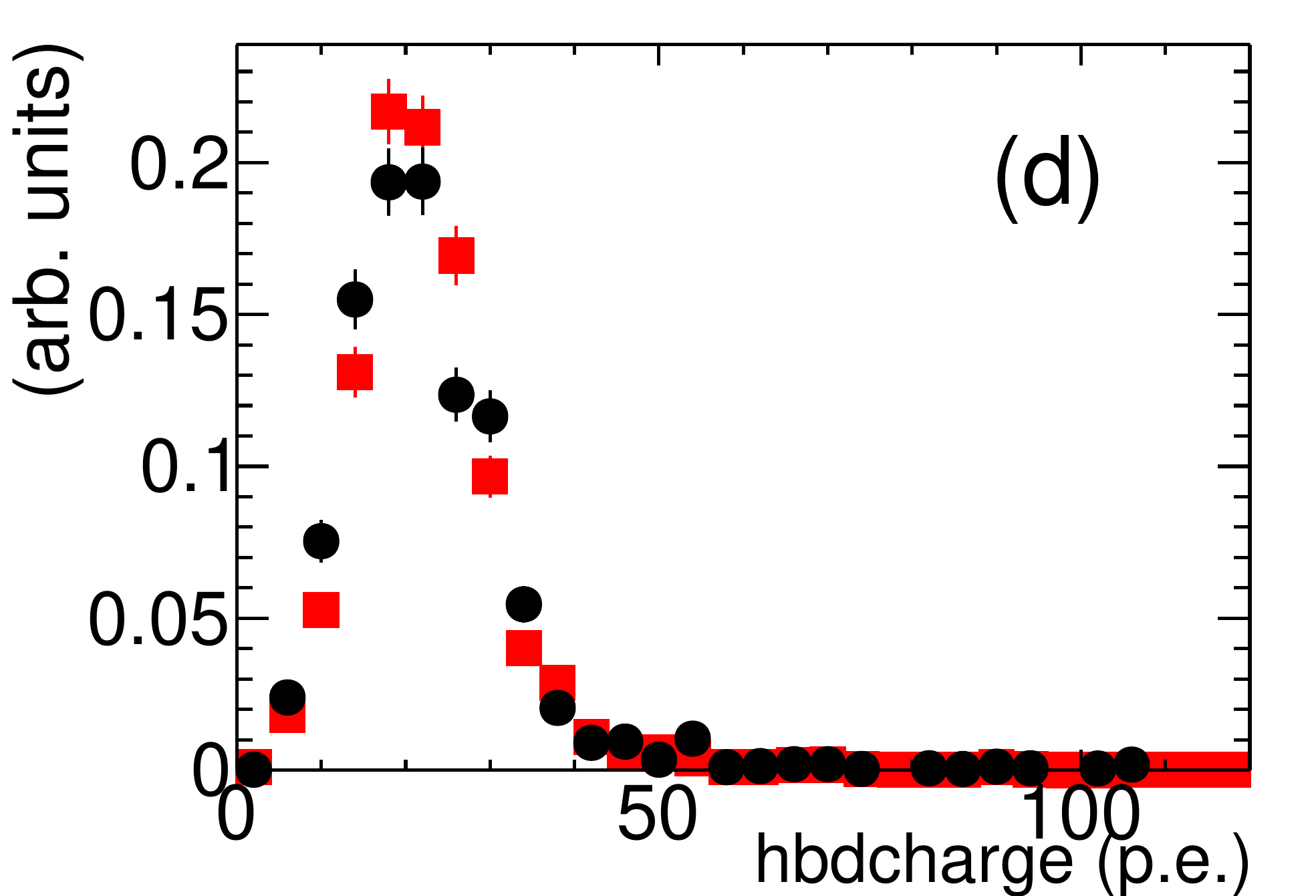}
\caption {(Color online) Comparison of electron identification variables 
in data (black) and in simulations (red). The variables are described in 
Section \ref{sec:electron_identification}. electrons in data and 
simulations are from fully reconstructed $\pi^0$ Dalitz decays with 
opening angle larger than 100 mrad.}
\label{fig:data_simulations} 
\end{figure}

Special care is taken to tune the simulations to the data to ensure that 
the detector response in the simulations is the same as in real data for 
all the subsystems involved in the analysis. As an example, Fig. 
\ref{fig:data_simulations} shows a comparison of a few electron 
identification variables in data and simulations. For this comparison we 
use a clean sample of electrons provided by fully reconstructed $\pi^0$ 
Dalitz decays with an opening angle larger than 100 mrad from the 
60\%--92\% centrality bin where the occupancy effects are very small and 
can be ignored. The eID variables of the two tracks from these pairs are 
compared to those of $\pi^0 \rightarrow$ \ee $\gamma$ simulations.

%%%%%%%%%%%%%%%%%%%%%%%%%%%%%%%%%%%%%%%%%%%%%%%%%%%%%%%%%%%% Fig_21

The HBD occupancy effects are taken into account by embedding the HBD hits 
from the cocktail simulation into real HBD events, and thus are included 
in the product 
$\epsilon^{eID}_{\rm pair}\cdot\epsilon^{live}_{\rm pair}\cdot\epsilon^{ghost}_{\rm pair}$. 
There are two other occupancy effects in the central arms that need to 
be taken into account and are included in Eq.~(\ref{eqn:effcor}) by the 
additional multiplicative factor $\epsilon^{mult}_{\rm pair}$. The first one 
is the decrease of track reconstruction efficiency as the detector 
occupancy increases with centrality. This loss is referred to as 
$\epsilon^{embed}_{\rm pair}$ and is determined by an embedding procedure.  
Electrons from $\phi$ decays that are reconstructed in single particle 
simulations, are embedded into real \auau events. Then the embedded events 
are run through the full reconstruction software chain and analyzed in 
exactly the same way as the data. The embedding efficiency for single 
tracks $\epsilon^{embed}_{single}$ is determined as the ratio of the 
number of reconstructed electron tracks from embedded data to the number 
of embedded tracks. The pair embedding efficiency is calculated as the 
square of the single track embedding efficiency, $\epsilon^{embed}_{\rm pair} 
= (\epsilon^{embed}_{single})^2$.

The second occupancy effect comes from the initial rejection of background 
electrons, discussed in Section \ref{sec:tagging_rich}, where PMTs fired 
by background electron tracks are removed. If such an electron is close to 
a signal electron in the RICH, the associated PMTs of the signal electron 
are also removed. The probability for this to happen is relatively small 
and increases with multiplicity. This loss is referred to as 
$\epsilon^{TPMT}_{\rm pair}$ and it is estimated by monitoring the yield of 
\ee pairs below 20~${\rm MeV}/c^2$ before and after erasing the PMTs 
for each centrality bin. This mass region is dominated by Dalitz decays 
and $\gamma$ conversions and provides a clean electron pair sample with a 
signal-to-background ratio of $\sim$200 even for the most central events.
Using these efficiency losses, $\epsilon^{mult}_{\rm pair}$ can be expressed as:
\begin{equation}
\epsilon^{mult}_{\rm pair} = \epsilon^{embed}_{\rm pair} \cdot \epsilon^{TPMT}_{\rm pair}
\end{equation}

Table \ref{tab:eff_emb} summarizes the values of $\epsilon^{embed}_{\rm pair}$ 
and $\epsilon^{TPMT}_{\rm pair}$ for the five centrality bins.

%===================================================== Table_IV
\begin{table}[hbt!]
\caption{Efficiency loss due to detector occupancy in the central arms $\epsilon^{embed}_{\rm pair}$ and to the tagging of RICH PMTs discussed in Section  \ref{sec:tagging_rich}  for the five centrality bins used in this analysis.}
\begin{ruledtabular} \begin{tabular}{lccccc}
& &  & Centrality &  &  \\
&0\%--10\% & 10\%--20\% & 20\%--40\% & 40\%--60\% & 60\%--92\% \\
\hline
$\epsilon^{embed}_{\rm pair}$& 0.53 & 0.65 & 0.76 & 0.86 & 0.95\\
$\epsilon^{TPMT}_{\rm pair}$& 0.88 & 0.92 & 0.94 & 0.98 & 1.00\\
\end{tabular} \end{ruledtabular}
\label{tab:eff_emb}
\end{table}

\begin{figure}[hbt!]
\includegraphics[width=1.0\linewidth]{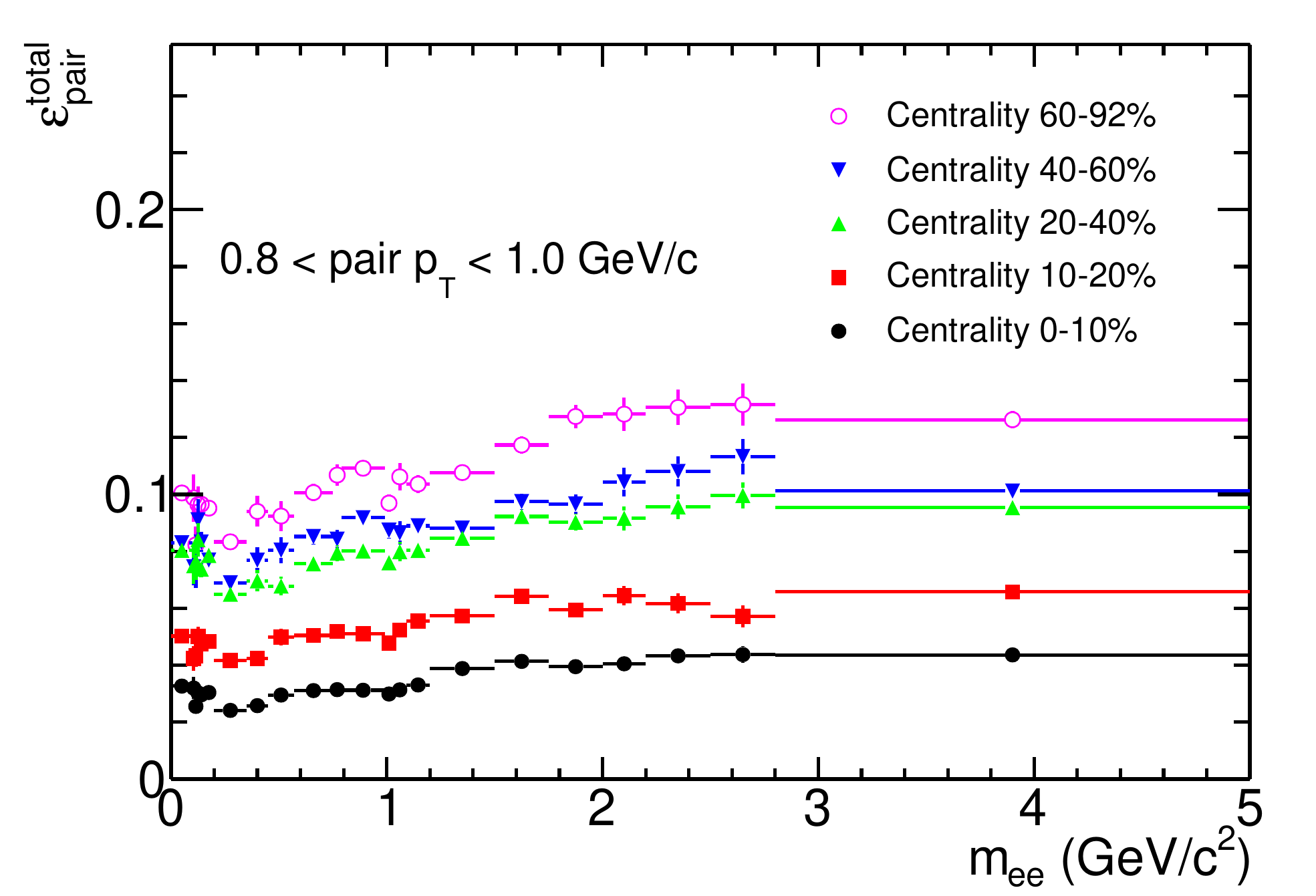}
\caption{(Color online) Pair efficiency correction for the pair $p_T$ range between 0.8 and 1.0~GeV/$c$ for each centrality bin. This represents the total efficiency including the eID selection cuts based on neural networks, losses in the acceptance due to detector inactive areas, losses induced by the pair cuts and occupancy effects in the central arm detectors.}
\label{fig:eff} 
\end{figure}

Figure \ref{fig:eff} shows the total pair reconstruction efficiency 
$\epsilon^{total}_{\rm pair}$ for pair \pt within 0.8-1.0~GeV/$c$ for each 
centrality bin.

%%%%%%%%%%%%%%%%%%%%%%%%%%%%%%%%%%%%%%%%%%%%%%%%%%%%%%%%%%%% Fig_22

          \subsection{Systematic Uncertainties}
          \label{sec:systematic_uncertainties}
           
The main systematic uncertainties on the corrected data arise from 
uncertainties on the electron identification, the acceptance and the 
background subtraction. They are discussed in detail below and summarized 
in Table \ref{tab:total_syst}. These uncertainties move all data points in 
the same direction but not by the same factor
 
%===================================================== Table_V
\begin{table*}[hbt!]
\caption{Summary of systematic uncertainties assigned to the corrected 
data for MB collisions. }
\begin{ruledtabular} \begin{tabular}{clccc}
& Component & Mass range  & Systematic uncertainty & \\
\hline
& eID + occupancy effects &    & $\pm$4\%          & \\
& Acceptance (time) &   & $\pm$8\%                 & \\
& Acceptance (MC) &    & $\pm$4\%                  & \\
& Combinatorial background & 0--5~GeV/$c^2$  &  $\pm$25\%  ($m_{ee} = 0.6$~GeV/$c^2$) & \\
& Residual yield & 0--0.08~GeV/$c^2$ & $-$5\% ($m_{ee} = 0.08$~GeV/$c^2$) & \\
& Residual yield & 1--5~GeV/$c^2$  & $-$15\%  ($m_{ee} = 1$~GeV/$c^2$) &  \\
\end{tabular} \end{ruledtabular}
\label{tab:total_syst}
\end{table*}

\subsubsection{Systematic uncertainty on electron identification and 
occupancy effects}

As described in Section \ref{sec:electron_identification}, electron 
identification is achieved using three neural networks. Different 
threshold cuts for the neural networks result in different electron 
identification efficiency and occupancy effects. The thresholds in the 
neural networks are varied by $\pm$20\% around the selected values and the 
variations of the electron pair yield in the mass region $m_{ee} <$ 150 
MeV/$c^2$, after applying the efficiency correction, are used to assess 
the systematic uncertainty of electron identification and occupancy 
effects.

By changing the thresholds by $\pm$20\% the raw electron pair yield 
changes by about $\pm$50\%. However, once the corresponding efficiency 
corrections are applied, the variations are below 4\% for all the 
centrality bins. Based on these results, we assign a $\pm$4\% systematic 
uncertainty on the electron identification.

\subsubsection{Systematic uncertainty on the acceptance}

We consider two sources of systematic uncertainties on the acceptance: 
variations of the pair acceptance vs time and variations of the pair 
acceptance between data and MC simulations.

The pair acceptance systematic uncertainty vs time is studied by 
considering the variations of the number of electron pairs per event for 
each run group. The weighted average of the $rms$ of the number of 
electrons 
per event in the five run groups is found to be 8\% and it is taken as the 
systematic uncertainty of the acceptance variation over time.

The systematic uncertainty on the data vs MC pair acceptance is studied by 
comparing the reconstructed $\pi^0$ yield in data and simulations. In data 
we select reconstructed pairs with $m_{ee} <$ 100~MeV/$c^2$, after 
subtracting the combinatorial and correlated components of the background, 
using data from one of the run groups.  In the MC simulations we use 
reconstructed pairs in the same mass range from $\pi^0$ Dalitz decays 
applying the fiducial cuts for the corresponding run group. The entire 
detector is divided into four sectors. Data and MC simulations are 
normalized in one sector. The variations of the yield ratios between data 
and MC simulations in the other sectors ranges between 1\% and 8\%. The 
weighted average of these variations is found to be 4\% and it is taken as 
the systematic uncertainty of the acceptance agreement between data and MC 
simulations.

\subsubsection{Systematic uncertainty on the background subtraction}

We consider two sources of systematic uncertainties on the background 
subtraction:
 
(i) Uncertainty on the combinatorial background subtraction. It is 
primarily due to the uncertainty in the normalization factor, and the 
latter is determined by the statistics in the normalization window, namely 
by 1/$\sqrt{N_{LS}}$ (see Section \ref{sec:background_normalization}). 
This translates into a relative uncertainty of the signal $\delta S/S = 
1/\sqrt{N_{LS}} \times B/S$. The ratio $B/S$ depends both on mass and 
centrality. In Table \ref{tab:total_syst} we quote the uncertainty at \mee 
= 0.6~GeV/$c^2$ which represents the worst case in mass, for MB 
events. The centrality dependence results in variations of the order of 
15\% from the MB values.
 
(ii) In the ideal case, the like-sign residual yield, i.e. the like-sign 
yield after subtracting all the background sources, should be zero. In 
practice it is not. As shown in Figs. \ref{fig:like-sign_0-92} and 
\ref{fig:like-sign}, there is a small residual yield. In this analysis, we 
assume that any residual yield is entirely due to unsubtracted background, 
and we take it as an additional source of systematic uncertainty, after 
transforming it into unlike-sign residual yield via the acceptance 
correction factor $\alpha$. This uncertainty takes into account any 
possible discrepancy in shape or magnitude of the various subtracted 
sources of background. The factor $\alpha$ accounts for the different 
acceptance of the PHENIX detector for like and unlike sign pairs. It is 
calculated as a function of pair mass and pair \pt using the mixed event 
background as:

\begin{equation}
\alpha(m,p_T)= \frac{MIX_{+-}(m,p_T)} {MIX_{++}(m,p_T)+MIX_{--}(m,p_T)}
\end{equation}

%%%%%%%%%%%%%%%%%%%%%%%%%%%%%%%%%%%%%%%%%%%%%%%%%%%%%%%%%%%% Fig_23
\begin{figure*}[hbt!]
\includegraphics[width=0.44\linewidth]{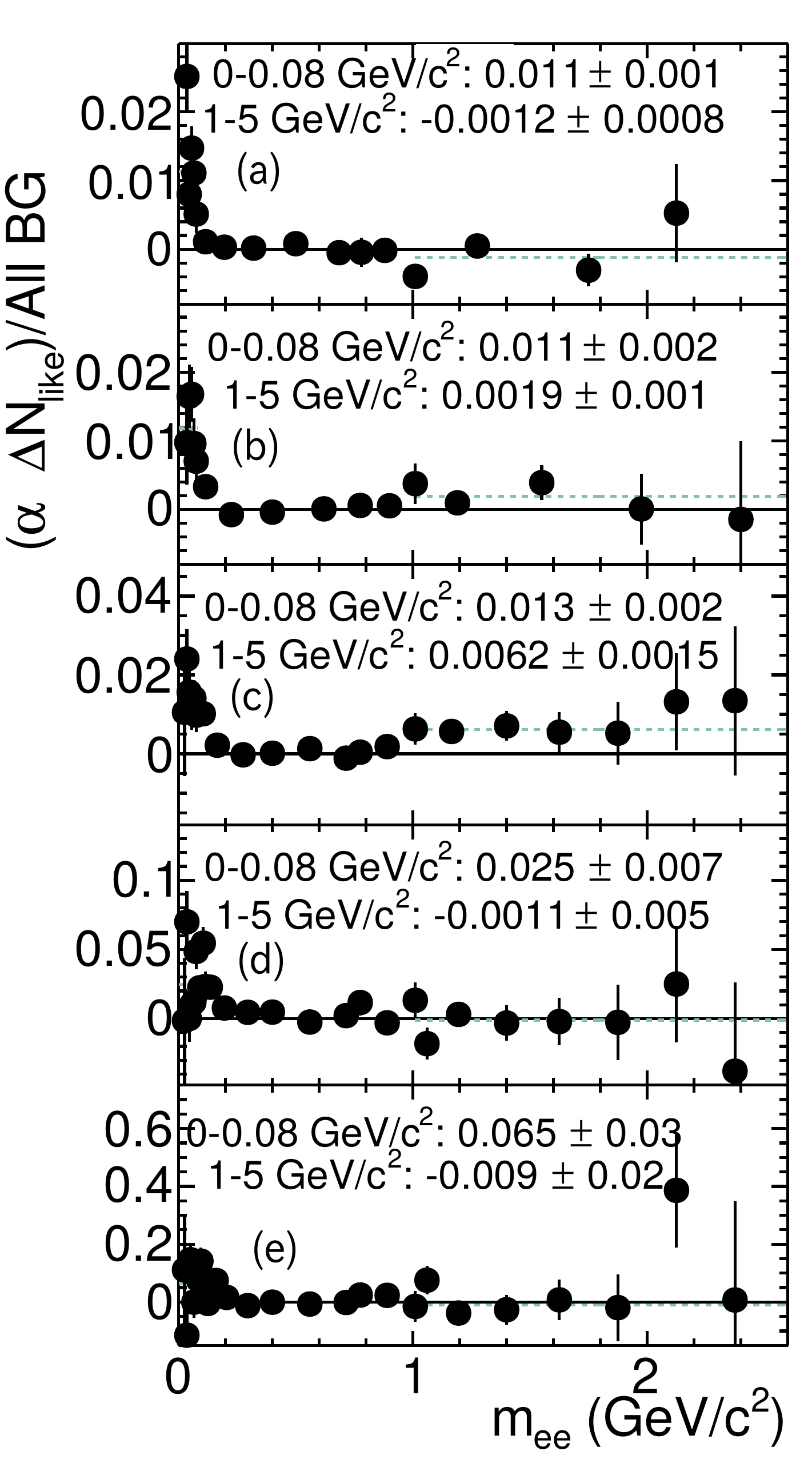}
\includegraphics[width=0.44\linewidth]{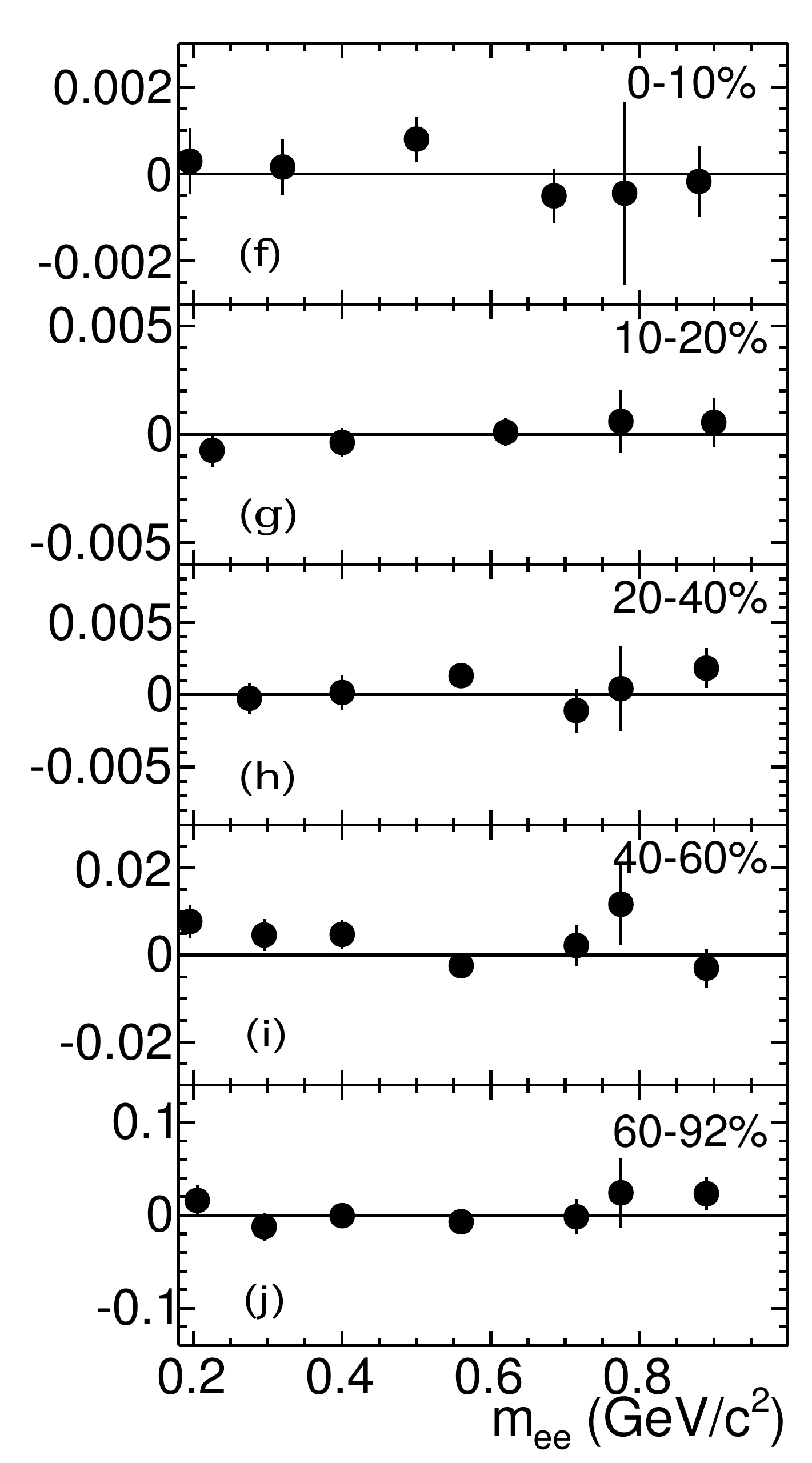}
\caption{(Color online) (a--e) Unlike-sign residual background yield 
derived from the like-sign residual yield, obtained after subtracting 
all background sources, via the acceptance correction factor $\alpha$ 
(see text). The legend and the dashed lines show the results of 
constant fits below 80~MeV/$c^2$ and above 1~GeV/$c^2$.  (f--j) Zoomed 
views in the vertical axis for the 0.2--1~GeV/$c^2$ mass range. }
\label{fig:residual_yield}
\end{figure*}

Figure~\ref{fig:residual_yield} panels (a)--(e) show $\alpha$ times the 
like-sign residual yield divided by the sum of all unlike-sign background 
sources as a function of mass for the five centrality bins, which 
represent the relative residual background yield in the unlike sign mass 
spectrum.  The mass regions $m_{ee}<0.08$~GeV/$c^2$, 
0.2~GeV/$c^2<m_{ee}<1.0$~GeV/$c^2$ and $m_{ee}>1$~GeV/$c^2$ are fitted to 
a constant to quantify the magnitude of the residual unlike-sign yield. 
The fit results are also shown.  Figure~\ref{fig:residual_yield} panels 
(f)--(j) show zoomed views in the vertical axis for the 0.2--1~GeV/$c^2$ 
mass range. The fits in the mass region $m_{ee}=0.2$--1.0~GeV/$c^2$ give 
results that are consistent with zero for all centrality bins.  For the 
other two mass ranges, the residual yields are considered as sources of 
systematic uncertainties if their significance is larger than 2$\sigma$.

The total systematic uncertainty in the background subtraction is obtained 
as the quadratic sum of the systematic uncertainties due to the 
combinatorial background subtraction and the residual yield. Both 
contributions are listed in Table \ref{tab:total_syst} for MB 
collisions. It is worth noting that the systematic uncertainty of the 
background subtraction is much lower than the required accuracy to measure 
a signal with the $S/B$ values shown in Section 
\ref{sec:quantitative_background}.

          \subsection{Cross checks}
          \label{sec:cross_checks}

A second independent analysis was performed as a cross check. The key 
features of the second analysis are discussed here. A more detailed 
description is given in Appendix~\ref{app:sb_analysis}. The second 
analysis is similar to the analysis described in Ref.~\cite{Adare:2009qk}, 
but it makes use of the HBD and includes all the important improvements 
developed in this work.  In particular, it makes use of the time-of-flight 
information for better hadron rejection, implements the shape distortion 
of the mixed event background due to elliptic flow (Section 
\ref{sec:combinatorial_background}), subtracts the correlated 
electron-hadron background (Section \ref{sec:electron_hadron_pairs}), and 
explicitly considers the away-side jet-pair component in the background 
subtraction (Section \ref{sec:jet_pairs}).

Important elements of the independent analysis are different from those of 
the main analysis. The most significant differences are: (i) The HBD 
underlying event subtraction is done using the average charge in the 
vicinity of a track as opposed to the average charge in a module as used 
in the main analysis. (ii) Electron identification is achieved by a 
sequence of independent one-dimensional cuts on each of the electron 
identification variables instead of the neural network approach. (iii) The 
normalization of each background source is determined from a fit to the 
like-sign spectra, in contrast to the main analysis where all the 
correlated background sources are absolutely normalized and only the 
combinatorial background is normalized to the like sign spectra.

The second analysis results in a factor of two smaller signal-to-background 
ratio and a 10\% reduction in purity of the electron sample in central 
collisions. However, once corrected for efficiency, the results of the 
second analysis are consistent within uncertainties with those obtained 
with the main analysis described in this section.
 
\section{COCKTAIL OF HADRONIC SOURCES}
    \label{sec:cocktail}

In this section we describe the procedures used to calculate the expected 
dielectron yield from hadronic decays, commonly referred to as the 
hadronic cocktail, that will be compared to the experimental results in 
Section \ref{sec:results}. The known \ee sources are calculated using the 
{\sc exodus}, {\sc pythia} and {\sc mc@nlo} event generators. {\sc exodus} 
is a phenomenological generator that simulates phase space distributions 
of the relevant electron sources and their decays~\cite{Adare:2008ac}. It 
generates the photonic sources, i.e. Dalitz decays of light neutral 
mesons: $\pi^{0}$, $\eta$, $\eta'$ $\rightarrow e^+e^-\gamma$ and $\omega 
\rightarrow e^+e^-\pi^0$ and the nonphotonic sources, i.e. dielectron 
decays of mesons: $\rho$, $\omega$, $\phi$, $J/\psi \rightarrow e^+e^-$. 
{\sc pythia}~\cite{Sjostrand:2000wi} and 
{\sc mc@nlo}~\cite{Frixione:2002ik,Frixione:2003ei} are used to generate 
the correlated pairs from semi-leptonic decays of heavy flavor (charm and 
bottom) mesons. The hadrons are assumed to have uniform pseudorapidity 
density within $|\eta| <$ 0.35 and uniform azimuthal distribution in 
2$\pi$. Once generated, the sources are filtered through the ideal 
acceptance of the PHENIX detector and smeared with the detector resolution 
for comparison to the measured invariant mass spectrum.

\subsection{Neutral pions} 

The dominant electron source as well as the fundamental input for {\sc 
exodus} is $\pi^{0}$. The shape of the $\pi^0$ \pt distribution is 
parameterized as:
\begin{eqnarray}
E\frac{d^{3}\sigma}{d^{3}p} \propto \frac{1}{(e^{-ap_{T} - bp_{T}^{2}} + p_{T}/p_{0})^{n}} 
\label{eqn:hagedorn} 
\end{eqnarray}
The parameters, $a$, $b$, $p_{0}$ and $n$, are obtained by a simultaneous 
fit of the PHENIX published results for $\pi^0$~\cite{Adler:2006bw, 
Adare:2008qa} and charged pions~\cite{Adler:2003cb}. The resulting fit 
parameters are shown in Table \ref{tab:hage_par} for the five centrality 
bins of this analysis. The absolute magnitude of the $\pi^0$ rapidity 
density, $dN_{\pi^0}/dy$, is obtained by fitting the cocktail to the data 
(see Section \ref{sec:cocktail_normalization}).

%===================================================== Table_VI
\begingroup  \squeezetable
\begin{table}[hbt!]
\caption{Fit parameters derived from the $\pi^0$ and charged pion $p_T$ 
distributions~\cite{Adler:2006bw,Adare:2008qa,Adler:2003cb} for different 
centralities using Eq. (\ref{eqn:hagedorn}). 
}
\begin{ruledtabular} \begin{tabular}{cllllllc}
&Parameter & 0\%--10\% & 10\%--20\%  & 20\%--40\%   & 40\%--60\%  & 60\%--92\% &\\ 
\hline
&$a$ [(GeV/$c$)$^{-1}$]    & 0.57  & 0.53 & 0.43   & 0.36   & 0.33  &\\
&$b$ [(GeV/$c$)$^{-2}$]    & 0.19  & 0.16 & 0.11   & 0.13   & 0.088 &\\
&$p_0$ [GeV/$c$]         & 0.74  & 0.75 & 0.79   & 0.76   & 0.74 &\\
&$n$                              & 8.4   & 8.3  & 8.5    & 8.4    & 8.4 &\\
\end{tabular} \end{ruledtabular}
\label{tab:hage_par}
\end{table}
\endgroup
 
\subsection{Other mesons}
\label{sec:other_mesons}

The \pt distributions of other light mesons are based on the 
parametrization of the pion spectrum assuming $m_{T}$ 
scaling~\cite{Adare:2009qk}, i.e. Eq. (\ref{eqn:hagedorn}) is used with 
$p_{T}$ replaced by $\sqrt{p_{T}^{2}+m_{meson}^2-m_{\pi^{0}}^{2}}$. This 
assumption reproduces well the measured light meson \pt distributions in 
Au$+$Au collisions as demonstrated in~\cite{Adare:2009qk}. The absolute 
normalization for each meson is provided by the ratio of the meson to 
$\pi^0$ invariant yield at high \pt (\pt $\geq$ 5~GeV/$c$). We use the 
values from Ref.~\cite{Adare:2010de}, summarized in Table 
\ref{table:meson_to_pi0}.

%===================================================== Table_VII
\begin{table}[hbt!]
\caption{Meson to $\pi^0$ ratio at high \pt (\pt $\geq$ 5~GeV/$c$)  
obtained from PHENIX data in \pp collisions~\cite{Adare:2010de}.}
\begin{ruledtabular} \begin{tabular}{ccccccc} 
&$\eta$/$\pi^0$ &  $\rho$/$\pi^0$ & $\omega$/$\pi^0$ & $\eta'$/$\pi^0$ & $\phi$/$\pi^0$ &\\ 
\hline
&0.48           &   1.0           &  0.90            & 0.25            & 0.40 &\\  
\end{tabular} \end{ruledtabular}
\label{table:meson_to_pi0}
\end{table}

The values were obtained from \pp collisions and are taken to be valid for 
\auau collisions because at high \pt the suppression of all mesons is 
found to be very similar to the $\pi^0$ suppression and consequently the 
meson/$\pi^0$ ratios in \auau collisions remain unchanged with respect to 
the ratios in \pp collisions~\cite{Adare:2011ht,Adare:2010pt,Adare:2010dc}.

For the \pt distribution of the $J/\psi$ we use the neutral pion \pt 
spectrum measured in \pp collisions~\cite{Adare:2007dg}, assuming $m_{T}$ 
scaling. Detector effects on the $J/\psi$ line shape are taken into 
account by passing the decay \ee through a {\sc geant} simulation of the 
PHENIX detector. The resulting \pt integrated invariant \ee mass 
distribution is then normalized to the measured cross section in \pp 
collisions~\cite{Adare:2009qk} and scaled to \auau collisions by the 
corresponding $\langle$\Ncollnospace$\rangle$ and the measured $R_{AA}$ 
for each centrality bin~\cite{Adare:2006ns}.

\subsection{Open heavy flavor}
\label{sec:open_heavy_flavor}

The correlated $e^+e^-$ yield from open heavy flavor decays is simulated 
using two different \pp event generators, {\sc pythia} and {\sc mc@nlo}, 
and measured $c\bar{c}$ and $b\bar{b}$ production cross sections.

{\sc pythia} simulations are used to calculate gluon fusion, the 
dominant process for heavy-quark production, in leading-order 
perturbative QCD.  Specifically, we use 
{\sc pythia}-{\footnotesize 6}~\cite{Sjostrand:2006za} \footnote{We use 
{\sc pythia}-{\scriptsize 6}~\cite{Sjostrand:2006za} with the 
following parameters MSEL[$c\bar{c}$]=4 or MSEL[$b\bar{b}$]=5, 
MSTP(91)=1, PARP(91)=1.5, MSTP(33)=1, PARP(31)=1.0, MSTP(32)=4, 
PMAS(4)=1.25, PMAS(5)=4.1.} and {\sc cteq{\footnotesize 5}l} as input 
parton distribution functions. The {\sc mc@nlo} package (vers. 
4.03)~\cite{Frixione:2002ik,Frixione:2003ei} is a next-to-leading order 
simulation that generates hard scattering events. These events are 
subsequently fed to {\sc herwig} (vers. 6.520)~\cite{Corcella:2000bw} 
for fragmentation in vacuum.

We use the $c\bar{c}$- and $b\bar{b}$-production cross sections measured by 
PHENIX~\cite{Adare:2014iwg}, by fitting the event generator ({\sc pythia} 
or {\sc mc@nlo}) output to the measured dielectron mass spectrum in \dau 
collisions for $m_{e^+e^-} >$ 1.15~GeV/$c^2$. These cross sections were 
scaled by the average number of \dau binary collisions 
($\langle$\Ncollnospace$\rangle$) to give the \pp equivalent cross 
section. For $b\bar{b}$, both generators gave within uncertainties the 
same result for the cross section extrapolated to zero invariant 
mass~\cite{Adare:2014iwg}: 
\begin{equation}
\frac{d\sigma^{pp }_{b\bar{b}}} {dy}\Big|_{y=0} = 1.36 \pm 0.32 ({\rm stat}) \pm 0.44({\rm syst})\ \mu{\rm b}
\label{eqn:sigma_bb}
\end{equation}

The $c\bar{c}$ cross section strongly depends on the 
event generator. The {\sc mc@nlo} yields the following cross 
section~\cite{Adare:2014iwg}:
\begin{equation}
\frac{d\sigma^{pp }_{c\bar{c}}} {dy}\Big|_{y=0} = 287\pm 29 ({\rm stat}) \pm 100({\rm syst})\ \mu{\rm b} \\
\label{eqn:sigma_cc_mcnlo} 
\end{equation}
whereas {\sc pythia} gives:
\begin{equation}
\frac{d\sigma^{pp }_{c\bar{c}}} {dy}\Big|_{y=0} = 106 \pm 9 ({\rm stat}) \pm 33({\rm syst})\ \mu{\rm b} \\
\label{eqn:sigma_cc_pythia} 
\end{equation}
This cross section, derived from \ee data in \dau collisions, is 
consistent within uncertainties with the cross section derived from 
measurements of single electrons from semileptonic decays of heavy flavor 
mesons in \pp collisions, extrapolated to \pt = 0~GeV/$c$ using {\sc 
pythia} simulations~\cite{Adare:2010de}. {\sc mc@nlo} was not used to 
derive the heavy flavor cross section from measurements of single 
electrons.

%%%%%%%%%%%%%%%%%%%%%%%%%%%%%%%%%%%%%%%%%%%%%%%%%%%%%%%%%%%% Fig_24
\begin{figure}[hbt!] 
\includegraphics[width=1.0\linewidth]{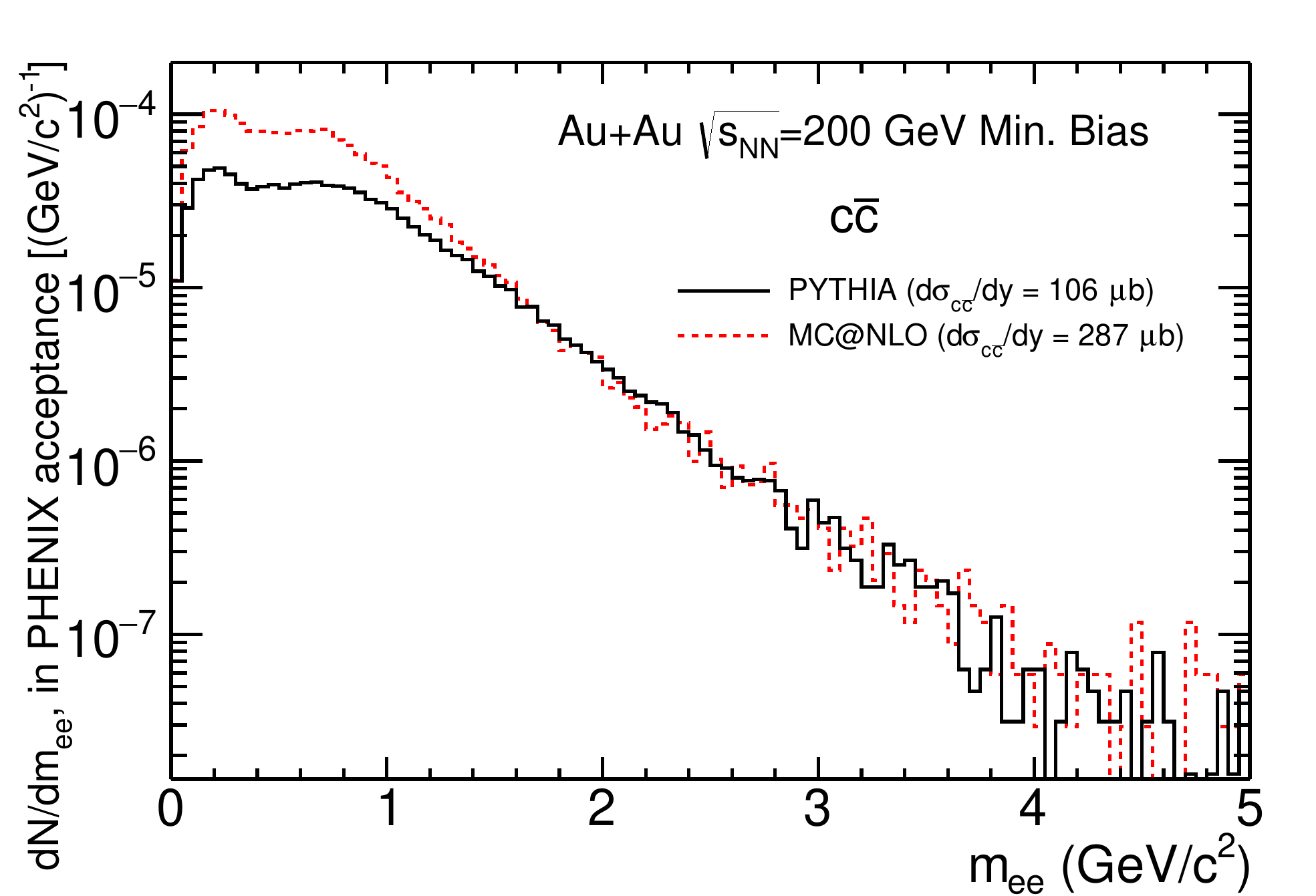}
\caption{(Color online) Comparison of the invariant dielectron yield from 
correlated heavy flavor meson decays for MB \auau collisions calculated 
with {\sc pythia} (solid line) and {\sc mc@nlo} (dashed line) using the 
$d\sigma_{c\bar{c}}^{pp}/dy$ cross sections of 106 $\mu$b and 287 $\mu$b, 
respectively~\cite{Adare:2014iwg}, scaled by 
$\langle$\Ncollnospace$\rangle$.}
\label{fig:pythia_mcnlo} 
\end{figure}

The two results, Eqs. (\ref{eqn:sigma_cc_mcnlo}) and 
(\ref{eqn:sigma_cc_pythia}), although consistent within $\sim$1.2 
$\sigma$, yield central values which differ by a factor of $\sim$2.5. This 
difference comes mainly from the extrapolation of the dilepton yield from 
\mee $>$ 1.15~GeV/$c^2$ to \mee= 0~GeV/$c^2$, as illustrated in 
Fig.~\ref{fig:pythia_mcnlo}.  Figure~\ref{fig:pythia_mcnlo} also shows an 
absolute comparison of the {\sc pythia} and {\sc mc@nlo} dielectron 
invariant yields from correlated heavy flavor meson decays in MB \auau 
collisions, obtained by \Ncoll scaling of the \pp cross sections quoted in 
Eqs. (\ref{eqn:sigma_cc_mcnlo}) and (\ref{eqn:sigma_cc_pythia}). At high 
masses, \mee $>$ 1.15~GeV/$c^2$, both generators give by construction the 
same yield, with a very small difference in shape.  However, at low masses 
there is a large discrepancy in the absolute yield.

The \dau (as well as the \pp) inclusive dilepton yield is not very 
sensitive to this variation of the cross section because the large effect 
at low masses is diluted by the contributions from light meson decays. The 
situation is quite different in \auau collisions. The yield from light 
meson decays scales approximately with \Npart, whereas the contribution 
from heavy flavor scales with \Ncoll making the latter dominant at 
low-masses in central collisions. The choice of the generator used to 
simulate the $c\bar{c}$ contribution will therefore affect the total 
cocktail yield at low masses and will influence the interpretation of the 
\auau data in terms of an excess with respect to the cocktail. The results 
will be presented in the next section using {\sc pythia} for an easier 
comparison with previously published results but both generators, {\sc 
pythia} and {\sc mc@nlo}, will be considered in the discussion.

\subsection{Cocktail normalization}
\label{sec:cocktail_normalization}

In the present analysis we use the precisely measured \ee data at low 
masses to derive the normalization of the cocktail of hadronic sources.  
In the restricted phase space defined by \mee$<$ 0.1~GeV/$c^2$ and 
\pt/\mee $>$ 5 the inclusive \ee yield is dominated by $\pi^0$ Dalitz 
decays with a small contribution of direct virtual photons and an even 
smaller contribution of $\eta$ Dalitz decays. To a very good approximation 
the mass spectrum of these three sources has a 1/\mee dependence and their 
relative magnitude is well known. The ratio of direct photons to $\pi^0$ 
is known from PHENIX measurements~\cite{Adare:2008ab, Adare:2014fwh} and 
the ratio of $\eta$ to $\pi^0$ can be easily obtained from the PHENIX 
measurement at high \pt~\cite{Adare:2010dc} and the $m_T$ scaling as 
described in Section \ref{sec:other_mesons}. By fitting the 
cocktail+direct virtual photons to the data in the restricted phase space 
defined above, one obtains the rapidity density $dN_{\pi^0}/dy$ that 
determines the normalization of the cocktail. The values are found to be 
consistent with measurements of neutral and charged pions 
~\cite{Adler:2006bw, Adare:2008qa, Adler:2003cb} within the systematic 
uncertainties of cocktail and data.

\begin{figure}[hbt!] 
\includegraphics[width=1.0\linewidth]{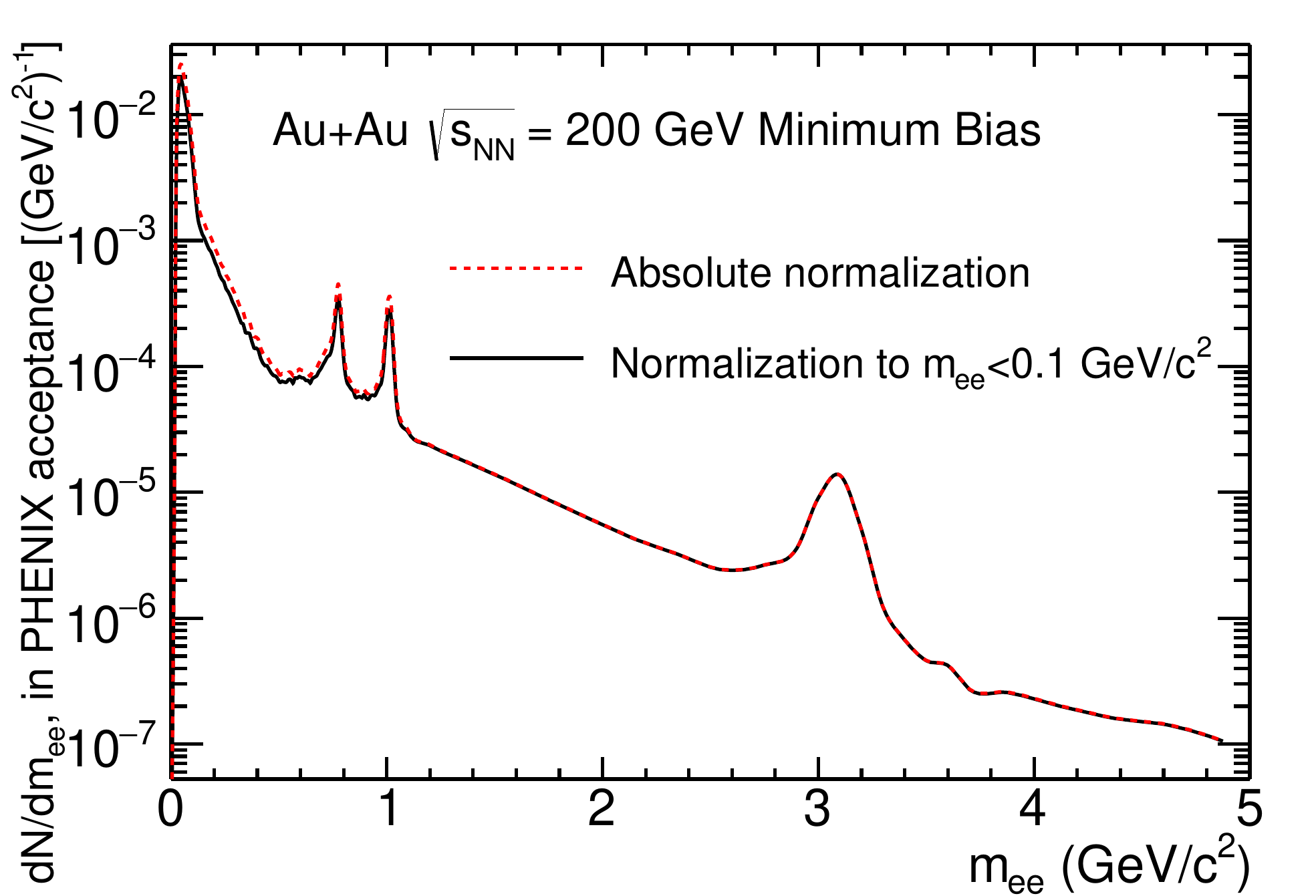}
\caption{(Color online) Cocktail of hadronic sources for the 2010 run with 
normalization provided by fitting to the present \ee invariant yield at 
masses \mee$<$ 0.1~GeV/$c^2$ (black line) or with absolute normalization 
to the $\pi^0$ rapidity density derived from measurements of neutral and 
charged pions~\cite{Adler:2006bw, Adare:2008qa, Adler:2003cb} (dashed 
line).}
\label{fig:cocktail_rescaled_absolute} 
\end{figure}

Alternatively, the cocktail can be absolutely normalized using the $\pi^0$ 
rapidity density $dN_{\pi^0}/dy$ derived from these measurements as done 
in Ref.~\cite{Adare:2009qk}. The cocktails obtained with these two 
procedures are compared in Fig. \ref{fig:cocktail_rescaled_absolute}. The 
results differ at masses \mee $<$ 100~MeV/$c^2$ by about 25\% which is 
approximately the contribution of the virtual direct photons. However, for 
the mass range of interest, that is typically 0.3--0.76~GeV/$c^2$, the 
difference is smaller and amounts to only 15\%. In this mass range, the 
yield is dominated by the contributions from correlated heavy flavor 
decays and changing $dN_{\pi^0}/dy$ by $\sim$25\% has a minor effect on 
the inclusive \ee yield. At even higher masses, \mee $>$ 1~GeV/$c^2$, the 
two procedures yield exactly the same results. The present procedure is 
adopted to be consistent with the known contribution of internal 
conversion.
 
%%%%%%%%%%%%%%%%%%%%%%%%%%%%%%%%%%%%%%%%%%%%%%%%%%%%%%%%%%%% Fig_25

\subsection{Systematic uncertainties on the cocktail}
\label{sec:cocktail_systematics}

The systematic uncertainties of the cocktail ingredients are estimated and 
propagated to determine the total cocktail systematic uncertainty. The 
following uncertainties are considered:

(i) Light meson to $\pi^{0}$ ratio: We adopt the same systematic 
uncertainties used in Ref.~\cite{Adare:2009qk}, namely $\pm$30\% for 
$\eta$, $\omega$ and $\phi$, $\pm$33\% for $\rho$ and $\pm$100\% for 
$\eta'$.

(ii) Direct photon: The systematic uncertainties in the direct photon 
$dN/dy$ are taken from Ref.~\cite{Adare:2014fwh}. They range from 
$\pm$24\% to $\pm$70\% from central to peripheral collisions, 
respectively.

(iii) Open heavy flavor ($c\bar{c}$, $b\bar{b}$): We use the systematic 
uncertainties of the open heavy flavor cross sections given in Eqs. 
(\ref{eqn:sigma_cc_mcnlo}) or (\ref{eqn:sigma_cc_pythia}) for $c\bar{c}$ 
and (\ref{eqn:sigma_bb}) for $b\bar{b}$, taken from 
Ref.~\cite{Adare:2014iwg}. The $\langle$\Ncollnospace$\rangle$ systematic 
uncertainties shown in Table \ref{tab:glauber} are added in quadrature 
when the \pp cross sections are scaled to \auau collisions.

(iv) $J/\psi$: The systematic uncertainty of the $J/\psi$ cross section in 
\pp collisions is estimated to be $\pm$14\%~\cite{Adare:2011vq}. The 
systematic uncertainties in $\langle$\Ncollnospace$\rangle$ and $J/\psi$ 
$R_{AA}$ are added in quadrature. The $R_{AA}$ uncertainties are taken 
from Ref.~\cite{Adare:2006ns}, ranging from $\pm$22\% to $\pm$35\% 
depending on centrality.

\begin{figure}[hbt!]
\includegraphics[width=1.0\linewidth]{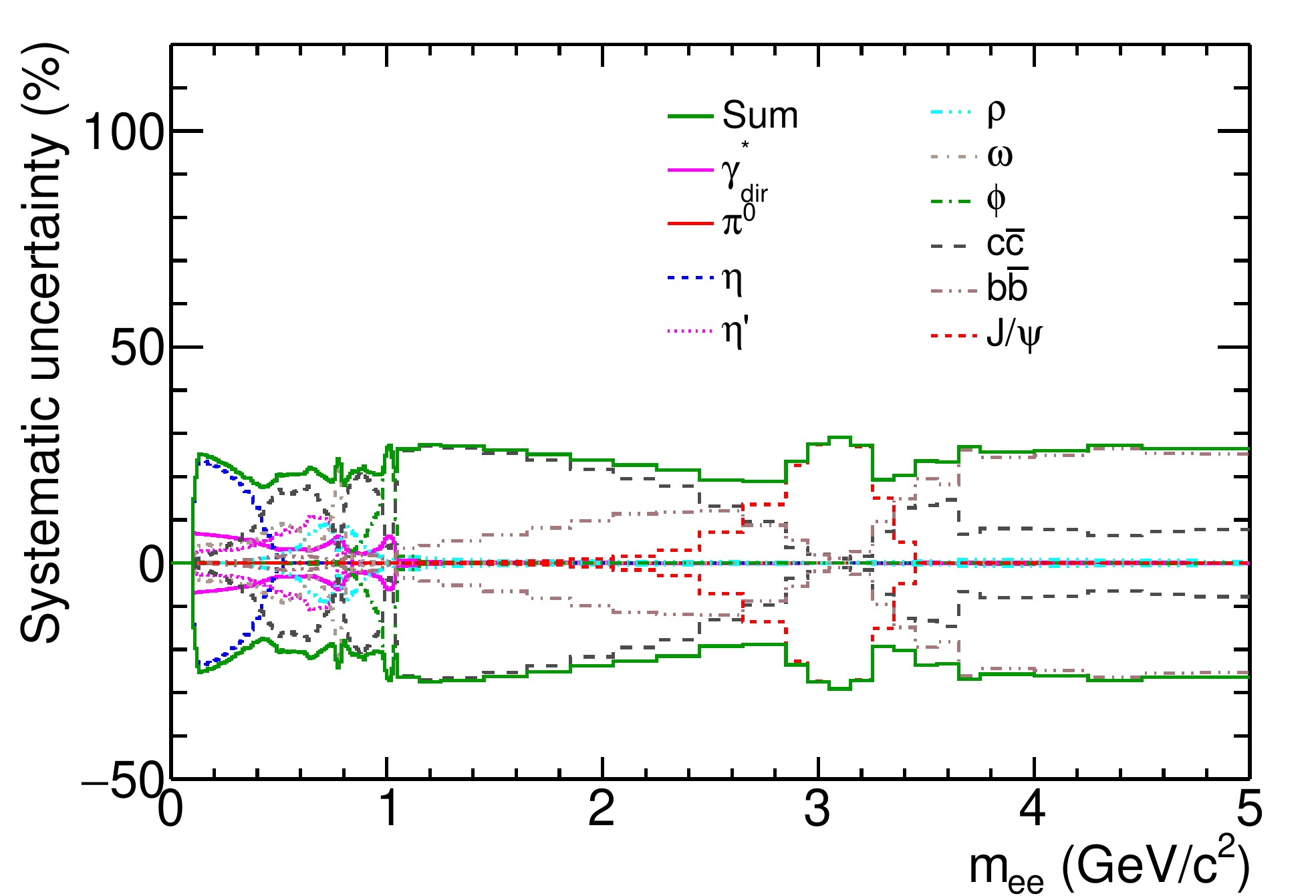} 
\caption{\label{fig:syst_cock}
(Color online) Systematic uncertainties assigned to each cocktail 
component and the total cocktail systematic uncertainty for MB events.}
\end{figure}

A summary of the cocktail systematic uncertainties is presented 
graphically in Fig. \ref{fig:syst_cock}, which shows the systematic 
uncertainty of each cocktail component together with the total cocktail 
systematic uncertainty, determined as their quadratic sum.

%%%%%%%%%%%%%%%%%%%%%%%%%%%%%%%%%%%%%%%%%%%%%%%%%%%%%%%%%%%% Fig_26

\subsection{The \auau Cocktail}
\label{sec:run10_cocktail}

\begin{figure}[hbt!] 
\includegraphics[width=1.0\linewidth]{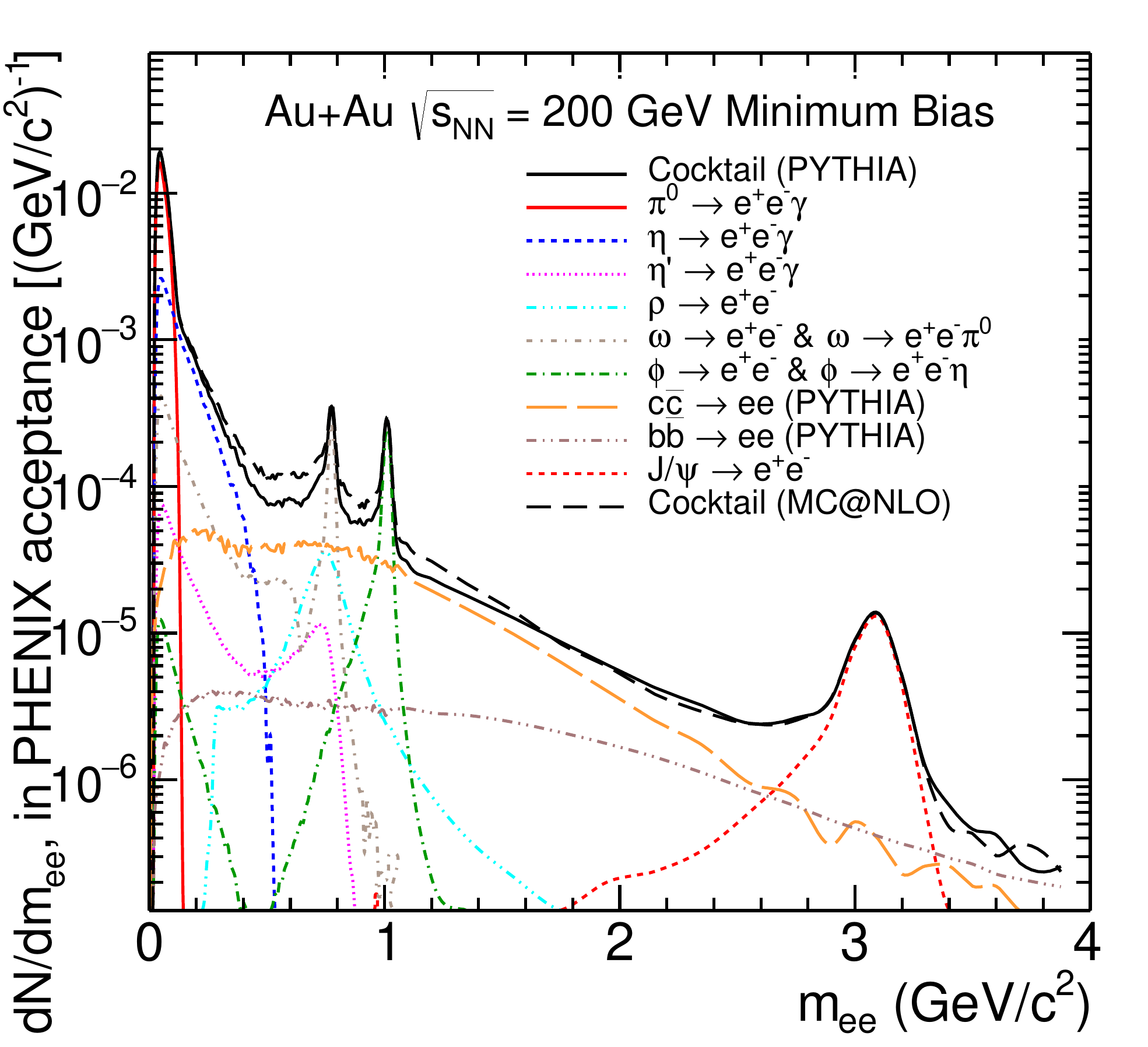}
\caption{(Color online) Cocktail of hadronic sources for the 2010 run 
(black solid line) using the {\sc pythia} generator for the open heavy 
flavor contributions. The individual components of the cocktail are also 
shown. For comparison, the total cocktail using {\sc mc@nlo} is shown 
(black dashed line).}
\label{fig:cocktail_pythia_mcnlo} 
\end{figure}

The cocktail, calculated as described above, using the {\sc pythia} 
generator for the open heavy flavor contributions, is presented in 
Fig.~\ref{fig:cocktail_pythia_mcnlo} for MB \auau collisions together with 
the individual components of the cocktail.  For comparison, 
Fig.~\ref{fig:cocktail_pythia_mcnlo} also shows the total cocktail using 
{\sc mc@nlo} for the open heavy flavor contributions. The differences 
discussed above in Section \ref{sec:open_heavy_flavor} are clearly 
reflected in this comparison.

%%%%%%%%%%%%%%%%%%%%%%%%%%%%%%%%%%%%%%%%%%%%%%%%%%%%%%%%%%%% Fig_27

\section{RESULTS and DISCUSSION}
      \label{sec:results}

      \subsection{Invariant mass spectra}
      \label{sec:invariant_mass}

\begin{figure}[hbt!]
\includegraphics[width=1.0\linewidth]{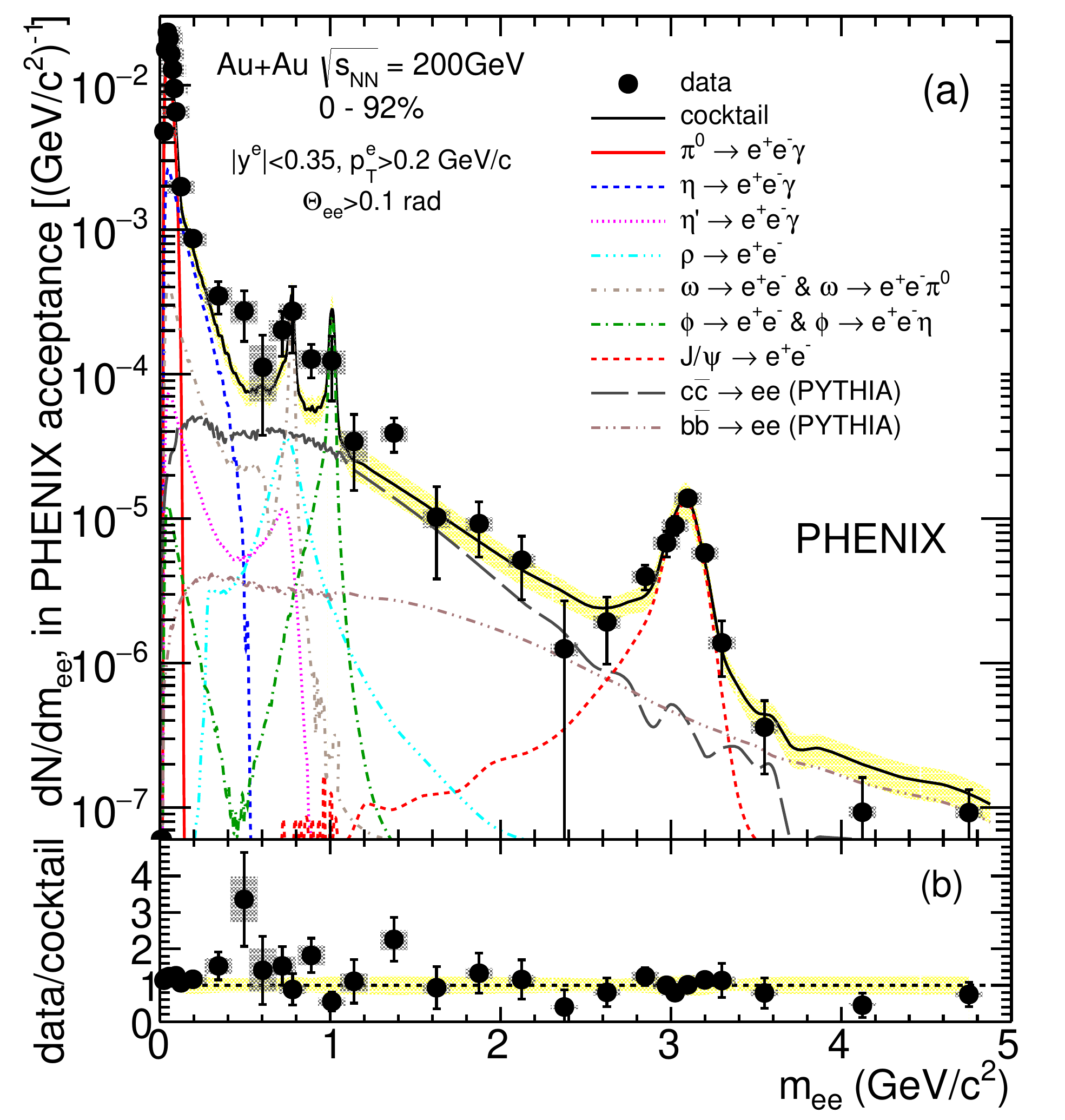}
\caption{(Color online) Invariant mass spectrum of \ee pairs in 
MB \auau collisions within the PHENIX acceptance compared to the 
cocktail of expected decays. }
\label{fig:mass_mb} 
\end{figure}

Figure \ref{fig:mass_mb} shows the invariant mass spectrum of \ee pairs 
within the PHENIX acceptance (as defined in Section 
\ref{sec:acceptance_run10}) for MB \auau collisions.  The spectra are 
subject to a \pt cut of 0.2~GeV/$c$ on the single electron tracks and to a 
100 mrad cut on the pair opening angle. Statistical and systematic 
uncertainties on the data points are shown separately by vertical bars and 
boxes, respectively.  Figure~\ref{fig:mass_mb} also compares the measured 
spectrum to the cocktail of expected \ee sources, where {\sc pythia} is 
used to calculate the correlated pairs from heavy flavor decays. The 
individual contributions to the cocktail are shown in the figure.

%%%%%%%%%%%%%%%%%%%%%%%%%%%%%%%%%%%%%%%%%%%%%%%%%%%%%%%%%%%% Fig_28

See Section \ref{sec:cocktail} for details about the cocktail calculation. 
The total systematic uncertainty of the cocktail is shown by the yellow 
band. The bottom panel shows the ratio of data to cocktail.

\begin{figure}[hbt!]
\includegraphics[width=1.0\linewidth]{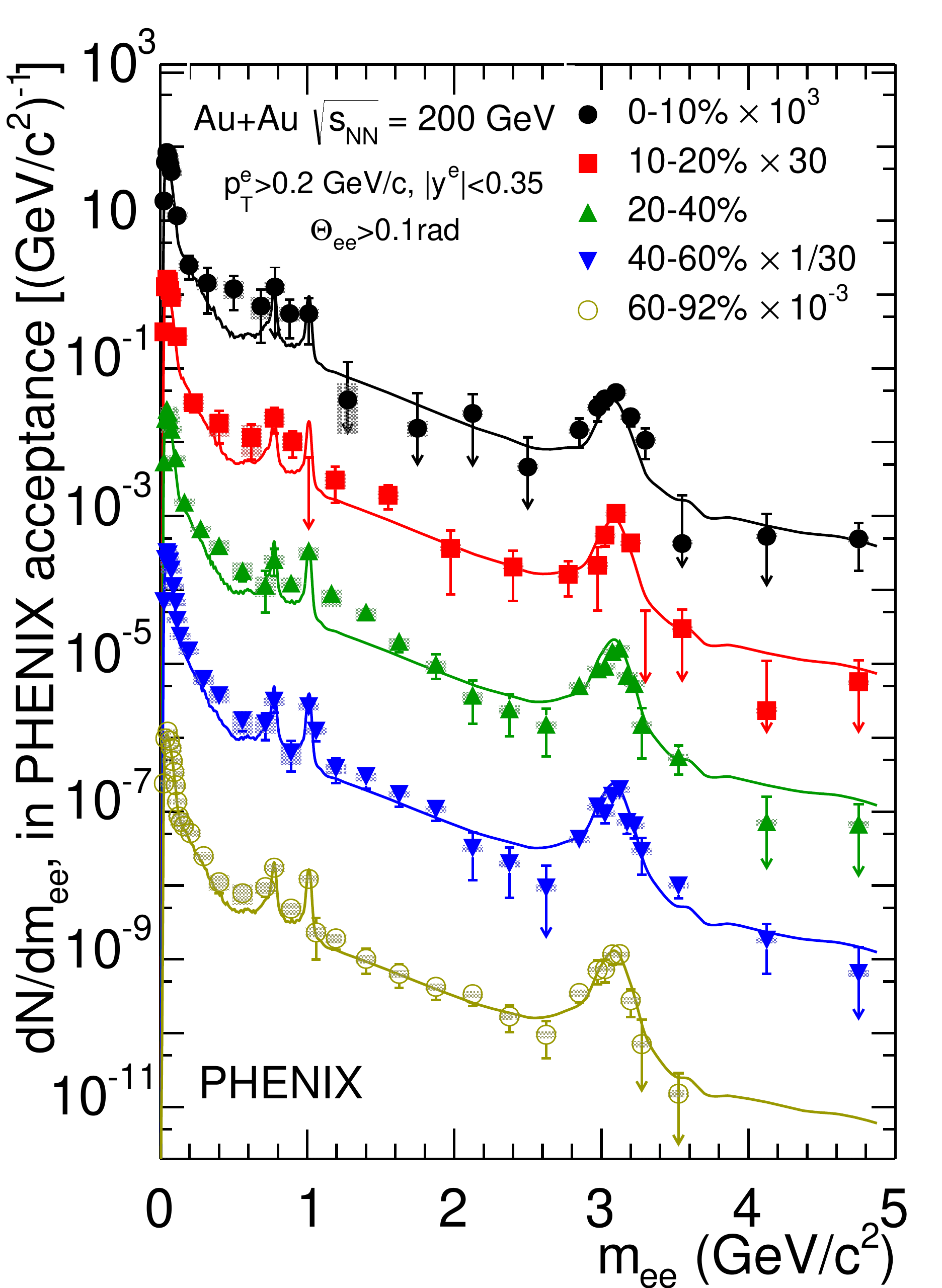}
\caption{(Color online) Invariant mass spectra of \ee pairs in \auau 
collisions within the PHENIX acceptance for the various centrality bins. 
The lines represent the total expected yield from all the sources 
indicated in Fig. \ref{fig:mass_mb}.}
\label{fig:centdep} 
\end{figure}

Figure \ref{fig:centdep} shows the invariant mass spectra of \ee pairs for 
the five centrality bins analyzed in this work, compared to the cocktail.
 
%%%%%%%%%%%%%%%%%%%%%%%%%%%%%%%%%%%%%%%%%%%%%%%%%%%%%%%%%%%% Fig_29
 
For a more detailed discussion of the centrality and transverse momentum 
dependencies of the dielectron yield, we consider three mass regions:\\

(a) the mass region \mee $<$ 0.10~GeV/$c^2$ that is dominated by the 
$\pi^0$ Dalitz decay. \\

(b) the low-mass region (LMR), 0.30 $<$ \mee $<$ 0.76~GeV/$c^2$, below the 
$\rho$ meson mass, that is the most sensitive region to in-medium 
effects.\\

(c) the intermediate-mass region (IMR), 1.2 $<$ \mee $<$ 2.8~GeV/$c^2$, 
that is dominated by the correlated pairs from the semi-leptonic decays of 
charm and bottom mesons. \\

\begin{figure}[hbt!]
\includegraphics[width=1.0\linewidth]{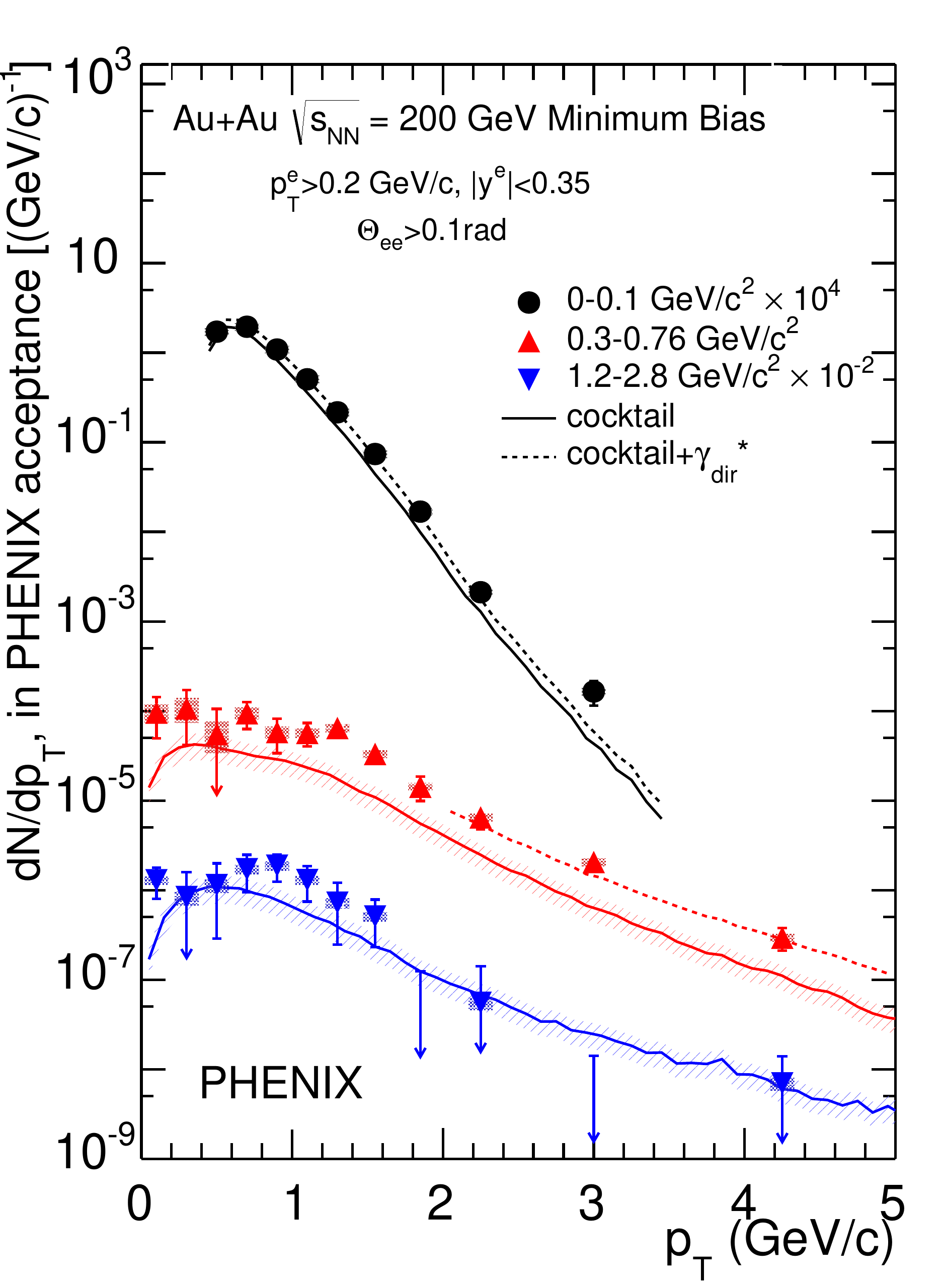}
\caption{(Color online) MB invariant \pt distributions for three mass 
windows as indicated in the legend. The solid lines represent the expected 
\pt distributions of the hadronic cocktail and the shadowed bands around 
the lines represent the cocktail systematic uncertainties. The dotted 
lines include the contribution from direct photons in the phase space 
region where they can reliably be calculated, i.e. \pt/\mee $>$ 5. }
\label{fig:pt_spectra_massbins} 
\end{figure}

Figure~\ref{fig:pt_spectra_massbins} shows the pair \pt distribution for 
these three mass intervals in MB collisions. In the following 
sections we discuss the results in these three mass intervals.

%%%%%%%%%%%%%%%%%%%%%%%%%%%%%%%%%%%%%%%%%%%%%%%%%%%%%%%%%%%% Fig_30

\subsection{$\pi^0$ Dalitz region}
\label{sec:pi0}

The mass region \mee $<$ 0.10~GeV/$c^2$ is dominated by the $\pi^0$ Dalitz 
decay with a small contribution of direct virtual photons of $\sim$20\% 
and an even smaller contribution of the $\eta$ Dalitz decay of $\sim$10\%. 
We discuss here only the shape of the \pt distribution because the 
integrated dielectron yield in this mass interval was used to normalize 
the cocktail for the five centrality bins as described in Section 
\ref{sec:cocktail}. Figure \ref{fig:pt_spectra_massbins} compares the 
measured dielectron \pt distribution for MB collisions in this 
mass interval to the \pt distribution of the hadronic cocktail that uses 
the parametrization for the $\pi^0$ and $\eta$ mesons [Eq. 
(\ref{eqn:hagedorn})]. The agreement between the two distributions, in 
shape and magnitude, is very good when adding the measured yield of direct 
virtual photons.

\subsection {Low-mass region (LMR)}
\label{sec:lmr} 

\begin{figure}[hbt!]
\includegraphics[width=1.0\linewidth]{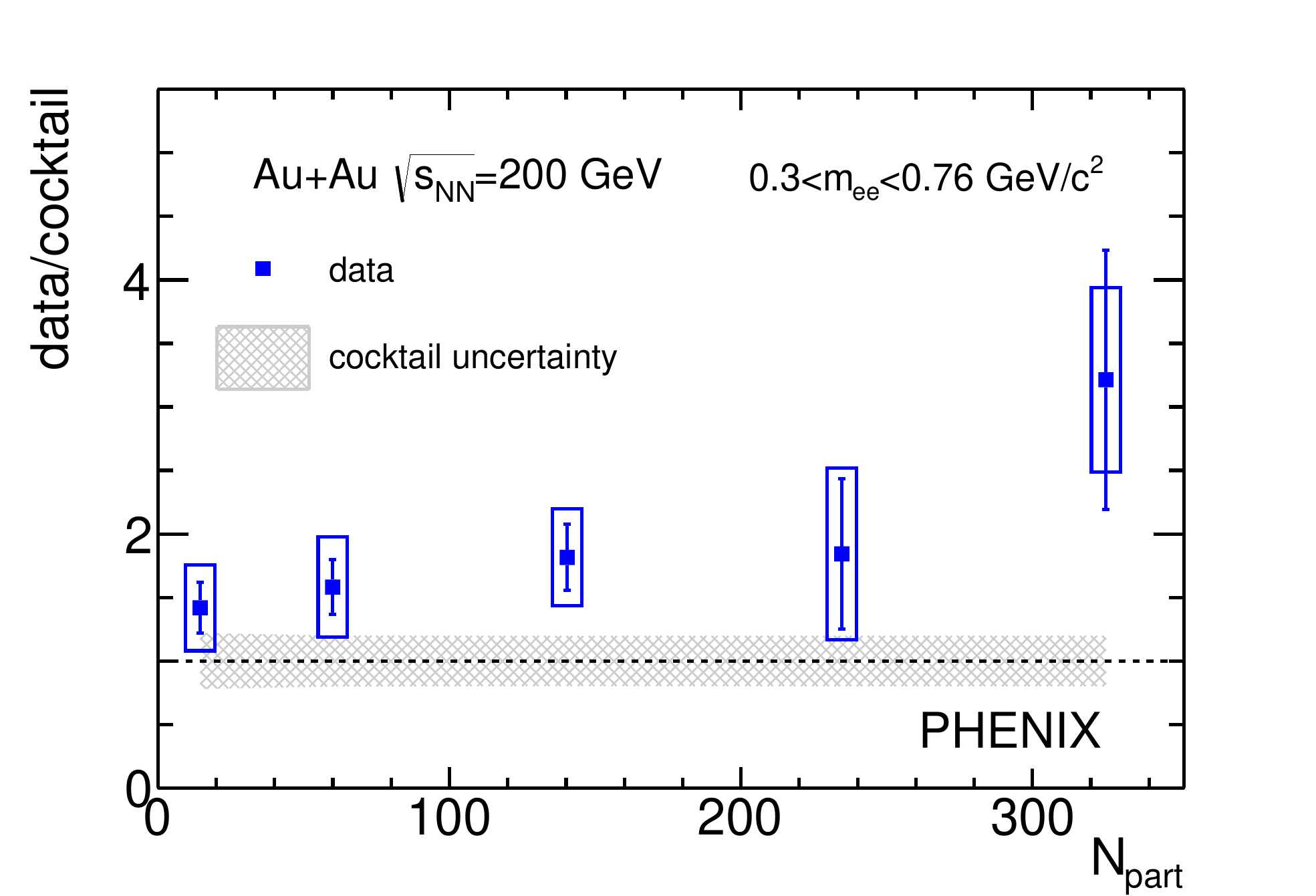} 
\caption{(Color online) Data to cocktail (using {\sc pythia} for heavy 
flavor contribution) ratio in the LMR versus centrality. The shaded band 
around one represents the cocktail systematic uncertainty.}
\label{fig:data_to_cocktail_lmr} 
\end{figure}

In the LMR, the yield is expected to be saturated by the light mesons 
($\eta, \rho$ and $\omega$) and the \ccbar contribution. Figure 
\ref{fig:mass_mb} shows an enhancement of \ee pairs with respect to the 
cocktail in MB  collisions. The enhancement develops with 
centrality as shown in Fig.  \ref{fig:centdep} and it appears to be 
distributed over the whole \pt range covered by the measurement, as can be 
seen in Fig. \ref{fig:pt_spectra_massbins}. We quantify the effect by the 
enhancement factor defined as the ratio of the measured over expected 
dilepton yield integrated in the LMR. As discussed in Section 
\ref{sec:open_heavy_flavor}, the cocktail yield in this mass region 
depends on the generator, {\sc pythia} or {\sc mc@nlo}, used to calculate 
the open heavy flavor contribution.  The enhancement factors obtained with 
{\sc pythia} are shown as a function of centrality in Fig. 
\ref{fig:data_to_cocktail_lmr} and they are listed in in Table 
\ref{tab:enhancement_factors} for the two cases. The enhancement factors 
are approximately 40\% higher when {\sc pythia} is used to calculate the 
open heavy flavor contribution instead of {\sc mc@nlo}.

%%%%%%%%%%%%%%%%%%%%%%%%%%%%%%%%%%%%%%%%%%%%%%%%%%%%%%%%%%%% Fig_31

%===================================================== Table_VIII
\begin{table}[hbt!]
\caption{Enhancement factors, defined as the ratio of measured over 
expected dilepton yield in the mass region \mee = 0.30--0.76~GeV/$c^2$, 
for the five centrality bins and for MB. The enhancement 
factors are quoted separately for the two cases where the correlated 
yield from $c\bar{c}$ decays is calculated with {\sc pythia} or {\sc 
mc@nlo}. The $\pm$model uncertainties represent the cocktail systematic 
uncertainties.
}
\begin{ruledtabular} \begin{tabular}{clccc} 
&Centrality   & \multicolumn{2}{c} {Enhancement factor $\pm$stat $\pm$syst $\pm$model} &\\
&                 & {\sc mc@nlo}   $c\bar{c}$                                      &  {\sc pythia}    $c\bar{c}$         &  \\
\hline
&MB            & 1.7 $\pm$0.3 $\pm$0.3 $\pm$0.2  &  2.3 $\pm$0.4 $\pm$0.4 $\pm$0.2 &\\ 
&0\%--10\%      & 2.3 $\pm$0.7 $\pm$0.5 $\pm$0.2 &  3.2 $\pm$1.0 $\pm$0.7$\pm$0.2 &\\
&10\%--20\%    & 1.3 $\pm$0.4 $\pm$0.5 $\pm$0.2  &  1.8 $\pm$0.6 $\pm$0.7 $\pm$0.2 &\\
&20\%--40\%    & 1.4 $\pm$0.2 $\pm$0.3 $\pm$0.2  &  1.8 $\pm$0.3 $\pm$0.4 $\pm$0.2 &\\
&40\%--60\%    & 1.2 $\pm$0.2 $\pm$0.3 $\pm$0.2  &  1.6 $\pm$0.2 $\pm$0.4 $\pm$0.2 &\\
&60\%--92\%    & 1.0 $\pm$0.1 $\pm$0.2 $\pm$0.2   & 1.4 $\pm$0.2 $\pm$0.3 $\pm$0.2 &\\
\end{tabular} \end{ruledtabular} 
\label{tab:enhancement_factors}
\end{table}  

\subsubsection{Comparison to previous PHENIX results}
\label{sec:comparison_run4}

The enhancement factors quoted above are significantly smaller than those 
previously reported by PHENIX~\cite{Adare:2009qk} in the same \auau 
collision system at the same energy of $\sqrt{s_{_{NN}}}$~=~200~GeV. There 
are a number of significant differences, both qualitative and 
quantitative, between the two analyses:
 
\begin{itemize}

\item Hadron contamination: The purity of the electron sample is very 
different in the two cases. In~\cite{Adare:2009qk} the hadron 
contamination was 30\% in central \auau collisions, whereas in the 
present analysis, the HBD enabled this contamination to be reduced to 
less than 5\% at all centralities.
 
\item Signal sensitivity: The signal sensitivity is usually quantified by 
the signal to background $S/B$ ratio. The $S/B$ values displayed in Fig. 
\ref{fig:signal_to_background} are similar to those quoted in 
Ref.~\cite{Adare:2009qk}. This is however, a misleading comparison, 
because in a situation of subpercent $S/B$ ratio, the magnitude of $S$ 
critically depends on the accuracy of the background subtraction. A better 
way to assess the sensitivity of the measurement is provided by the 
cocktail/background, $C/B$, ratio. From the signal/background ratio and 
the enhancement factors quoted in Ref.~\cite{Adare:2009qk}, we estimate an 
average value of $C/B$ over the mass range \mee = 0.15--0.75~GeV/$c^2$ of 
$\sim$1/600 in MB collisions. In the present analysis the same 
ratio is found to be $\sim$1/250. In addition to that, one should take 
into account that in the 2010 run with the $+ - $ field configuration 
there is a larger track acceptance of $\sim$20\%. This rough estimate 
indicates that at the same multiplicity the signal sensitivity in the 
present analysis is larger by a factor of $\sim$3.5 compared to the 
previous one.

\item Pair cuts: Loose pair cuts were applied in Ref.~\cite{Adare:2009qk} 
compared to the cuts used in this analysis. The cuts used in 
Ref.~\cite{Adare:2009qk} are found to leave a sizable amount of detector 
induced correlation in the mass region \mee = 0.4--0.6~GeV/$c^2$.

\item Flow: As demonstrated in Section \ref{sec:combinatorial_background} 
the collective flow that is inherent to nuclear collisions, affects the 
shape of the combinatorial component of the background and violates the 
square root relation [Eq. (\ref{eq:2sqrt})]. These two effects were not 
taken into account in the data analysis of Ref.~\cite{Adare:2009qk}.

\item Electron-hadron pairs: As shown in Section 
\ref{sec:electron_hadron_pairs}, the $e$-$h$ pairs originate in the 
central arm detectors and in particular in the RICH detector. This source 
of correlated pairs was not considered in~\cite{Adare:2009qk}.

\item Away-side jet component: The away-side jet component of the 
correlated background was found to be negligible in~\cite{Adare:2009qk} 
and only the near-side jet component was considered. In the present 
analysis, both components are absolutely calculated. The away-side 
component is indeed relatively small but both components are considered 
and subtracted.

\item Background subtraction procedure: In Ref.~\cite{Adare:2009qk}, the 
shapes of the three components of the background (combinatorial 
background, cross pairs and near-side jet) were calculated whereas their 
absolute scales were obtained by fitting to the like-sign spectra. In the 
present analysis, all components of the correlated background (cross 
pairs, jet pairs and electron-hadron pairs) are calculated and subtracted 
in absolute terms. There is only one free parameter in the background 
subtraction procedure, namely the normalization factor of the 
combinatorial background.

\end{itemize}

In conclusion, we do not confirm our previous report of a large excess 
seen in the LMR~\cite{Adare:2009qk}. 
The differences listed above affect the yield in the mass region where the excess was reported but not always in the same direction. For example, the loose pair cuts lead to under subtraction of the background whereas neglecting the flow modulation, has the opposite effect namely it leads to over subtraction in the mass region where the excess was observed. These differences also do not affect the unlike-sign yield by a similar magnitude. The hadron contamination, the loose pair cuts and the electron-hadron pairs are the most significant ones in this respect.  Taking all the differences together, the present analysis is much improved compared to the previous one and we thus consider the previous result on the low-mass excess to be superseded by the results presented here. 

\subsubsection{Comparison to STAR results}

Recently, STAR published results on \ee production in \auau collisions at 
\sqnr ~\cite{Adamczyk:2013caa,Adamczyk:2015lme}. In the same mass range of 
\mee = 0.30--0.76~GeV/$c^2$, STAR observes an excess of dielectrons and 
quotes a value of 1.77$\pm 0.11^{stat} \pm 0.24^{syst} \pm 0.33^{model}$ 
in MB collisions, for the ratio of the dielectron yield to the 
hadronic cocktail excluding the $\rho$ meson contribution. There are two 
factors that should be taken into account when comparing the STAR results 
with those quoted in Table \ref{tab:enhancement_factors}. First, excluding 
the $\rho$ contribution results in an increase of about 10\% of the data 
to cocktail ratio. Second, STAR uses {\sc pythia} with a charm cross 
section $d\sigma_{c\bar{c}}/dy$ = 171 $\pm$26 
$\mu$b~\cite{Adamczyk:2013caa} which is between the PHENIX cross sections 
quoted in Section \ref{sec:cocktail} for {\sc pythia} and {\sc mc@nlo}.  
Taking those two differences into account, as well as the experimental uncertainties, we find that the results of the two experiments are consistent in the LMR. The centrality and $p_T$ dependencies of the 
enhancement reported in~\cite{Adamczyk:2015lme} are also consistent with 
our results.
 
\subsection {Intermediate-mass region (IMR)}
\label{sec:imr}

The IMR is dominated by correlated pairs from the semi-leptonic decays of 
$D\overline D$ mesons, with a small contribution from $B\overline B$ 
mesons and an even smaller contribution from Drell Yan. The latter is 
neglected in the cocktail calculation. This mass interval is singled out 
by theory as the most sensitive window to identify the thermal radiation 
of the QGP in the dilepton spectrum~\cite{Ruuskanen:1991au, 
Turbide:2003si}.

\begin{figure}[hbt!]
\includegraphics[width=1.0\linewidth]{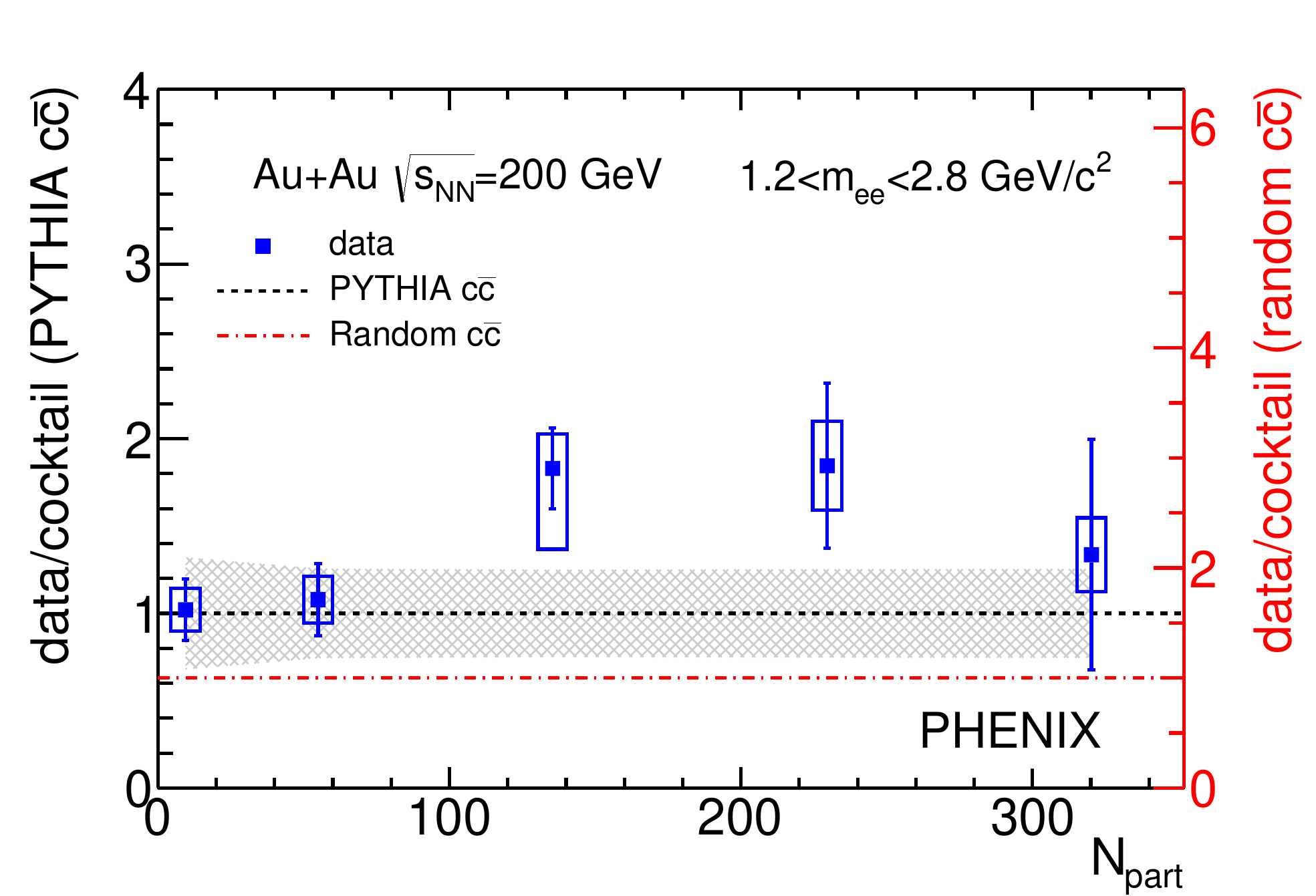}
\caption{(Color online) Data to cocktail ratio in the IMR versus 
centrality. The cocktail uses {\sc pythia} for the $c\bar{c}$ contribution 
(left scale) or random $c\bar{c}$ contribution (right scale). The shaded 
band represents the {\sc pythia} cocktail systematic uncertainty. The same 
uncertainty applies also to the random $c\bar{c}$ cocktail.}
\label{fig:data_to_cocktail_imr} 
\end{figure}

The results displayed in Figs. \ref{fig:mass_mb} and \ref{fig:centdep} 
show a small enhancement of dileptons in the IMR with respect to the yield 
from $c\bar{c}$ decays calculated using {\sc pythia}. The enhancement 
factors are shown in Fig. \ref{fig:data_to_cocktail_imr} as a function of 
centrality and the values are listed in Table 
\ref{tab:enhancement_factors_imr}. 
The results are consistent with those of Ref. \cite{Adare:2009qk} within the large experimental uncertainties of the latter. 
There is very little difference in the 
dilepton yield in this mass interval if {\sc mc@nlo} is used instead of 
{\sc pythia}, as demonstrated in Fig. \ref{fig:cocktail_pythia_mcnlo}. The 
shapes are very similar and the integral yields in the IMR differ by less 
than 10\% in the two cases.

%%%%%%%%%%%%%%%%%%%%%%%%%%%%%%%%%%%%%%%%%%%%%%%%%%%%%%%%%%%% Fig_32

%===================================================== Table_IX
\begin{table}[hbt!]
\caption{Enhancement factors, defined as the ratio of measured to expected 
dilepton yield in the mass region \mee = 1.2--2.8~GeV/$c^2$, calculated 
using {\sc pythia} for the five centrality bins and for minimum bias. The 
last line gives the enhancement factor assuming random correlation (see 
text).}
\begin{ruledtabular} \begin{tabular}{clcc} 
&Centrality   &  Enh. factor $\pm$stat $\pm$syst $\pm$model &\\
\hline
{\sc pythia} $c\bar{c}$ & \\
&0\%--10\%      & 1.3 $\pm$0.7 $\pm 0.2$ $\pm$0.3 & \\
&10\%--20\%    & 1.8 $\pm$0.5 $\pm 0.3$ $\pm$0.3  &\\
&20\%--40\%    & 1.8 $\pm$0.2 $_{-0.5}^{+0.2}$ $\pm$0.3 & \\
&40\%--60\%    & 1.1 $\pm$0.2 $\pm 0.1$  $\pm$0.3 & \\
&60\%--92\%    & 1.0 $\pm$0.2 $\pm 0.1$  $\pm$0.3 &\\
&MB            & 1.5 $\pm$0.3 $\pm 0.2$  $\pm$0.3 &\\ 
\\
MB (random $c\overline c$)     & 2.5 $\pm$0.5 $\pm 0.3$  $\pm$0.3 \\ 
\end{tabular} \end{ruledtabular} 
\label{tab:enhancement_factors_imr}
\end{table}  
 
\begin{figure}[hbt!]
\includegraphics[width=1.0\linewidth]{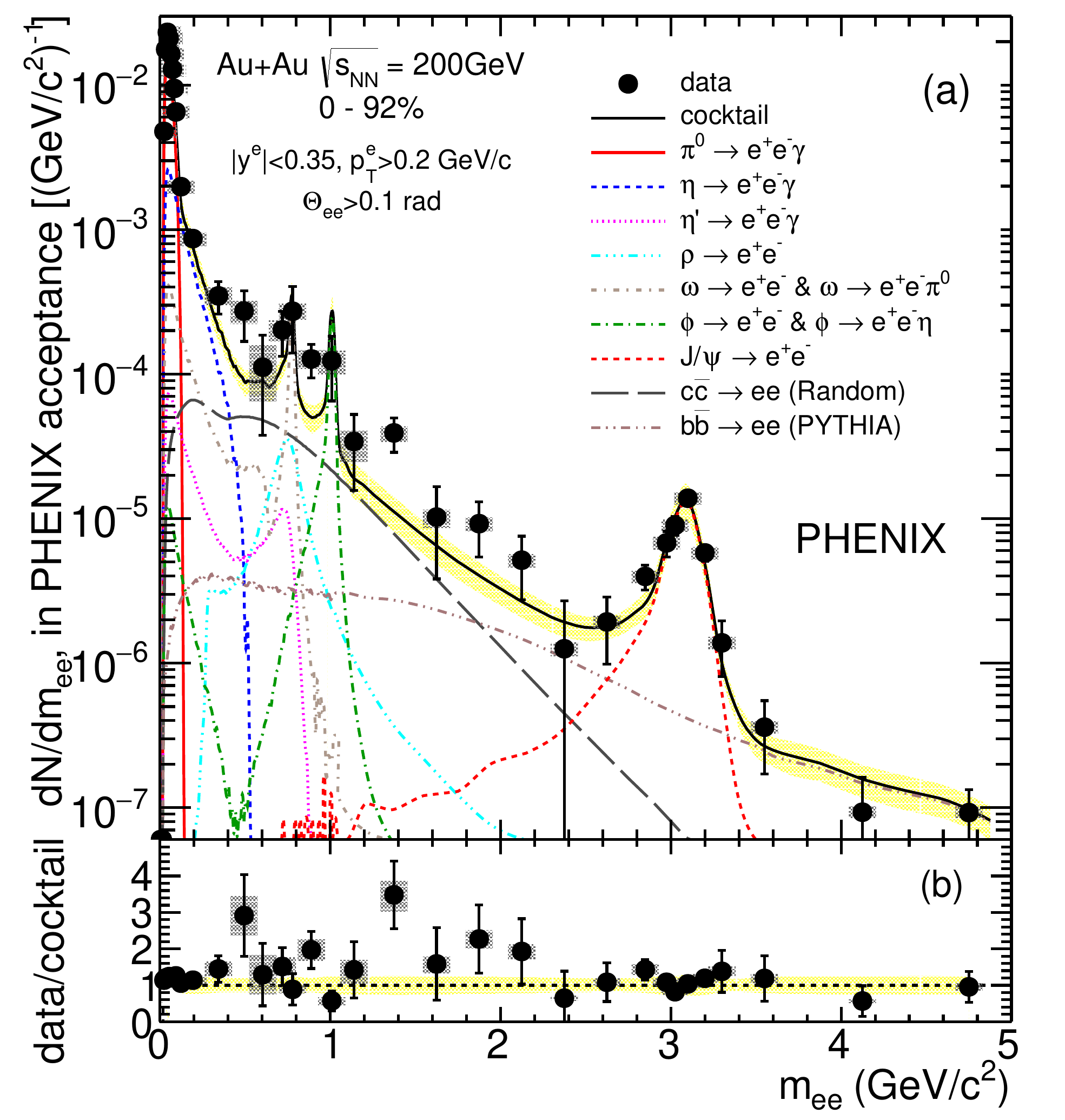}
\caption{(Color online) Invariant mass spectrum of \ee pairs in 
MB \auau collisions within the PHENIX acceptance compared to the 
cocktail of expected decays when the $c\bar{c}$ decay component is 
calculated assuming no correlation between the $c$ and $\bar{c}$. }
\label{fig:mass_mb_wrandom} 
\end{figure}

Using {\sc pythia}, the enhancement factor in MB events is $\sim$1 
standard deviation away from unity. However, the data to cocktail 
comparison discussed above, represents an extreme case in which it is 
assumed that the correlations between the $c\bar{c}$ pairs in \auau 
collisions are the same as in \pp collisions. It is however, well known 
that heavy flavor quarks exhibit energy loss and collective flow in the 
medium formed in \auau collisions, as manifested for example in 
measurements of single electrons~\cite{Adare:2010de,Adare:2006nq}. This 
should affect the correlation between the \ee pairs from $c\bar{c}$ 
decays. Lacking a suitable generator to model this effect, we consider 
also the opposite extreme approach in which we assume that the pair is 
totally decorrelated. The invariant mass is calculated using two electrons 
randomly selected from the measured \pt distribution of single electrons 
from heavy flavor decays~\cite{Adare:2010de}, with uniform distributions 
in pseudorapidity and azimuthal angle. The pair is filtered through the 
ideal PHENIX acceptance and the integral is normalized to the calculated 
{\sc pythia} yield from $c\bar{c}$ decays. This extreme case results in a 
softer mass distribution in the IMR as can be seen in Fig. 
\ref{fig:mass_mb_wrandom}.

%%%%%%%%%%%%%%%%%%%%%%%%%%%%%%%%%%%%%%%%%%%%%%%%%%%%%%%%%%%% Fig_33

There is a small yield depletion at high masses compensated by a higher 
yield at low masses. The integral in the IMR is lower resulting in 
enhancement factors that are $\sim$70\% larger compared to those derived 
from {\sc pythia}. The enhancement factor in MB collisions is 
quoted in the last line of Table \ref{tab:enhancement_factors_imr} and the 
centrality dependence is seen by comparing the data points to the 
dot-dashed line in Fig. \ref{fig:data_to_cocktail_imr}.

\subsection{Comparison to theory}
\label{sec:theory}

In this section we compare our results to the model originally developed 
by Rapp and Wambach~\cite{Rapp:1999us,Rapp:2000pe}.  The model uses an 
effective Lagrangian and a many body approach to compute the 
electromagnetic spectral function which is the main factor in the 
calculation of the dilepton production rates. In the LMR, the spectral 
function is saturated via vector meson dominance, by the light vector 
mesons, in particular the $\rho$ meson, whereas at larger masses it is 
dominated by multipion states or equivalently, via quark-hadron duality, 
by $q {\overline q}$ annihilation. The dilepton yields are obtained by an 
appropriate integration of the thermal rates over the space-time evolution 
of the fireball. This model was very successful in reproducing the 
low-mass dilepton enhancement discovered at SPS by the CERES experiment 
and later further studied by the NA60 experiment. In the comparison below, 
we use an improved version of the model that incorporates recent 
developments, a nonperturbative QGP equation of state and QGP emission 
rates, i.e. $q {\overline q}$ annihilation at temperatures higher than the 
critical temperature, both based on lattice QCD~\cite{Rapp:2013nxa}.  It 
is important to note that this updated version preserves the agreement 
with the SPS data and also reproduces the RHIC results from STAR.
 
\begin{figure}[hbt!]
\includegraphics[width=1.0\linewidth]{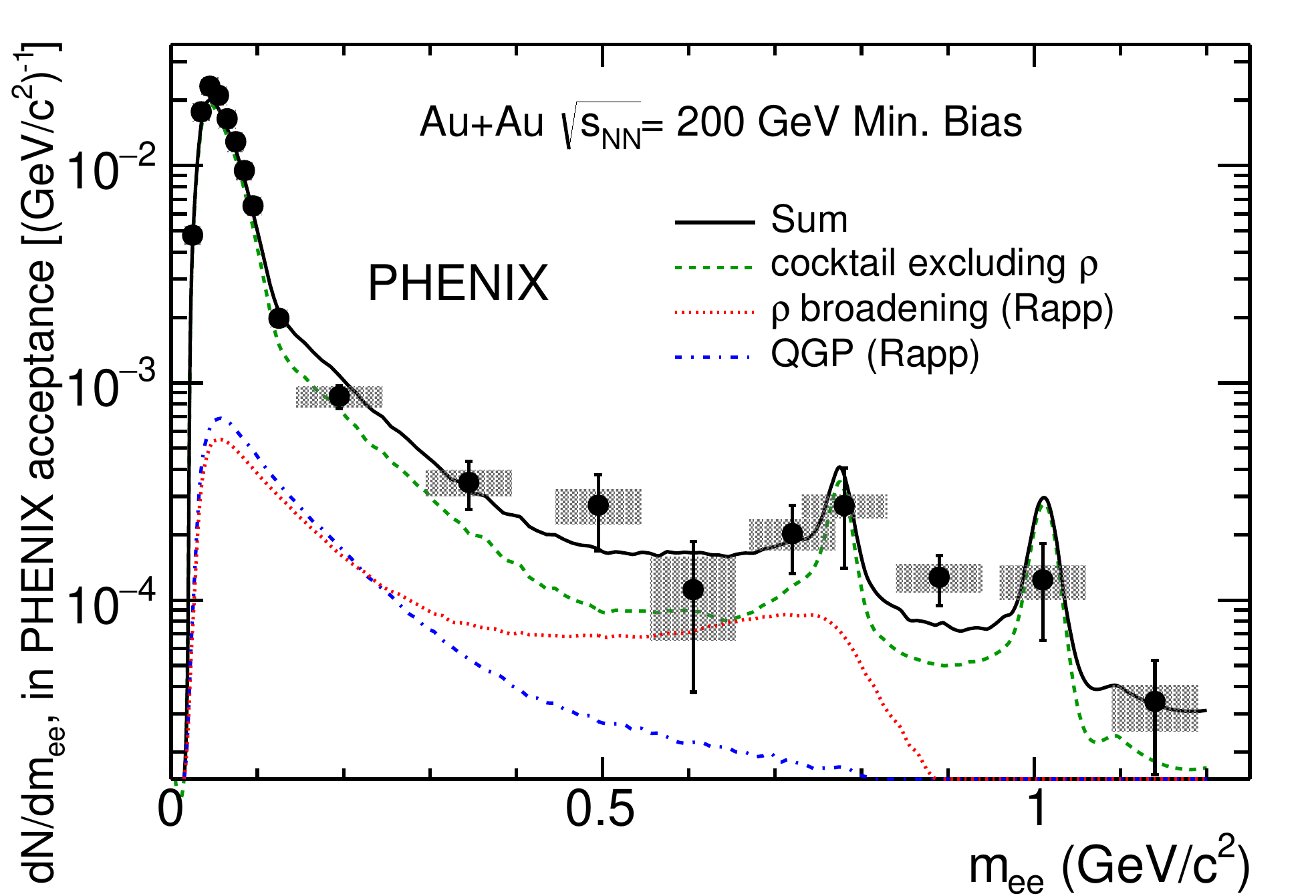}
\caption{(Color online) MB invariant mass spectrum compared to 
the model calculations of Rapp (solid line)~\cite{rapp}. The main 
contributions, the in-medium $\rho$ broadening (dotted line), the QGP 
thermal radiation (dot-dashed line) and the cocktail excluding the $\rho$ 
(dashed line) are also shown.}
\label{fig:rapp_mass_mb} 
\end{figure}

\begin{figure}[hbt!]
\includegraphics[width=1.0\linewidth]{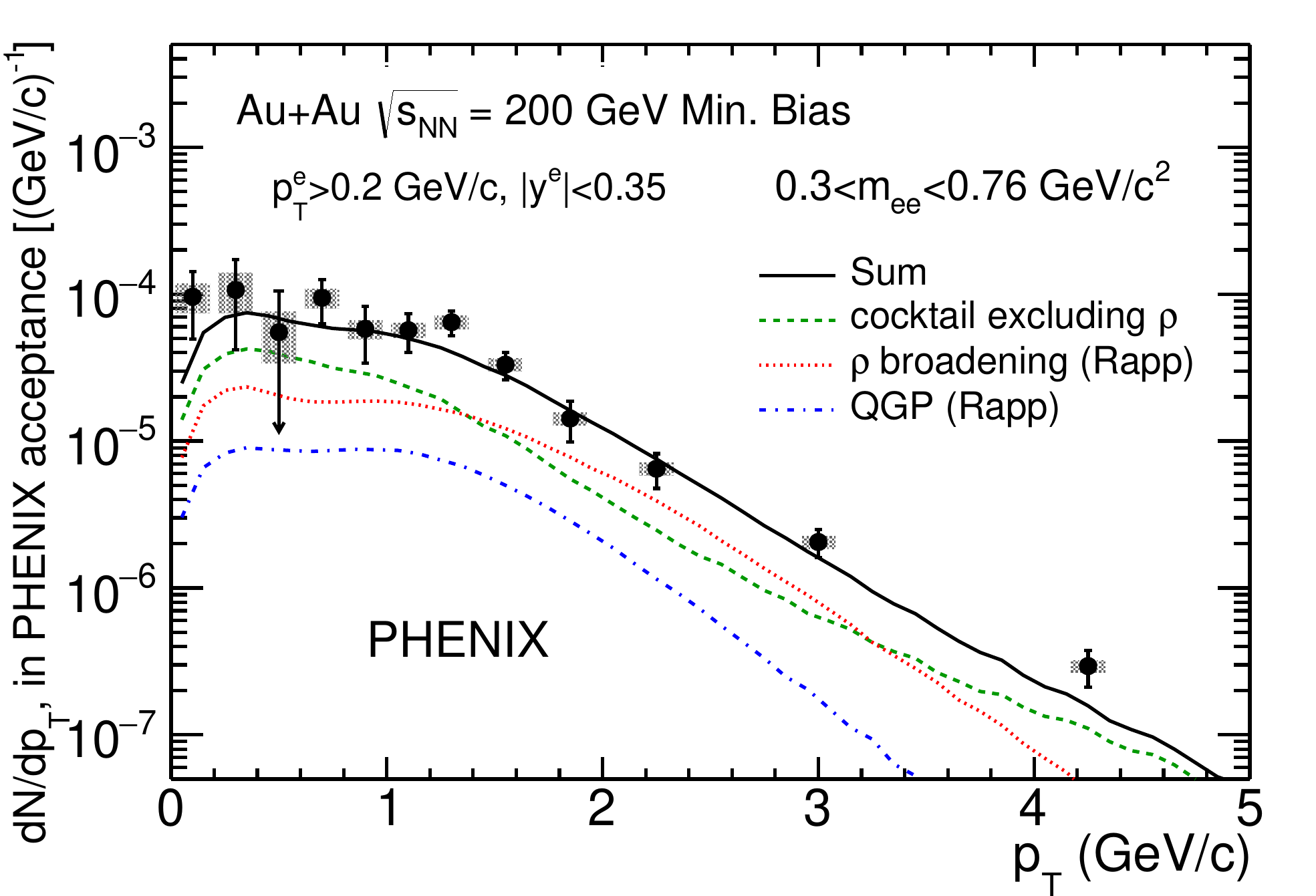}
\caption{(Color online) Dielectron \pt distribution in the LMR compared to 
model calculations (solid line)~\cite{rapp}. The main contributions, the 
in-medium $\rho$ broadening (dotted line), the QGP thermal radiation 
(dot-dashed line) and the cocktail excluding the $\rho$ (dashed line) are 
also shown.}
\label{fig:rapp_pt_lmr} 
\end{figure}

Figures~\ref{fig:rapp_mass_mb} and \ref{fig:rapp_pt_lmr} compare the 
invariant mass spectrum and the LMR pair \pt distribution with the model 
calculations for MB collisions~\cite{rapp}.  The main components, 
in-medium $\rho$ broadening, QGP thermal radiation and cocktail excluding 
the $\rho$, together with their sum, are shown separately.

%%%%%%%%%%%%%%%%%%%%%%%%%%%%%%%%%%%%%%%%%%%%%%%%%%%%%%%%%%%% Fig_34

%%%%%%%%%%%%%%%%%%%%%%%%%%%%%%%%%%%%%%%%%%%%%%%%%%%%%%%%%%%% Fig_35

In both figures the data are consistent with the calculations. Within this 
model, the enhancement in the LMR originates from the in-medium $\rho$ 
broadening, i.e. the thermal radiation of the hadronic phase, with a very 
small contribution from the QGP.

%%%%%%%%%%%%%%%%%%%%%%%%%%%%%%%%%%%%%%%%%%%%%%%%%%%%%%%%%%%% Fig_36
\begin{figure}[hbt!]
\includegraphics[width=1.0\linewidth]{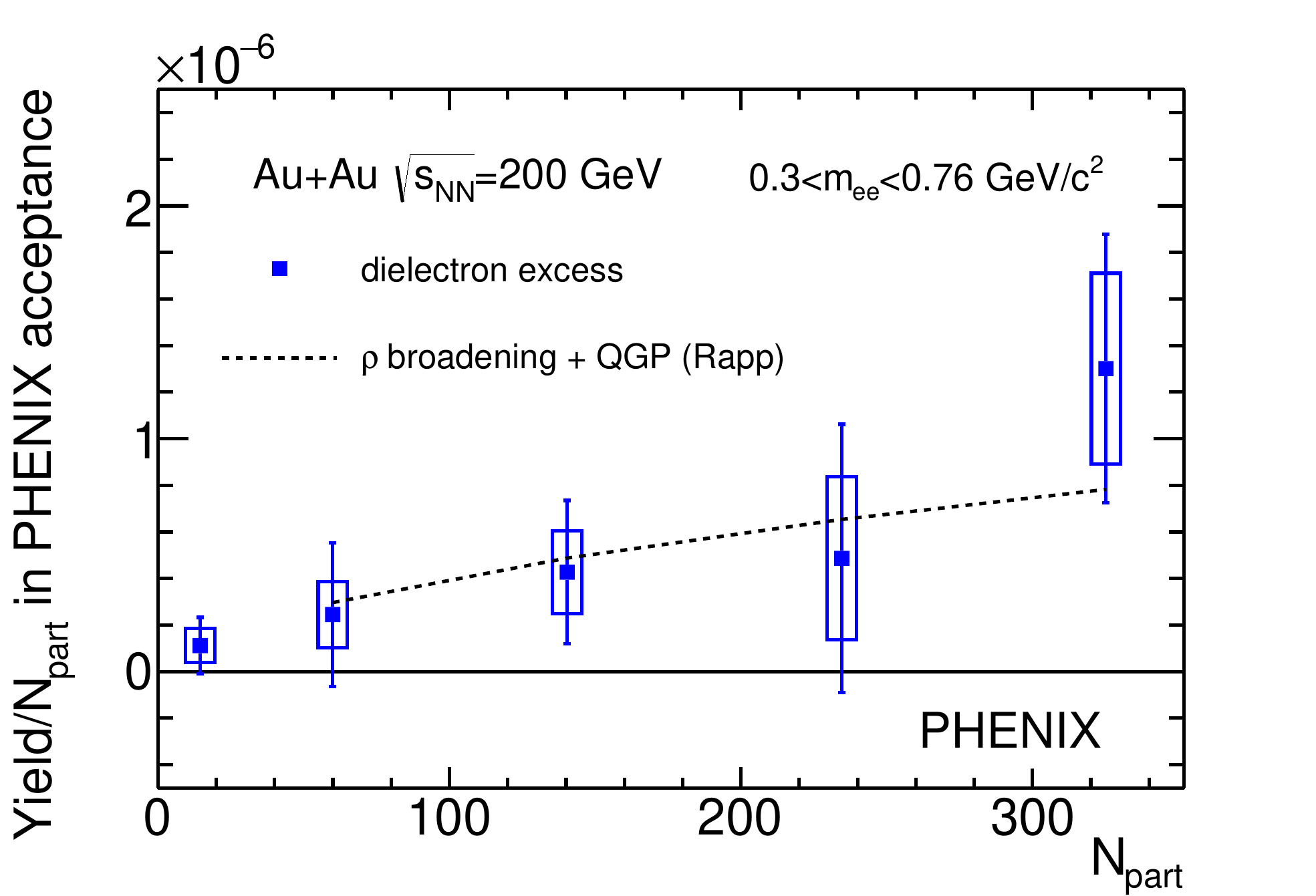}
\caption{(Color online) 
Centrality dependence of the dielectron excess, defined as (data 
$-$ cocktail excluding $\rho$)  compared to the thermal radiation from the 
hadronic ($\rho$ broadening) and QGP phases from model calculations 
(dashed line)~\cite{rapp}.}
\label{fig:integral_lmr_wrapp} 
\end{figure}

In the model, the centrality dependence of the thermal radiation is 
reasonably well described, within an uncertainty of $\sim$10\%, by a 
power-law scaling of the charged particle rapidity density 
($dN_{ch}/dy$)$^\alpha$, with $\alpha \simeq$ 1.45~\cite{Rapp:2013nxa}, 
very similar to the scaling of the thermal photon 
yield~\cite{Turbide:2003si, Adare:2014fwh}. Within uncertainties, the 
present data are consistent with this scaling as illustrated in Fig. 
\ref{fig:integral_lmr_wrapp}, which also shows the centrality dependence 
of the excess, i.e. the data after subtracting the cocktail without the 
vacuum $\rho$, together with the expected power-law scaling (dashed line).

\section{SUMMARY AND CONCLUSIONS}
     \label{sec:summary}   
       
PHENIX has measured invariant mass spectra, \pt distributions and the 
centrality dependence of the \ee pair production in \auau collisions at 
$\sqrt{s_{_{NN}}}$~=~200~GeV. The use of the HBD provided additional 
electron identification to the central arm detectors, additional hadron 
rejection and increased rejection of the combinatorial background.
 
A new analysis procedure based on neural networks has been developed that 
combines in an efficient way the information from the HBD and the central 
arm detectors, RICH, TOF and EMCal.  This results in three independent 
parameters for electron identification, hadron rejection and close pair 
rejection, instead of the fourteen parameters of the four detectors 
involved in these tasks. A quantitative understanding of the total 
background at the subpercent level is achieved in the most central 
collisions. This is realized by a precise evaluation of all the background 
sources. The combinatorial background is determined by the event mixing 
technique together with an exact weighting procedure to take into account 
the flow effects that are inherent in the foreground events and cannot be 
reproduced in the mixed events. All the correlated background sources are 
calculated in absolute terms using simulations and published results.

The results are compared with a cocktail of the known \ee sources. The 
contributions from light hadron decays that dominate the \ee yield at low 
masses \mee$<$ 1~GeV/$c^2$, are determined using PHENIX measurements for 
pions and $m_T$ scaling for other mesons. The contributions from 
semileptonic decays of heavy flavor (charm and bottom) mesons are 
calculated with the {\sc pythia} or {\sc mc@nlo} generators using 
$\langle$\Ncollnospace$\rangle$ scaled \pp cross sections. Both generators 
give very similar yields in the IMR. However, they predict very dissimilar 
results that differ from each other by a factor of $\sim$2 in the LMR.  
Precise measurements of the charm cross section over the entire phase 
space are needed to resolve this discrepancy.
 
A small enhancement of \ee is observed in the LMR with respect to the 
cocktail.  The enhancement is distributed over the entire \pt range 
measured (\pt$<$ 5~GeV/$c$). It increases with centrality and amounts to 
$2.3\pm0.4({\rm stat})\pm0.4({\rm syst})\pm0.2^{\rm model}$ for MB collisions when 
{\sc pythia} is used to calculate the open heavy flavor 
contribution. If instead {\sc mc@nlo} is used, the enhancement factors are 
$\sim$40\% smaller and for MB collisions it is found to be 
$1.7\pm0.3({\rm stat})\pm0.3({\rm syst})\pm0.2^{\rm model}$. The large enhancement 
of \ee pairs in the LMR previously reported by PHENIX, in \auau collisions 
at \sqnr~\cite{Adare:2009qk}, is not confirmed by the results of the 
present improved analysis. In particular, the concentration of the excess 
at low \pt (\pt $<$ 1~GeV/$c$) is not observed here.  The present results 
are consistent with those recently published by the STAR 
Collaboration~\cite{Adamczyk:2013caa} within the uncertainties of the two 
experiments.

In the IMR, the results are compared with calculations of the expected 
yield from the semileptonic decays of heavy flavor mesons in two extreme 
scenarios. In the first scenario, the heavy flavor contribution is 
calculated assuming that the correlations between the $c\bar{c}$ are the 
same in \auau as in \pp collisions, ignoring decorrelation effects 
produced by the interactions of heavy flavor quarks with the medium. A 
small enhancement is observed with respect to the yield predicted by {\sc 
pythia}. It amounts to $1.5\pm0.3({\rm stat})\pm0.2({\rm syst})\pm0.3^{\rm model}$ for 
MB collisions. In the other scenario, the opposite extreme approach is 
adopted where the pair is assumed to be totally decorrelated. In this 
case, the enhancement factor becomes 
$2.5\pm0.5({\rm stat})\pm0.3({\rm syst})\pm0.3^{\rm model}$. The reality is somewhere 
between these two extreme cases and we conclude that there is room in the 
data for a significant additional contribution, for example of thermal 
radiation, in the IMR. The nature of the IMR pairs will be studied with 
high statistics Au$+$Au data in 2014 data taking with the silicon vertex 
tracker (VTX) installed in PHENIX.

The results in the LMR are compared to calculations based on the model 
originally developed by Rapp and Wambach~\cite{Rapp:1999us,Rapp:2000pe} 
with subsequent improvements that incorporate recent 
developments~\cite{Rapp:2013nxa}. The model includes thermal radiation 
emission from the QGP phase ($q\bar{q}$ annihilation)  as well as from the 
hadronic phase (mainly from the $\rho$ meson copiously produced by pion 
annihilation, $\pi^+ \pi^- \rightarrow \rho \rightarrow e^+e^-$). The 
invariant mass and \pt distributions as well as the centrality dependence 
are well reproduced by the calculations. The enhancement observed in the 
LMR from SPS up to RHIC energies is thus consistently reproduced by a 
single model.   Within this model, the enhancement originates from the 
melting of the $\rho$ meson resonance as the system approaches chiral 
symmetry restoration.

\section*{ACKNOWLEDGMENTS}  

% Run-14 long form for all journals

We thank the staff of the Collider-Accelerator and Physics
Departments at Brookhaven National Laboratory and the staff of
the other PHENIX participating institutions for their vital
contributions.   We also thank R. Rapp for providing us the 
results of his model calculations and for helpful discussions.  
We acknowledge support from the
Office of Nuclear Physics in the
Office of Science of the Department of Energy,
the National Science Foundation,
Abilene Christian University Research Council,
Research Foundation of SUNY, and
Dean of the College of Arts and Sciences, Vanderbilt University
(U.S.A),
Ministry of Education, Culture, Sports, Science, and Technology
and the Japan Society for the Promotion of Science (Japan),
Conselho Nacional de Desenvolvimento Cient\'{\i}fico e
Tecnol{\'o}gico and Funda\c c{\~a}o de Amparo {\`a} Pesquisa do
Estado de S{\~a}o Paulo (Brazil),
Natural Science Foundation of China (People's Republic of~China),
Croatian Science Foundation and Ministry of Science, Education, 
and Sports (Croatia),
Ministry of Education, Youth and Sports (Czech Republic),
Centre National de la Recherche Scientifique, Commissariat
{\`a} l'{\'E}nergie Atomique, and Institut National de Physique
Nucl{\'e}aire et de Physique des Particules (France),
Bundesministerium f\"ur Bildung und Forschung, Deutscher
Akademischer Austausch Dienst, and Alexander von Humboldt Stiftung 
(Germany),
National Science Fund, OTKA, K\'aroly R\'obert University College,
and the Ch. Simonyi Fund (Hungary),
Department of Atomic Energy and Department of Science and Technology 
(India),
Israel Science Foundation (Israel),
Basic Science Research Program through NRF of the Ministry of Education 
(Korea),
Physics Department, Lahore University of Management Sciences (Pakistan),
Ministry of Education and Science, Russian Academy of Sciences,
Federal Agency of Atomic Energy (Russia),
VR and Wallenberg Foundation (Sweden),
the U.S. Civilian Research and Development Foundation for the
Independent States of the Former Soviet Union,
the Hungarian American Enterprise Scholarship Fund,
and the US-Israel Binational Science Foundation.

%\clearpage

\appendix

\section{Introducing Flow in the Mixed Events}
\label{app:flow}

In this section, we analytically derive the weighting factor introduced in 
Eq. (\ref{eq:v2_weight}). We start from the azimuthal distribution of a 
particle that follows the expression:
\begin{equation}
P(\phi-\Psi) = \epsilon(\phi) (1 + 2 v_{2} \cos 2 (\phi - \Psi))
\label{eqn:app_v2_single}
\end{equation}
where $\phi$ is the azimuthal angle of the particle, $\Psi$ is the 
reaction plane azimuthal angle of the event and $\epsilon(\phi)$ is the 
detection efficiency of the spectrometer at $\phi$.

The $\Delta \phi$ distribution of any two particles in the same event 
(foreground pairs) can be calculated as:

\begin{eqnarray}
\lefteqn{P_{FG}(\Delta \phi)} \\ 
&=& \frac{1}{\pi}\int^{\pi/2}_{-\pi/2}{\rm d}\Psi \int_{\phi_{1}-\phi_{2}=\Delta \phi}{\rm d}\phi_{1} {\rm d}\phi_{2} P (\phi_{1}-\Psi) P(\phi_{2}-\Psi) \nonumber \\
&=& \frac{1}{\pi}\int^{\pi/2}_{-\pi/2}{\rm d}\Psi \int^{\pi}_{-\pi}{\rm d}\phi_{1}  P (\phi_{1}-\Psi) P(\phi_{1}+\Delta \phi -\Psi) \nonumber
\end{eqnarray}

Replacing $P(\phi-\Psi)$ by its expression in (\ref{eqn:app_v2_single}) 
allows one to write $P_{FG}$ as the sum of four integrals:

\begin{eqnarray}
P_{FG}(\Delta \phi) = \frac{1}{\pi}\int^{\pi/2}_{-\pi/2}{\rm d}\Psi \int^{\pi}_{-\pi}{\rm d}\phi_{1}  (A+B+C+D) 
\end{eqnarray}

\begin{eqnarray}
A=  \epsilon(\phi_{1})\epsilon(\phi_1+\Delta\phi)
\end{eqnarray}
\begin{eqnarray}
B=2v_2\epsilon(\phi_1)\epsilon(\phi_1+\Delta\phi)\cos2(\phi_1-\Psi)
\end{eqnarray}
\begin{eqnarray}
C=2v_2\epsilon(\phi_1)\epsilon(\phi_1+\Delta\phi)\cos2(\phi_1+\Delta\phi-\Psi)
\end{eqnarray}

\begin{eqnarray}
D&=&4v_{2}v_{2}\epsilon(\phi_{1})\epsilon(\phi_1+\Delta\phi)(\cos 2 
(\phi_{1} - \Psi)) \\
&& \times (\cos 2 (\phi_{1} + \Delta \phi- \Psi) \nonumber
\end{eqnarray}

It is easy to show that the integrals of $B$ and $C$ are equal to 0 and 
the integral of $D$ leads to:

\begin{eqnarray}
 \frac{1}{\pi}\int^{\pi/2}_{-\pi/2}{\rm d}\Psi \int^{\pi}_{-\pi}{\rm d}\phi_{1}  
D&=&2v_2v_2\cos2\Delta\phi \\
&& \times \int_{-\pi}^{\pi}\epsilon(\phi_1)\epsilon(\phi_1+\Delta\phi) \nonumber
\end{eqnarray}

Therefore,
\begin{eqnarray}
P_{FG}(\Delta\phi) &=& \Bigl(\int^{\pi}_{-\pi}{\rm d}\phi_1 \epsilon(\phi_{1})\epsilon(\phi_1+\Delta\phi)\Bigr) \\
&& \times (1+2v_{2}v_{2}\cos 2 \Delta \phi) \nonumber
\label{eqn:pfg_dist}
\end{eqnarray}

In a similar way one can calculate the $\Delta\phi$ distribution of mixed 
BG pairs produced without reaction plane binning:

\begin{eqnarray}
\lefteqn{P_{MIX}(\Delta \phi)} \\ 
&=&\frac{1}{\pi^{2}} \int^{\pi/2}_{-\pi/2}{\rm d}\Psi_1 \int^{\pi/2}_{-\pi/2}{\rm d}\Psi_2 \int_{\phi_{1}-\phi_{2}+\Delta\phi} \nonumber \\
&& \times {\rm d}\phi_1{\rm d}\phi_2 P(\phi_1-\Psi_1)P(\phi_2-\Psi_2) \nonumber 
\end{eqnarray}
where $\phi_{1(2)}$ and $\Psi_{1(2)}$ represents the azimuthal angle of 
particle 1(2) and the reaction plane azimuthal angle of the events from 
which the particles are taken. Replacing $P(\phi-\Psi)$ by 
(\ref{eqn:app_v2_single}):

\begin{eqnarray}
\lefteqn{P_{MIX}(\Delta \phi)} \\ 
&=&\frac{1}{\pi^{2}} \int^{\pi/2}_{-\pi/2}{\rm d}\Psi_1 \int^{\pi/2}_{-\pi/2}{\rm d}\Psi_2 \int_{\phi_{1}-\phi_{2}+\Delta\phi} \nonumber \\
&& \times {\rm d}\phi_1{\rm d}\phi_2 (E+F+G+H) \nonumber
\end{eqnarray}
\begin{eqnarray}
E=  \epsilon(\phi_{1})\epsilon(\phi_1+\Delta\phi)
\end{eqnarray}
\begin{eqnarray}
F= 2v_{2}\epsilon(\phi_{1})\epsilon(\phi_1+\Delta\phi)\cos 2 (\phi_{1} - \Psi_1)
\end{eqnarray}

\begin{eqnarray}
G=2v_{2}\epsilon(\phi_{1})\epsilon(\phi_1+\Delta\phi)\cos 2 (\phi_{1} +\Delta\phi- \Psi_2)
\end{eqnarray}
\begin{eqnarray}
H&=& 4v_{2}v_{2}\epsilon(\phi_{1})\epsilon(\phi_1+\Delta\phi)\cos 2 (\phi_{1} - \Psi_1) \\
&\ & \times \cos 2 (\phi_{1} +\Delta\phi - \Psi_2 ) \nonumber 
\end{eqnarray}

Because $F$, $G$ and $H$ are again easily proved to be 0, 
$P_{MIX}(\Delta\phi)$ can now be written as:

\begin{equation}
\label{eqn:pbg_dist}
P_{MIX}(\Delta\phi) = \int^{\pi}_{-\pi}{\rm d}\phi_1 \epsilon(\phi_{1})\epsilon(\phi_1+\Delta\phi)
\end{equation}

The weighting factor to introduce the flow correlation into the mixed BG 
pairs is then given by the ratio between Eq. (\ref{eqn:pfg_dist}) and Eq. 
(\ref{eqn:pbg_dist}):

\begin{eqnarray}
w(\Delta \phi) &=& \frac{P_{FG}(\Delta \phi)}{P_{MIX}(\Delta \phi)} \\
&=& 1+2v_{2}v_{2}\cos 2 \Delta \phi \nonumber
\end{eqnarray}

\section{Violation of $CB_{+-}=2\sqrt{CB_{++}CB_{--}}$ due to flow}
\label{app:sqrt_relation}

In this appendix, we demonstrate that the combination of elliptic flow and 
nonuniform detection efficiency violates the well-known relation between 
unlike-sign and like-sign combinatorial background:

\begin{equation}
\label{eqn:app_2sqrt}
\langle CB_{+-}\rangle=2\sqrt{\langle CB_{++}\rangle\langle CB_{--}\rangle}
\end{equation}
where $\langle CB_{+-/++/--}\rangle $ are the unlike-sign and like-sign 
integral yields or average numbers of pairs per event.

We start from the case without elliptic flow.  Then, as proven in 
Ref~\cite{Adare:2009qk}, if $e^{+}$ and $e^-$ are always produced in pairs 
independent of each other, the average number of unlike-sign and like-sign 
combinatorial pairs can be calculated as:
\begin{eqnarray}
\label{eqn:app_bg12}
\langle CB_{+-}\rangle &=& [\varepsilon_p+\varepsilon_+(1-\varepsilon_p)][\varepsilon_p+\varepsilon_-(1-\varepsilon_p)] \\ 
&\ & \times (\langle N^2\rangle-\langle N\rangle) \nonumber \\
\label{eqn:app_bg11}
\langle CB_{++}\rangle &=& \frac{1}{2}[\varepsilon_p+\varepsilon_+(1-\varepsilon_p)]^2 (\langle N^2\rangle-\langle N\rangle) \\
\label{eqn:app_bg22}
\langle CB_{--}\rangle &=& \frac{1}{2}[\varepsilon_p+\varepsilon_-(1-\varepsilon_p)]^2 (\langle N^2\rangle-\langle N\rangle) 
\end{eqnarray}
where $\varepsilon_p$ is the probability to reconstruct both tracks of a 
pair, $\varepsilon_{+/-}$ is the probability to reconstruct only a single 
track and $N$ is the number of pairs in an event.

If $\varepsilon_{p/+/-}$ are assumed to be constants, Eq. 
(\ref{eqn:app_2sqrt}) can easily be proven from Eqs. 
(\ref{eqn:app_bg12}-\ref{eqn:app_bg22}). However, in the presence of 
elliptic flow, the probabilities $\varepsilon_{p/+/-}$ depend on the 
reaction plane angle:
\begin{equation}
\label{eqn:reactiondep_eff}
\varepsilon_{p/+/-} (\psi)= \int {\rm d}\phi\ \varepsilon_{p/+/-}(\phi) (1+2v_2\cos(\phi - \psi))
\end{equation}

\begin{eqnarray}
\langle CB_{+-}(\psi) \rangle &=& [A(\psi)B(\psi)]\times (\langle N^2\rangle-\langle N \rangle) \\
\langle CB_{++}(\psi) \rangle &=& \frac{1}{2}[A(\psi)]^2\times (\langle N^2\rangle-\langle N \rangle) \\
\langle CB_{--}(\psi) \rangle &=& \frac{1}{2}[B(\psi)]^2\times (\langle N^2\rangle-\langle N \rangle) \\
A(\psi)&=&\varepsilon_p(\psi)+\varepsilon_+(\psi)(1-\varepsilon_p(\psi))  \\
B(\psi)&=&\varepsilon_p(\psi)+\varepsilon_-(\psi)(1-\varepsilon_p(\psi)) 
\end{eqnarray}
Taking the average over $\psi$ within [$-\frac{\pi}{2}, \frac{\pi}{2}$] 
gives:
\begin{eqnarray}
\langle CB_{+-}\rangle   &=&  (\langle N^2\rangle-\langle N \rangle) \int {\rm d}\psi\ A(\psi)B(\psi) \\
\langle CB_{++}\rangle  &=& \frac{1}{2} (\langle N^2\rangle-\langle N \rangle) \int {\rm d}\psi\ A(\psi)^2 \\
\langle CB_{--}\rangle   &=& \frac{1}{2} (\langle N^2\rangle-\langle N \rangle) \int {\rm d}\psi\ B(\psi)^2 
\end{eqnarray}
Using the Cauchy-Schwarz inequality, one obtains:
\begin{eqnarray} 
\left[\int {\rm d}\psi\ A(\psi)B(\psi)\right]^2 &\leq&  \int {\rm d}\psi\ A(\psi)^2 \\
&& \cdot \int {\rm d}\psi\ B(\psi)^2 \nonumber 
\end{eqnarray}
and consequently, 
\begin{equation}
\langle CB_{+-}\rangle \leq 2\sqrt{\langle CB_{++}\rangle\langle CB_{--}\rangle}
\end{equation} 
\section{A second, independent analysis}
\label{app:sb_analysis}

A subset of the data, $4.8\times10^9$ MB events, was analyzed by 
a second independent team. The second analysis follows the analysis 
strategy presented in Ref.~\cite{Adare:2009qk}, but includes the 
information provided by the HBD and other important improvements developed 
in this work.

In this appendix we present the key features of the second analysis with 
an emphasis on the most important differences to the main analysis: (i) 
the HBD underlying event subtraction and cluster algorithm, (ii) the 
electron identification cuts and (iii) the background normalization. All 
analysis steps not explicitly mentioned are identical between the two 
analyses. In particular, identical cuts on the acceptance and inactive 
detector areas, and the same pair cuts are applied.  At the end of this 
appendix we discuss the efficiency correction and compare the results of 
both analyses.

The net number of photo electrons in an HBD cluster was calculated 
with a different algorithm than discussed in Section \ref{sec:hbd}, 
using a local estimate of the scintillation background rather than a 
module average. As an electron typically fires three HBD readout 
cells, 3-cell triplets are used to initiate the cluster search. All 
possible triplets are formed. The photo-electron background due to 
scintillation light is estimated by the median amplitude in the first 
and second neighboring cells around the triplet. The background 
subtracted triplet charge is calculated as:
\begin{equation}
\label{eq:hbd_lbs_triplet_charge}
q_{net} = q_t - A_t \times \frac{\langle q_{fn}\rangle + \langle q_{sn}\rangle}{2}
\end{equation}
where $q_t$ is the total charge in the triplet, $A_t$ the number of cells 
with charge in the triplet, and $\langle q_{fn}\rangle$, $\langle 
q_{sn}\rangle$ are the median charge in the first and second neighboring 
cells, respectively. Only triplets with $0<q_{net}<60$ p.e. are recorded.

Electron candidates are projected to the HBD, and triplets within 1.5~cm 
of the track are merged to form a cluster. The net charge of the cluster 
$q_r$ is calculated starting from the sum of the charge of all cells in 
the cluster:
\begin{equation}
\label{eq:hbd_lbs_cluster_charge}
q_r = q_{totclust} - A_{clust} \times \frac{\langle q_{fn} \rangle + \langle q_{sn} \rangle}{2}
\end{equation} 
where $q_{totclust}$ is the sum of the charge of all cells in the cluster, 
$A_{clust}$ is the number of cells in the cluster, $\langle q_{fn} 
\rangle$, $\langle q_{sn} \rangle$ are again the median charge per cell in 
the first and second neighbors but now around the cluster.

This analysis uses a number of sequential one-dimensional cuts to identify 
electrons. The variables used for the electron identification are defined 
in Section~\ref{sec:eid_variables}. The following cuts are used:

\begin{itemize}
\item n0 $ >$ 2: The exclusion of RICH photo-multipliers fired by 
background electrons (Section~\ref{sec:tagging_rich}) is not used in this 
analysis.
\item disp $<$ 5.5 cm  
\item chi2/npe0$\bf <$ 20  
\item emcsdr $<$ 3 
\item $|$dep$|$ $<$ 2 
\item $m^2_{\rm TOF}<$1.5$\sigma$: Calculated based on the 
time-of-flight measured by either the EMCal or the TOF-E detectors.
\item 10 $<$ $q_r$ $<$ 40 p.e.: Cluster charge as defined in Eq. 
(\ref{eq:hbd_lbs_cluster_charge})
\end{itemize}

With these cuts, a purity of the electron sample of 86\% is achieved for 
the most central bin, which quickly increases to above 99\% for the most 
peripheral collisions.

The combinatorial background is calculated by event mixing. We use the 
method outlined in~\cite{Adare:2009qk}, but included the weighting for the 
azimuthal anisotropy as implemented in the main analysis and described in 
Section \ref{sec:combinatorial_background}. For the correlated background 
both analyses use the same MC simulations. For cross-pairs and jet-pairs 
the simulated pairs were reanalyzed with the track selection cuts and HBD 
cluster algorithm mentioned above. The shapes of the mass spectra are 
consistent within systematic uncertainties for the two analysis methods. 
For the electron-hadron and $B\bar{B}$ contributions the simulated pairs 
were not reanalyzed.

%%%%%%%%%%%%%%%%%%%%%%%%%%%%%%%%%%%%%%%%%%%%%%%%%%%%%%%%%%%% Fig_37
\begin{figure}[htb!]
\includegraphics[width=1.0\linewidth]{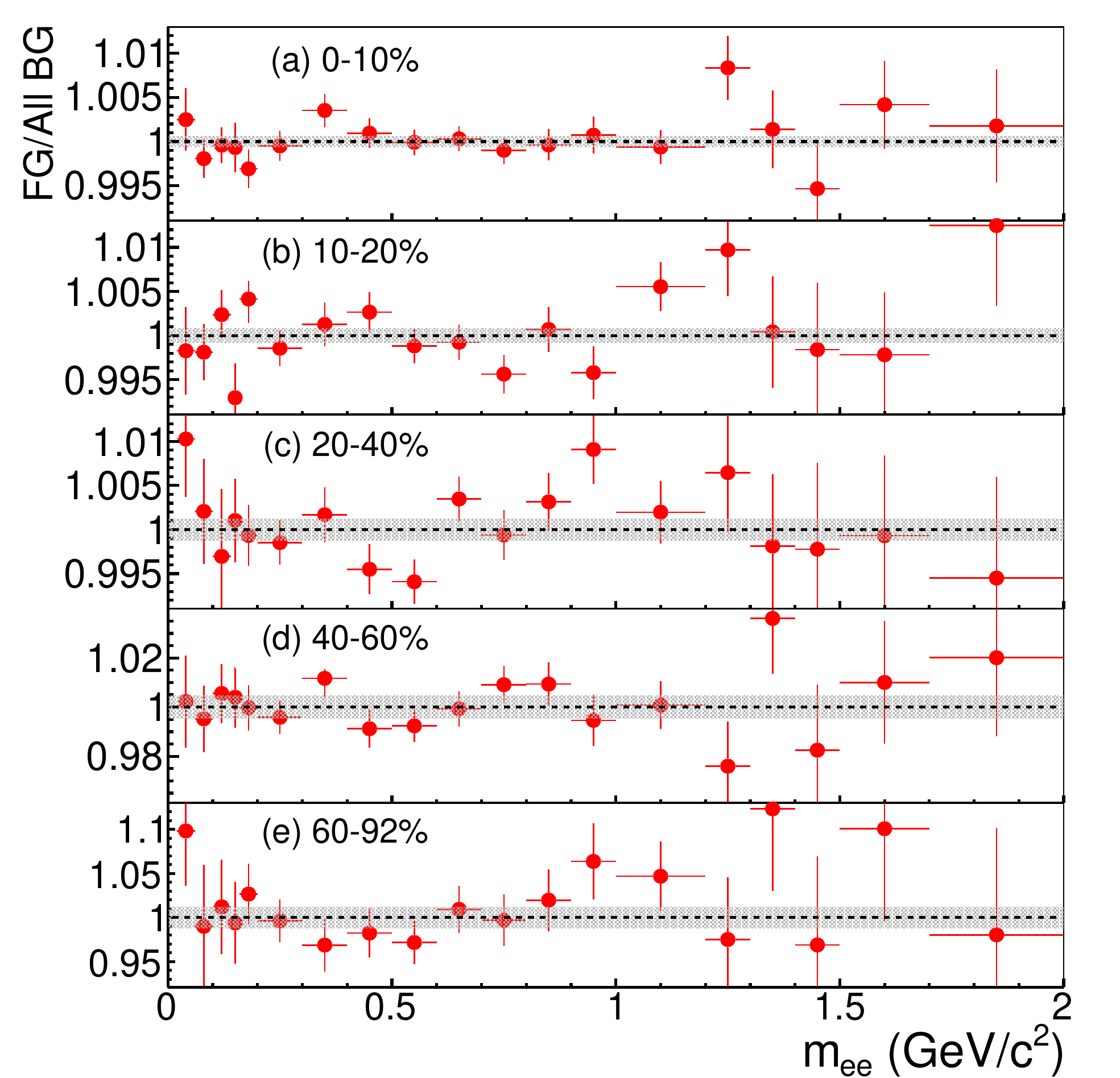}
\caption{(Color online) The ratio of the foreground like-sign pairs 
to the sum of combinatorial and correlated pair sources in centrality 
bins 0\%--10\%, 10\%--20\%, 20\%--40\%, 40\%--60\% and 60\%--92\%.}
\label{fig:sb_likesign_ratio}
\end{figure}

%%%%%%%%%%%%%%%%%%%%%%%%%%%%%%%%%%%%%%%%%%%%%%%%%%%%%%%%%%%% Fig_38
\begin{figure}[hbt!]
\includegraphics[width=1.0\linewidth]{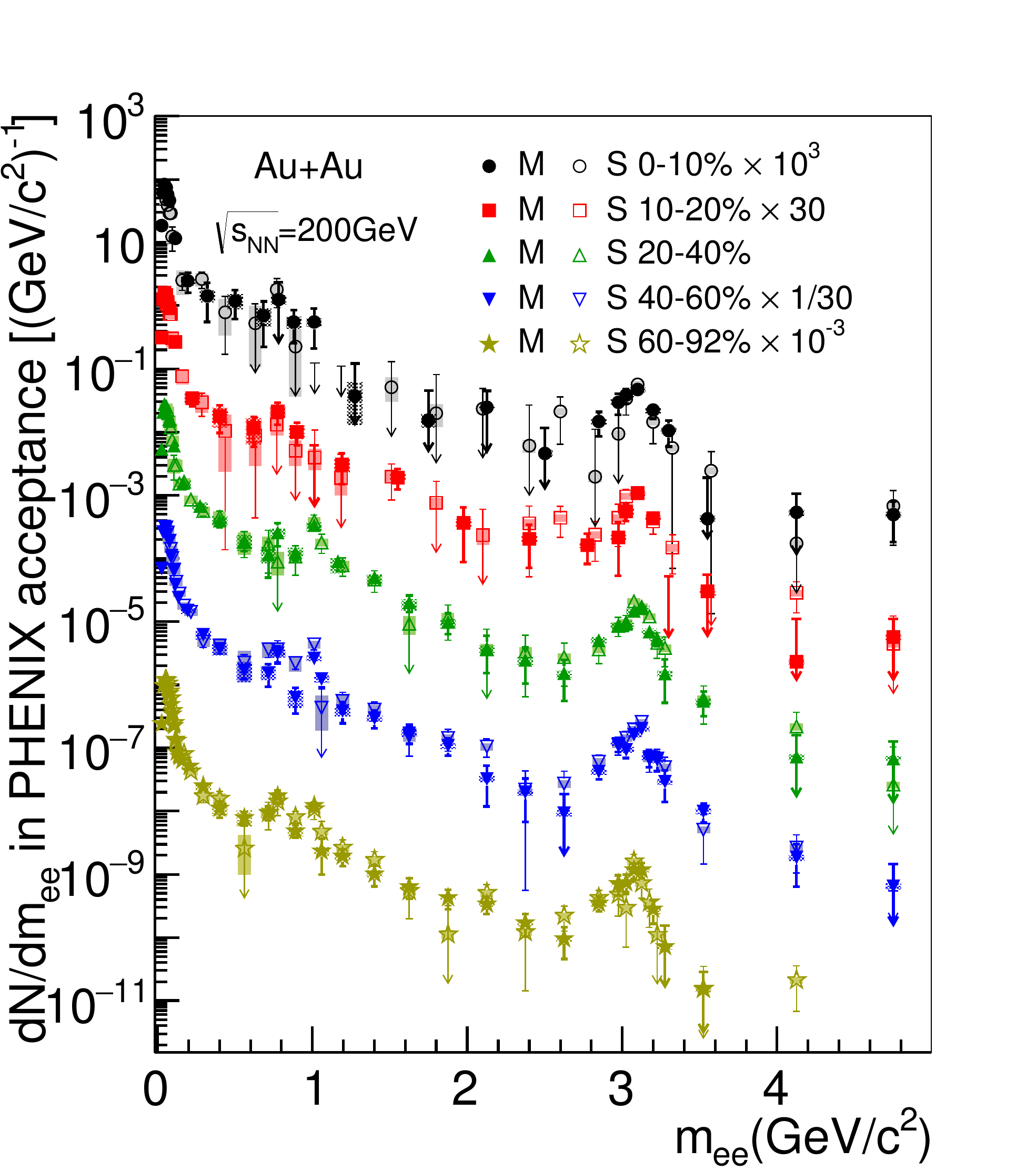}
\caption{(Color online) 
Comparison of final spectra from the main (M) and second (S) analyses.}
\label{fig:compare_spectra}
\end{figure}

The normalizations of all the background components were fitted 
simultaneously to the full mass and \pt range of the like-sign spectra:
\begin{eqnarray}
FG_{++--} &=& a_0 BG_{++--} + a_1 CP_{++--}  \\ 
   & & \ \  + a_2 JP^{\rm same}_{++--} + a_3 JP^{opposite}_{++--} \nonumber \\
 & &\ \ \ \ \ \        + a_4 EH_{++--} + a_5 BB_{++--} \nonumber
\label{eq:bkgnorm}
\end{eqnarray}
The parameters $a_i$ are the individual normalization constants. 
Figure~\ref{fig:sb_likesign_ratio} shows the like-sign foreground divided 
by the sum of all background sources for the five centrality classes. The 
uncertainty on the combinatorial background normalization is shown as a 
gray band on each panel.  No systematic deviation from unity 
is observed, indicating that the sum of the different background 
components gives a sufficiently accurate description over the mass range 
up to 2~GeV/$c^2$ with no indication of any shape variation within the 
shown uncertainties. Above 2~GeV/$c^2$ the statistical significance makes 
a comparison at the shown scale meaningless.

After fixing the normalization of all background sources so that a 
satisfactory description of the like-sign pairs is achieved, the analysis 
is extended to unlike-sign pairs. The normalizations for the unlike-sign 
cross-pairs, jet-pairs and electron-hadron pairs are taken from Eq. 
(\ref{eq:bkgnorm}). For the combinatorial unlike-sign pairs we use 
unlike-sign mixed event pairs. The normalization is also taken from Eq. 
(\ref{eq:bkgnorm}), but needs to be corrected to account for the different 
effect of the pair cuts on like- and unlike-sign pairs as done in 
Ref.~\cite{Adare:2009qk}.

To estimate the uncertainty on the raw yield due to the background 
subtraction one needs to consider the signal-to-background ratio $S/B$.  
The uncertainties on the $a_i$ are multiplied by $B/S$ and added in 
quadrature. This results in $\sim 55\%$ systematic uncertainties at 
0.6~GeV/$c^2$ for MB collisions.

We factorize the efficiency into 3 terms, which are determined separately. 
\begin{equation}
\label{eqn:sb_effcor}
\epsilon^{total}_{\rm pair} = \epsilon^{}_{\rm pair}\cdot\epsilon^{\rm TOF}_{\rm pair}\cdot\epsilon^{embed}_{\rm pair}
\end{equation}
The first factor describes the effect of all reconstruction algorithms and 
cuts except for the time-of-flight cut and the centrality dependence of 
the reconstruction efficiency in the central arms, which are treated 
separately. It is obtained by a MC simulation of $e^+e^-$ pairs 
that are processed through the full PHENIX detector simulation, including 
the HBD. The simulated HBD hits are embedded into real HBD data as 
discussed in Section~\ref{sec:raw_spectra}. These events are then analyzed 
with the same electron identification, fiducial, and pair cuts used in the 
independent analysis, with exception of the time-of-flight cut. The 
systematic uncertainty of $\epsilon_{\rm pair}$ is about 12\%. It was 
determined from the measured yield of pairs in the $\pi^0$ Dalitz decay 
region when varying electron identification cuts in a way that changes the 
raw pair yields by factors between 0.5 and 1.5.

The efficiency $\epsilon^{\rm TOF}_{\rm pair}$ is determined from tracks 
measured in peripheral collisions, where the hadron contamination is 
negligible, by comparing data obtained with a 1.5 $\sigma$ cut to the case 
with no time-of-flight cut. We find that on average the TOF efficiency for 
tracks is 93\% above 0.4~GeV/$c$, but drops to 80\% at 0.2~GeV/$c$ 
independent of centrality. This drop results from a failure of the 
electronics to properly record time for low amplitude signals. In the main 
analysis this issue was avoided by treating tracks with no time 
information separately. The systematic uncertainty due to this cut is a 
few percent at 0.6~GeV/$c^2$.

The efficiency $\epsilon^{embed}_{\rm pair}$ was determined by embedding 
MC-simulation tracks into the data of all used central arm detectors and 
analyzing these embedded tracks using the same cuts as used in the data. 
The values are found to be very similar to those derived in the main 
analysis. For central collisions an additional 8\% systematic uncertainty 
is added.

Compared to the main analysis, the total reconstruction efficiency 
$\epsilon^{total}_{\rm pair}$ is a factor of $\sim$2 smaller for central 
collisions. The difference drops to $\sim$30\% for the most peripheral 
collisions.

The fully corrected mass spectra from the independent analysis are 
compared to those from the main analysis in Fig. \ref{fig:compare_spectra} 
for all five centrality bins. The results are consistent within 
uncertainties.
 
%\clearpage

%\bibliography{ppg177x1}

 
\end{document}